\documentclass[aps,prb,floats]{revtex4}
\usepackage{color,ulem,verbatim,float,wrapfig}
\usepackage{amssymb,amsbsy,amsmath,mathrsfs}
\usepackage{graphicx}
\usepackage{times}
\usepackage{color}
\usepackage{subfigure}
\usepackage{braket}
\usepackage{bbold}
\usepackage{commath}
\usepackage{setspace}
\usepackage{bm} 
\newcommand\bea{\begin{eqnarray}}
\newcommand\eea{\end{eqnarray}}
\newcommand\beq{\begin{equation}}
\newcommand\eeq{\end{equation}}

\newcommand{\non}{\nonumber}
\newcommand{\al}{\alpha}

\newcommand{\si}{\sigma}

\newcommand{\om}{\omega}

\usepackage[colorlinks = true,
linkcolor = blue,
urlcolor = blue,
citecolor = blue,
anchorcolor = blue]{hyperref}
\usepackage{soul}
\usepackage[dvipsnames]{xcolor}
\begin{document}

\title{Effects of topological and non-topological edge states on 
information propagation and scrambling in a Floquet spin chain}
\author{Samudra Sur$^1$ and Diptiman Sen$^{1,2,}
\footnote{Author to whom any correspondence should be addressed.
~~E-mail: diptiman@iisc.ac.in}$}
\affiliation{$^1$Center for High Energy Physics, Indian Institute of Science, Bengaluru 560012, India \\
$^2$Department of Physics, Indian Institute of Science, Bengaluru 560012, India}

\begin{abstract}
The action of any local operator on a quantum system propagates through the system carrying the information of the operator. This is usually 
studied via 
the out-of-time-order correlator (OTOC). We 
numerically study the information propagation from one end of a periodically driven spin-1/2 $XY$ chain with open boundary conditions using the Floquet 
infinite-temperature OTOC. We calculate the OTOC for two different spin operators,
$\si^x$ and $\si^z$.
For sinusoidal driving, the model can be shown to host different types of
edge states, namely, topological (Majorana) edge states 
and non-topological edge states. 
We observe a localization 
of information at the edge for both $\sigma^z$ and $\sigma^x$ OTOCs whenever 
edge states are present. In addition, in the case of non-topological edge states, we see oscillations of the OTOC in time near the edge, the oscillation period being 
inversely proportional to the gap between the Floquet eigenvalues of the edge states. We provide an analytical understanding of these effects due to the edge
states. 
It was known earlier that the OTOC for the spin operator which is local in terms of Jordan-Wigner fermions
($\si^z$) shows no signature of information scrambling inside the light cone of propagation, while the OTOC for the spin operator which is non-local in 
terms of Jordan-Wigner fermions ($\si^x$) shows signatures of scrambling.
We report a remarkable `unscrambling effect' in the $\sigma^x$ OTOC after reflections from the ends of the system. Finally, we demonstrate that the 
information propagates into the system mainly via the bulk states with the 
maximum value of the group velocity, and we show 
how this velocity is controlled by the driving frequency and amplitude.
\end{abstract}

\maketitle

\section{Introduction}
\label{sec1}

Out-of-time ordered correlators (OTOC) have been a topic of growing interest in recent times and it has been studied extensively both theoretically~\cite{Roberts,Luitz,Wei,iyoda2018,von2018,niknam2020,shukla2022,alavirad2019,hosur2016,swingle2017,schleier2017,pappalardi2018,klug2018,khemani2018,maldacena2016,shenker2014,kitaev2018,fan2017,huang2017out} and experimentally~\cite{Swingle,Zhu,Garttner2017,Li,Lewis,Nie,Braumuller2022,zhao2022}.
The OTOC was first introduced by Larkin and Ovchinnikov in a semiclassical 
description to understand superconductivity \cite{larkin1969}. It is now widely 
used as a 
measure for understanding quantum chaos and spreading of local quantum information in both quantum many-body systems and black holes. The idea of chaos is best understood in classical physics as the exponential deviation of the trajectory of 
a particle for an infinitesimally small difference of the initial condition, also known as the "butterfly effect". This deviation is measured using $\partial x(t) /\partial x(0) $, where $x(t)$ is the position of the particle. This can be re-written as the Poisson bracket $\{x(t),p\}$, $p$ being the momentum. In the case 
of quantum many-body systems, the pertinent question to ask is how the local quantum information, like action of a local operator $O$ in a state, spreads through the 
system. Borrowing the idea from classical chaos, this information spreading is measured using the 
expectation value of the squared commutator, $\langle [O(t), O]^{2} \rangle$, or equivalently the OTOC, $\langle O(t) O O(t) O\rangle$, which are linearly related to 
each other. In brief, the OTOC is used to quantify the delocalization of local information, known as scrambling in a quantum system. 

Quantum information scrambling has been studied extensively in various systems ranging from the 
interior of a black hole~\cite{maldacena2016,maldacena2017} and the Sachdev-Ye-Kitaev~\cite{maldacena2016remarks,sachdev1993gapless} model by 
the high energy physics community to various models in condensed matter physics~\cite{patel2017quantum1,patel2017quantum,dora2017out,shen2017out,heyl2018detecting,sahu2019scrambling,lin2018out,mcginley2019slow,daug2020topologically}, as well 
as in statistical physics~\cite{campisi2017thermodynamics,chenu2018quantum}. 
To name a few, the OTOC has been studied in fermionic systems with a critical Fermi surface~\cite{patel2017quantum}, Luttinger liquids~\cite{dora2017out}, weakly 
diffusive metals~\cite{patel2017quantum1}, Ising and $XY$ spin models~\cite{heyl2018detecting,bao2020out,lin2018out,odell2019,daug2020topologically}, the 
symmetric Kitaev chain~\cite{mcginley2019slow}, and many-body localized systems~\cite{riddell2019out,lee2019typical,fan2017,huang2017out}. Moreover, it has been 
proposed recently that the OTOC can be used to dynamically detect phase transitions, such as many-body localization transitions~\cite{huang2017out}, equilibrium phase 
transitions~\cite{heyl2018detecting,dag2019detection,wei2019dynamical}, and 
dynamical quantum phase transitions in quenched systems~\cite{Nie,heyl2018detecting}.

However, OTOC has been studied relatively less in Floquet systems~\cite{shukla2021out,shukla2022characteristic,zamani2022out,nizami2020quantum}, which has also been a topic of intense research for several years. Periodic 
driving of quantum systems can give rise to many interesting effects such as 
the generation of effective models with novel properties~\cite{kitamura2017probing,claassen2017dynamical,sriram2022light,sur2022driven,decker2020floquet}, non-trivial band structures and edge states~\cite{rudner2013anomalous,nathan2015topological,kitagawa2010topological,kitagawa2011transport,lindner2011floquet,liu,kundu2013transport,kundu2014effective,thakurathi2013floquet,thakurathi2014majorana,deb2017generating,saha2017generating,balabanov,yates2017,molignini,lili,muller,seshadri2019generating,seshadri2022engineering,sur2021floquet}, time crystals~\cite{else2016floquet,zeng2017prethermal,zhang2017observation,russomanno2017floquet,surace2019floquet,yarloo2020homogeneous,lazarides2020time,natsheh2021critical}, Floquet dynamical quantum phase transition~\cite{yang2019floquet,zamani2020floquet,jafari2021floquet,jafari2022floquet,asboth2012symmetries,asboth2013bulk,naji2022dissipative}, and dynamical freezing~\cite{dunlap1986dynamic,das2010exotic,bhattacharyya2012transverse,nag2015maximum,mondal2013dynamics,divakaran2014dynamic,agarwala2016effects,agarwala2017effects,iubini2019dynamical,mukherjee2020collapse,haldar2021dynamical}. Recently, it has been shown that the time-averaged OTOC can be used to detect dynamical phase transitions in Floquet $XY$ spin chains and synchronized Floquet $XY$ spin 
chains~\cite{zamani2022out}. Furthermore, the OTOC has been studied in the
transverse field Ising model with a periodic binary drive to obtain the
dynamical phase structure of the model~\cite{shukla2021out}.

In this work, we study the effects of edge states on information propagation and scrambling from one end of a $XY$ spin chain which is subjected to a sinusoidal 
drive of the transverse field with a driving frequency $\om$. (To the best of our
knowledge, OTOCs for a periodically driven system with open boundary conditions 
and edge states have not been studied earlier). For an undriven model, if the 
exchange couplings for the $x$ and $y$ components of the spin, $J_x$ and 
$J_y$, are unequal, and the transverse field satisfies $h_0 < (J_x + J_y)/2$, a system with open boundary conditions is known to have a symmetry-protected zero energy topological (Majorana) state at each edge. 
This can be understood by analyzing the model in a fermionic language using the Jordan-Wigner transformation from spin-$1/2$ operators to spinless fermions. 
When the transverse field is sinusoidally varied, we find that for a certain 
range of parameters the system can host two types of edge states. The first
type is a topological edge state whose quasienergy (rather than the energy) is 
equal to zero or $\om/2$. There can be another type of edge state whose 
quasienergy has a value different from zero or $\omega/2$; these
states have no topological significance. We call these non-topological edge states. 
We numerically study the information propagation using the
spatio-temporal evolution of the infinite-temperature Floquet OTOC in both the undriven and driven model for the operators $O = \sigma^{z}$ and $\sigma^{x}$. In the absence of any types of edge state, the $\sigma^{z}$ OTOC shows that information 
just propagates into the system with a velocity given by the maximum velocity of the bulk states and is reflected back
upon reaching the ends. The $\sigma^{x}$ OTOC also shows a
similar propagation of information into the bulk of the system and reflection upon reaching the ends. However, in this case the
information gets scrambled inside the light cone of propagation; this has been reported earlier in a system without boundaries~\cite{lin2018out}. Rather remarkably, we find that the scrambled information gets {\it unscrambled} after a reflection at the other 
end of the system; this striking effect is entirely due to the presence of 
open boundary conditions. The scrambling can be attributed to the non-locality of the $\sigma^{x}$ operator in terms of Jordan-Wigner fermions
which is absent in the case of the $\sigma^{z}$ OTOC~\cite{lin2018out}.

If there are edge states present in the model, the OTOCs shows distinct signatures 
of them at the edge of the system. We find 
that a major part of information stays localized at the edge, given by a non-trivial value of the OTOC and a small fraction of information leaks into the system due to the bulk 
states. This fraction depends on the total contribution of the bulk states at the edge site. Furthermore, we see an interesting oscillatory behavior in both $\sigma^{z}$ and $\sigma^{x}$ OTOC at the edge in the presence of 
non-topological edge states. We 
find that the oscillation period depends on the quasienergies of the 
non-topological edge states. We provide analytical expressions of the
OTOC to describe all these effects at the edge due to the presence of 
edge states. Finally, we study the variation of the velocity of information propagation through the 
bulk states, which is simply the maximum velocity in the bulk state dispersion, as a 
function of the driving parameters, namely, the frequency and amplitude of driving.

The plan of the paper is as follows. In Sec.~\ref{sec2}, we introduce the model 
which is a spin-1/2 chain with $xx$ and $yy$ couplings between nearest neighbors
and Zeeman coupling to a magnetic field in the $z$ direction which is driven 
periodically in time. This is followed in Sec.~\ref{sec2a} by a discussion of 
the Jordan-Wigner transformation which maps the spin model to a system of 
non-interacting spinless Majorana fermions known as the Kitaev model. 
In Sec.~\ref{sec2b} we describe the OTOC in detail and discuss the construction of the infinite-temperature Floquet OTOC in terms of Majorana fermions for our model. 
In Sec.~\ref{sec3} we present all our numerical results. First we describe the edge 
states and Floquet bulk bands of the driven Kitaev chain in Sec.~\ref{sec3b}. Then 
in Sec.~\ref{sec3c} we present our numerical results of the effects of edge 
states on the
OTOCs for both the undriven and the driven model. This is followed by Sec.~\ref{sec3d} where we provide analytical expressions which explain our main 
numerical results. In Sec.~\ref{sec3e}, we study the behavior of the
$\sigma^x$ OTOC over a very long time period at different sites of
the system. In Sec.~\ref{sec3f} we discuss how the velocity of information 
propagation via
the bulk states varies with the driving parameters. In Sec.~\ref{sec3g} we briefly discuss the effects of integrability-breaking interactions between the
fermions on the OTOCs. Finally, we conclude by summarizing our results in 
Sec.~\ref{sec4} and mentioning possible
systems where OTOCs can be experimentally measured.

\section{Periodically driven spin-1/2 $XY$ chain in a transverse field}
\label{sec2}

We consider the Hamiltonian for a spin-1/2 $XY$ chain with a transverse field on a 
one-dimensional lattice with $N$ sites and open boundary conditions.
\bea H ~=~ J_{x}~\sum^{N-2}_{j=0} S^{x}_{j} S^{x}_{j+1} ~+~ J_{y}~\sum^{N-2}_{j=0} S^{y}_{j} S^{y}_{j+1} ~+~ h~\sum^{N-1}_{j=0} S^{z}_{j}, \label{ham1} \eea
where $J_x$, $J_y$ are the nearest-neighbor exchange couplings for the $x$ and $y$ components of the spins, $h$ denotes the magnetic field in the transverse direction,
and $S_i^a = \si_i^a /2$ for $a=x,y,z$. (We will set $\hbar =1$ throughout this
paper). We will vary the transverse magnetic field sinusoidally in time
as $h(t) = h_0 + h_1 \sin (\omega t)$, where $h_{0}$ is the static part of the magnetic field, and $h_1$ and $\omega$ are the amplitude and frequency of the periodic drive.
Note that we are considering a sinusoidal drive as opposed to $\delta$-function kicks or square wave pulses~\cite{thakurathi2013floquet, rodriguez2018universal, sreejith2016parafermion}. The periodically driven Hamiltonian commutes at all times $t$
with the fermion parity operator defined as $\Pi_{j=0}^{N-1} \si^{z}_{j}$. 
Next, if we make a $\pi/2$ rotation about the $z$-axis in the spin space at alternate lattice sites, which is a unitary transformation, we obtain $S^{x}_{i} \rightarrow 
(-1)^i S^{x}_{i}$ and $S^{y}_{i} \rightarrow (-1)^i S^{y}_{i}$, while $S^{z}_{i}$ remains unchanged. This transformation preserves the commutation relations between the spin operators and effectively changes the couplings $J_x \rightarrow -J_x$ 
and $J_y 
\rightarrow -J_y$. Thus $H(J_x, J_y)$ and $H(-J_x, -J_y)$ are unitarily related.
In the absence of driving, the model can be solved exactly by mapping it to a model 
of non-interacting spinless fermions known as the Kitaev chain. In the next section, 
we will follow this procedure to obtain a periodically driven version of the 
Kitaev chain.

\subsection{Jordan-Wigner transformation and the Floquet operator}
\label{sec2a}

Using the $(J_x, J_y) \rightarrow (-J_x, -J_y)$ symmetry of the model, we
rewrite the Hamiltonian in Eq.~\eqref{ham1} as 
\beq H(t) ~=~ - ~\frac{J_{x}}{4}~\sum^{N-2}_{j=0} \sigma^{x}_{j} \sigma^{x}_{j+1} 
~-~ \frac{J_{y}}{4}~\sum^{N-2}_{j=0} \sigma^{y}_{j} \sigma^{y}_{j+1}
~+~ \frac{h(t)}{2}~\sum^{N-1}_{j=0} \sigma^{z}_{j}, \label{ham2} \eeq
The Jordan-Wigner transformation of the spin operators to fermionic 
operators is given by
\begin{eqnarray}
\sigma_{j}^{+} &=& e^{i \pi \sum_{l<j} {\hat n}_l} ~c_j^\dagger, \non \\
\sigma_{j}^{-} &=& e^{i \pi \sum_{l<j} {\hat n}_l} ~c_{j}, \non \\
\sigma_{j}^{z} &=& 2 {\hat n}_j ~-~ 1, \label{jw} \end{eqnarray}
where $\sigma_{j}^{\pm} = (\sigma^{x}_{j} \pm i \sigma^{y}_{j})/2$, and
${\hat n}_l = c_l^\dagger c_l$ is the fermion number operator at site $l$. At each site $l$, the spin-up (spin-down) state is mapped to a fermionic filled (empty) 
state corresponding to $n_l = 1 ~(0)$ respectively, and the string 
of operators $e^{i \pi {\hat n}_l} = - \si_l^z$ is needed to have anticommutation relations between the fermionic operators. 
This transforms the Hamiltonian in Eq.~\eqref{ham2} to
\beq H(t)~=~ - \left(\frac{J_x+J_y}{4}\right) ~\sum^{N-2}_{j=0}(c_{j}^{\dagger} 
c_{j+1} + {\rm H.c.})
~-~ \left(\frac{J_x-J_y}{4}\right) ~\sum^{N-2}_{j=0}(c_{j}^{\dagger} c_{j+1}^{\dagger} + {\rm H.c.})
~+~ h(t)~\sum^{N-1}_{j=0} (\hat{n}_{j} -1/2). \label{ham3} \eeq
The second term on the right-hand side of Eq.~\eqref{ham3} creates or annihilates
fermions in pairs and hence does not conserve the total fermion number $\sum_{l=0}^{N-1} {\hat n}_l$, but it conserves the fermion parity $e^{i 
\pi \sum_{l} \hat{n}_{l}}$. We are assuming open boundary conditions in 
Eq.~\eqref{ham3} since we wish to look at the effects of edge states.

To study the properties of the model described in Eq.~\eqref{ham3}, 
it is useful to introduce Majorana fermion
operators. These operators are defined as $a_{2j} = c_{j}^{\dagger} +c_{j}$ and $a_{2j+1} = i(c_{j}^{\dagger} - c_{j}) $, where $j = 
0,1,\ldots, N-1$. We have $2N$ Majorana fermions for $N$ Dirac fermions. 
These Majorana operators obey the 
anticommutation relation $\{a_{m},a_{n}\} = 2 \delta_{mn}$. The 
Hamiltonian then takes the form
\beq H(t) ~=~ \frac{i}{4} ~\sum_{j=0}^{N-2} ~\left( J_y a_{2j}a_{2j+3} ~-~ 
J_x a_{2j+1} a_{2j+2}\right)
~+ ~\frac{i~h(t)}{2} ~\sum_{j=0}^{N-1} ~a_{2j} a_{2j+1}. \label{ham4} \eeq

To analyze a periodic Hamiltonian satisfying $H(t+T) = H(t)$, 
where $T = 2\pi /\omega$ is 
the time period of the drive, we will compute the unitary Floquet operator $U = \mathcal{T} e^{~-i\int_{0}^{T} dt H(t)}$ numerically, where $\mathcal{T}$ denotes time-ordering. 
To construct it numerically we divide the time interval $T$ into $\mathcal{N}$ steps of 
size $\Delta t$ such that $\mathcal{N} \Delta t = T$. We define $t_{j} = (j-1/2) 
\Delta t$, where $j= 1,2,3,\cdots,\mathcal{N}$. In terms of these 
discrete time instants, the Floquet operator is defined as 
\begin{equation} U ~=~ e^{-i \Delta t ~H(t_\mathcal{N})} ~\cdots~ e^{-i \Delta t ~
H(t_{2})} ~e^{-i \Delta t ~H(t_1)}. \label{ut} \end{equation}
We obtain the exact result for $U$ in the limit $\mathcal{N} \to \infty$ and $\Delta t \to 0$, keeping $\mathcal{N} \Delta t$ fixed. Numerically, we consider a value of $\mathcal{N}$ to be 
large enough if the result for $U$ does not change if we increase $\mathcal{N}$ further.
Having constructed $U$, we calculate its eigenvalues and eigenstates called the Floquet
eigenvalues and Floquet eigenstates; these are labelled by an integer $k$ which 
can take values from 1 to $2N$. The Floquet eigenvalues have the form
$e^{-i \theta_k}$ which can be written as $e^{- i \epsilon_k T}$. Here $\epsilon_k$
is called the quasienergy and it can be chosen to lie in the range $-\omega /2 < 
\epsilon_k \le \omega /2$.

Before ending this section, we note that the Hamiltonian is 
quadratic in terms of fermions at all times. Hence the model is integrable even in
the presence of driving, although the driving effectively gives rise to couplings
between pairs of Majorana operators separated by arbitrarily long distances.

\subsection{Construction of Floquet OTOC in Majorana language}
\label{sec2b}

Let us consider a system with Hamiltonian $H$ and an initial state given by a density matrix $\hat{\rho}$. {The effect of a time-evolved 
local operator
$W_j (t)$ can be probed by another local operator $V_{i}$ using the
expectation value of the square of a commutator
defined as}
\begin{equation}
C_{i,l}(t) ~=~ \frac{1}{2}\langle [W_{i+l}(t), V_{i}]^{\dagger} [W_{i+l}(t), V_{i}] \rangle. \end{equation}
Here the time-evolved operator is defined as $W_{j}(t)=U^{\dagger}(t)W_{j} U(t)$, where $U(t) = \mathcal{T} e^{~-i\int_{0}^{t} H(t') dt'}$, and the expectation
value is defined as $\langle O \rangle = \text{Tr}(\hat{\rho} O)/\text{Tr}(\hat{\rho})$. This quantity probes how local information (i.e., the 
effect of a local operator) spreads through the system under a unitary time evolution starting from an initial state. The initial density matrix can be a pure state 
like the ground state or a mixed state like a thermal ensemble. Here we consider 
$\hat{\rho}$ to be a thermal density matrix at infinite temperature ($\beta =0$).

The out-of-time-ordered correlator (OTOC) is defined as 
\begin{equation}
F_{i,l}(t) = \langle W_{i+l}(t) V_{i}W_{i+l}(t) V_{i} \rangle, 
\label{otoc1} \end{equation}
where we assume both $W$ and $V$ to be Hermitian. This is related to the squared commutator as $C_{i,l}(t) = 1~-~ \text{Re}[F_{i,l}(t)]$. The OTOC is an equivalent probe of the action of a local operator on the system. For a Floquet system, 
there is no notion of a definite 
energy state or even a minimum energy state (ground state). Hence, to calculate the OTOC, we take into account all the eigenstates of the Floquet 
operator with equal weights. Such an ensemble of states is called the 
infinite-temperature ensemble, and the Floquet OTOC takes the form
\begin{eqnarray} F_{i,l}(t) &=& \frac{\text{Tr}( e^{~-\beta H} W_{i+l}(t) V_{i}W_{i+l}(t) V_{i})}{\text{Tr}(e^{~-\beta H})} \Big|_{\beta \to 0} \non \\
&=& \frac{1}{M} \sum_{p=1}^{M} \bra{\psi_p} W_{i+l}(t) V_{i}W_{i+l}(t) V_{i} \ket{\psi_p}, \label{otoc2}
\end{eqnarray}
where $M$ is the dimension of the Hilbert space, and $\ket{\psi_p}$ are the many-body Fock states in the basis of eigenstates of the Floquet operator. 

In this work, we will be interested in calculating the Floquet OTOC of the Pauli spin operators 
at different lattice points. So we look for quantities like $F_{i,l}(t) = \langle \sigma^{\alpha}_{i+l}(t) \sigma^{\alpha}_{i} \sigma^{\alpha}_{i+l}(t) \sigma^{\alpha}_{i}\rangle$, where $\alpha = x$ and $z$. 
We will make use of the non-interacting Majorana fermion picture for convenience. The many-body Fock states are constructed as tensor products of the occupation number states of single-particle states which are eigenstates of the Floquet 
operator, and the spin operators are re-written as Majorana 
operators using the Jordan-Wigner transformation. We choose $i=0$ to study how the local information 
from one end of the system propagates in the presence or absence of edge states. In terms of Majorana operators, we have
\begin{subequations}
\bea
F^{zz}_{l}(t) &=& \langle a_{2l}(t)a_{2l+1}(t)a_{0} a_{1} a_{2l}(t)a_{2l+1}(t) a_{0} a_{1}\rangle, \label{subeq1} \\
F^{xx}_{l}(t) &=& \langle \left(\prod_{j<l} i a_{2j}(t)a_{2j+1}(t) \right) 
a_{2l}(t) a_{0}
~ \left( \prod_{j<l} i a_{2j}(t)a_{2j+1}(t) \right) a_{2l}(t) a_{0} \rangle.
\label{subeq2} \eea
\end{subequations}
We observe that $F^{zz}_{l}(t)$ in Eq.~\eqref{subeq1} is made up of local operators
and is therefore much easier to compute than $F^{xx}_{l}(t)$ in Eq.~\eqref{subeq2}. For a general $l$, $F^{xx}_{l}(t)$ consists of $4l+4$ Majorana operators. The 
calculation of all these multi-operator expectation values reduce to calculating two-point correlators like $\langle a_{n}(t) a_{m}(t') \rangle$ due to Wick's theorem. For a Floquet system, the correlators are evaluated only at times $t$ 
which are equal to integer multiples of the driving time period $T$.

We now describe the procedure for obtaining the two-point Majorana correlators for the Floquet system. Since we are working with open boundary conditions, we cannot use 
translation symmetry to perform the calculations in the momentum basis. The Floquet operator $U$ is therefore constructed numerically as described in Sec.~\ref{sec2a} from 
the Hamiltonian in Eq.~\eqref{ham4} in the Majorana fermion language in the position basis. Since $U$ is a $2N \times 2N$ matrix, we obtain $2N$ eigenvalues and 
eigenvectors. The Floquet eigenvalues lie on the complex unit circle and are denoted as $e^{-i\theta_j}$. Since the Hamiltonian is purely imaginary in the Majorana basis,
the Floquet operator is purely real, implying that its eigenvalues come in complex conjugate pairs.

If $\mathbf{v}^{\pm}_{j}$ denotes the eigenvector corresponding to an eigenvalue $e^{\pm i\theta_j}$, then we can define $N$ fermionic operators which diagonalize 
the Floquet unitary operator.
\bea{} \alpha_{j} = \frac{1}{\sqrt{2}} \sum_{n=0}^{2N-1} (\mathbf{v}^{+}_{j})_{n} ~a_{n}, \non \\
\alpha^{\dagger}_{j} = \frac{1}{\sqrt{2}} \sum_{n=0}^{2N-1} (\mathbf{v}^{-}_{j})_{n} ~a_{n},
\label{alphaop}
\eea{}
where $(\mathbf{v}^{\pm}_{j})_n$ denotes the $n$-th component of the eigenvector with eigenvalue $e^{\pm i \theta_j}$. There are $N$ such pairs of equations. The time evolution of the $\alpha$-fermions is easy to calculate since $\alpha_j(t=pT) = (U^{\dagger})^p \alpha_j (U)^p = e^{-i p\theta_j} \alpha_j$ and $\alpha^{\dagger}_j(t=pT) = e^{i p\theta_j} \alpha^{\dagger}_j$. Inverting Eq.~\eqref{alphaop} and using the above time-evolution equations, we obtain 
\beq a_{m}(pT) = \sum_{j=0}^{N-1}( (\mathbb{A}^{-1})_{m,2j} ~e^{-ip \theta_j}
\alpha_{j} + (\mathbb{A}^{-1})_{m,2j+1} ~e^{ip \theta_j}\alpha^{\dagger}_{j}).
\label{am} \eeq
The matrix $\mathbb{A}$ is made out of the Floquet eigenvectors as $\mathbb{A}_{2n,m} = \frac{1}{\sqrt{2}} (\mathbf{v}^{+}_{n})_m$ and $\mathbb{A}_{2n+1,m} = \frac{1}{\sqrt{2}} (\mathbf{v}^{-}_{n})_m $, where $n = 0,1, \ldots N-1$ and $m = 
0,1, \ldots 2N-1$. There are some subtleties in the numerical calculation of $\mathbb{A}^{-1}$ in the case of degenerate eigenvalues which are explained 
in Appendix A. 

We can now proceed to calculate the infinite-temperature two-point correlation function of the form $\langle a_{m}(pT) a_{n}(qT) \rangle = \frac{1}{M} \sum_{p=1}^{M}
\bra{\psi_p} a_{m}(pT) a_{n}(qT) \ket{ \psi_p }$. Since we have decomposed the Majorana operators in terms of the $\alpha$ fermions, it will be convenient to write 
the many-body Fock states $\ket{\psi_p}$ in the basis of occupation number of the $\alpha$ fermion states. The calculation for the two-point correlation is then 
straightforward (see Appendix B), and we obtain
\bea
\langle a_{m}(pT) a_{n}(qT) \rangle &=& \frac{1}{2^N} ~\text{Tr}(a_{m}(pT) a_{n}(qT)) \non \\
&=& \frac{1}{2} ~\sum_{j=0}^{N-1} ~[ (\mathbb{A}^{-1})_{m,2j} (\mathbb{A}^{-1})_{n,2j+1} e^{-i(p-q)\theta_j} ~
+ ~(\mathbb{A}^{-1})_{m,2j+1} (\mathbb{A}^{-1})_{n,2j} e^{i(p-q)\theta_j} ].
\label{2point}
\eea
The OTOCs in Eqs.~\eqref{subeq1} and \eqref{subeq2} are then found using the 
Pfaffian method for Wick's theorem~\cite{lin2018out}.
If $\gamma_{i}$ is the $i$-th 
element inside the thermal average of the OTOC, then the average is given by $\text{Pf}(\mathbb{\Phi})$, where $\mathbb{\Phi}$ is an antisymmetric matrix whose 
elements are defined as $[\mathbb{\Phi}]_{i,j} = \langle \gamma_{i} \gamma_{j} \rangle$. We stress that $F^{xx}_{l}(t)$ for a general $l$ is composed of $4l+4$ 
fermionic operators due to the non-locality of $\sigma^{x}_{l}$ in the Jordan-Wigner transformation. Hence we have to calculate the Pfaffian of a $(4l+4) \times (4l+4)$
antisymmetric matrix. On the other hand, $F^{zz}_{l}(t)$ contains only eight fermionic operators irrespective of the value of $l$. 

\section{Results}
\label{sec3}

\subsection{Edge states in an undriven Kitaev chain}
\label{sec3a}

We will begin by recalling some topological properties of an {\it undriven} Kitaev chain~\cite{kitaev2001}. The Hamiltonian is given by
\beq H(t) ~=~ - \left(\frac{J_x+J_y}{4}\right) ~\sum^{N-1}_{j=0}(c_{j}^{\dagger} 
c_{j+1} + {\rm H.c.}) ~-~ \left(\frac{J_x-J_y}{4}\right) ~\sum^{N-1}_{j=0}(c_{j}^{\dagger} c_{j+1}^{\dagger} + {\rm H.c.})
~+~ h_0 ~\sum^{N-1}_{j=0} (\hat{n}_{j} -1/2). \label{ham5} \eeq
For convenience, we will take the number of sites $N$ to be even and will
assume antiperiodic boundary conditions (instead of
open boundary conditions), namely, $c_N = - c_0$ and $c_N^\dagger = - c_0^\dagger$.
{(Note that the original spin model described in Eq.~\eqref{ham1} has 
open boundary conditions. But we are studying the corresponding fermionic
model with closed boundary conditions because that makes it easier to study
its bulk properties).} Upon transforming to momentum space,
Eq.~\eqref{ham5} takes the form 
\beq H = \sum_{0<k<\pi} \left(c^{\dagger}_{k} ~~c_{-k}\right)
\Tilde{h}_{k} \begin{pmatrix}
c_k \\[0.2cm] 
c^{\dagger}_{-k} \end{pmatrix},
\label{ham-ka} \eeq
where
\beq \Tilde{h}_{k} ~=~ \left[ - \frac{J_x + J_y}{2} \cos k + h_0 \right]\tau^{z} 
+ \frac{J_x -J_y}{2} \sin k ~\tau^{y}. \label{ham-kb} \eeq
Here $k$ takes the values from $\pi/N$ to $\pi - \pi/N$ in steps of $2\pi/N$ and $\tau^{\alpha}$ are Pauli matrices. (An advantage of assuming $N$ to be even
and antiperiodic boundary conditions is that $k$ cannot take the values zero and $\pi$. Hence the momenta $k$ and $-k$ appearing in Eq.~\eqref{ham-ka} are 
necessarily different from each other). 

The model described above is known to be in a topological phase if $J_x \neq J_y$ 
and the transverse field satisfies $|h_0| < (J_x+ J_y)/2$~\cite{kitaev2001,thakurathi2013majorana}. 
A long system with open boundary conditions then hosts edge states with zero energy.
By long we mean that the length of the system is much
larger than the decay lengths of the wave functions of the edge states. If this
condition is not satisfied, the edge states at the two ends of the system
hybridize with each other and this shifts their energies away from zero. The
topological phase is characterized by a non-zero topological invariant which is a 
winding number for this model~\cite{thakurathi2013majorana}. On the 
other hand, if $|h_0| > (J_x + J_y)/2$, the system is in a non-topological phase,
there are no edge states at the ends of an open system, and the winding number is
zero.


\subsection{Edge states and Floquet bands in a driven Kitaev chain}
\label{sec3b}

We will now consider a driven system in which the transverse field varies in 
time as 
\beq h(t) ~=~ h_0 ~+~ h_1 \sin (\omega t). \label{ht} \eeq 
For antiperiodic boundary conditions, the Hamiltonian in momentum space 
has a form similar to Eqs.~(\ref{ham-ka}-\ref{ham-kb}),
except that $\Tilde{h}_k$ is now time-dependent and is given by
\beq \Tilde{h}_{k} (t) ~=~ \left[- \frac{J_x + J_y}{2} \cos k + h(t) \right] 
\tau^{z} + \frac{J_x -J_y}{2} \sin k ~\tau^{y}. \label{ham-k2} \eeq
For each value of $k$, we can numerically calculate the Floquet operator $U_k$ (this 
is a $SU(2)$ matrix since ${\tilde h}_k (t)$ is a sum of two Pauli matrices), and then
find the quasienergies given by $\pm \epsilon_k$ from the two eigenvalues of $U_k$.
In addition, we can consider an open system, calculate its Floquet operator $U$
(this is a $2N \times 2N$ matrix) and find if any of its eigenvectors are localized
near the ends of the system. We discover that there are two kinds of
edge states, topological states whose quasienergies are equal to zero or 
$\om /2$ (for a long system), and non-topological states whose 
quasienergies are not equal to zero or 
$\omega/2$.~\cite{liu,saha2017generating,balabanov,molignini,lili,muller} 
The total number of topological states (with quasienergies zero and $\om /2$
combined) at each end of a system is given by a winding number. 
The non-topological states are so named because they do not seem to have any 
topological significance. As far as 
we know, there is no topological invariant which gives the number of 
these states at each end of a system.

In the rest of the section we will assume that the static part of the transverse field $h_0 = 0$ unless otherwise specified. Our results for the
Floquet quasienergy bands and edge states are as follows.
We find numerically that when the driving frequency $\omega$ is 
much smaller than the other parameters in the Hamiltonian ($J_x, ~J_y$ and $h_1$), 
there are numerous non-topological edge states and their number increases rapidly as 
$\omega$ decreases. We will therefore focus on a range of $\omega$ 
and $h_1$ where the non-topological edge states are stable, i.e., the number of such states 
(two pairs in this case, each end of the system having one pair) stays unchanged as 
$\omega$ varies. It turns out that the region for the non-topological 
states to exist is approximately given by $(J_x + J_y)/2 < \omega < J_x + J_y$. We can understand this by looking at the quasienergy bulk bands $\epsilon_k$ as
follows.

We find that non-topological edge states exist only when the quasienergy spectrum is
gapped at the boundary of the Floquet Brillouin zone (FBZ) where the quasienergy is $\omega /2$. (Since the quasienergies are
defined modulo $\omega$, we define the quasienergy gap as the minimum of the gap 
between the two Floquet bands either across the middle of the Floquet Brillouin zone 
(FBZ) or across the boundary of the FBZ).
Along the line $\omega = (J_x + J_y)/ 2$ the bulk Floquet bands become gapless at 
$k = 0$ and $\pi$ with quasienergy $\epsilon_k = 0$ for $h_0 = 0$. Along the line
$\omega = J_x + J_y$ the gap closes at $k = 0$ and $\pi$ with quasienergy $\epsilon_k = \omega /2$, i.e., at the boundary of the FBZ. These lines 
for the special values of $\om$ can be derived analytically from the Floquet 
operator $U_k$, since the $\tau^{y}$ component vanishes at $k=0$ and $\pi$. 
We see from Eq.~\eqref{ham-k2} that for $k=0$ and $\pi$ and
$h(t) = h_1 \sin (\om t)$, the Floquet operator is $U_{k} = e^{i(J_x + J_y) T/2}$. This is equal to 1 when $\om = (J_x + J_y)/2$ 
and $-1$ when $\om = J_x + J_y$. There are also some additional lines in the 
$h_1 -\omega$ plane where the spectrum becomes gapless at either $\epsilon_k = 0$ or $\epsilon_k = \omega /2$. Along these lines the gap closes at values of $k \neq 0$ or
$\pi$, and these lines (and the corresponding values of $k$) are difficult to find analytically from the structure of the
Floquet operator $U_k$. The non-topological states do not appear on the lines in the $h_1
-\omega$ plane when the gap closes at $\epsilon_k = \omega /2$.

On the other hand, we find that the topological edge states at 
$\epsilon = 0$ are always present as long as the spectrum is gapless at quasienergy 
$\epsilon =0$ and $J_x \neq J_y$. These edge states are absent if there is a constant 
part $h_0$ in $h(t)$ which exceeds $(J_x + J_y) /2$.

\begin{figure}[H]
\centering
\includegraphics[width=0.5\textwidth]{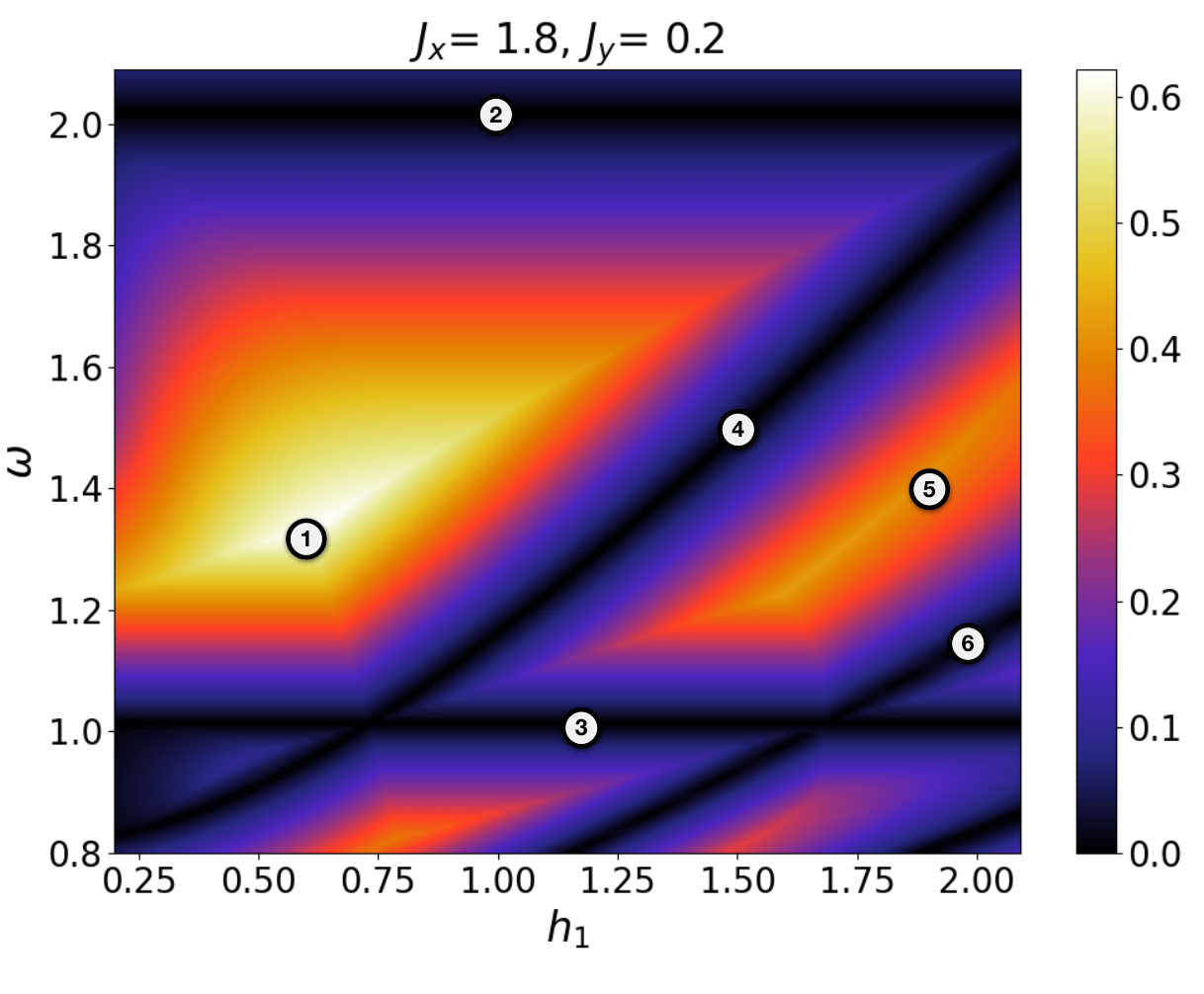}
\caption{Plot of Floquet quasienergy gap as a function of $h_1$ and $\omega$. The other parameters are chosen $J_x =1.8, ~J_y = 0.2$ and $h_0 =0$. Along the horizontal dark lines at $\omega =1$ and $2$, the dispersion is gapless at momenta $k=0$ and $\pi$. The bands touch at
the quasienergies $\epsilon= 0$ and $\omega/2$ for $\omega =1$ and 2 respectively. 
The slanted dark lines in the plot are where the dispersion is gapless at momenta 
other than $0$ or $\pi$.} \label{fig1} \end{figure}

In Fig.~\ref{fig1} we show the Floquet quasienergy gap as a function of $h_1$ and $\omega$ for $J_x = 1.8, ~J_y =0.2$ and $h_0 = 0$. (Since the quasienergies are
defined modulo $\omega$, the quasienergy gap is defined as the minimum of the gap between the two Floquet bands either across the middle of the FBZ or across 
the boundary of the FBZ).
The horizontal dark lines along $\omega = 1~\& ~ \omega = 2$ mark the gapless boundaries in the phase space where we see the two degenerate pairs of non-topological edge states. The slant dark lines mark the other gapless lines discussed previously. 

\begin{figure*}[ht]
\centering
\includegraphics[width=\textwidth, height = 0.8\textwidth]{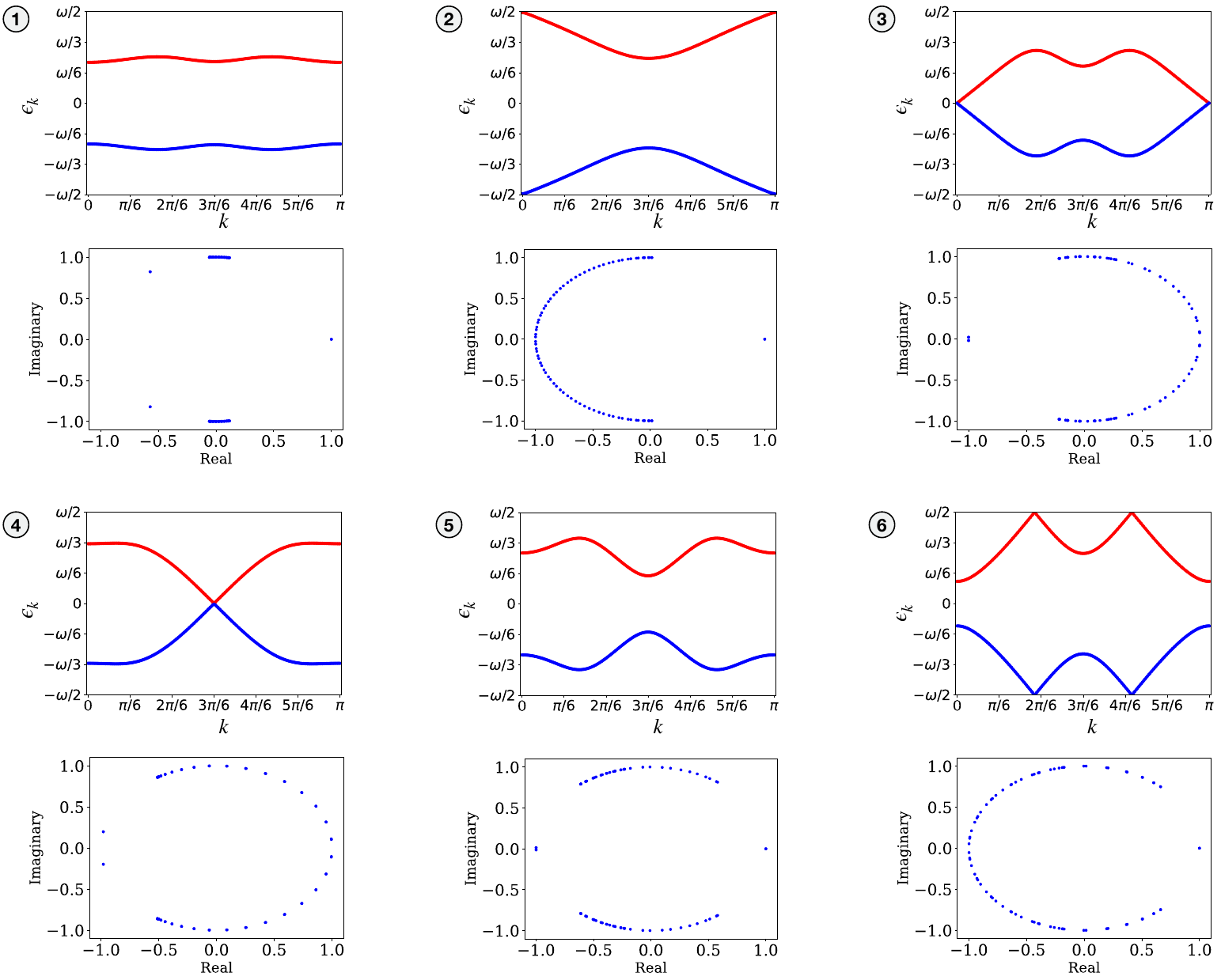}
\caption{Bulk bands of Floquet quasienergies $\epsilon_k$ and eigenvalues 
$e^{-i \theta_k}$ (shown on a unit circle) for different values of the driving parameters $(h_1, \omega)$, for $J_x =1.8, J_y =0.2$ and $h_0 = 0$. Figure 1: $h_1 = 0.6,~\omega 
=1.3$. Top: almost flat bulk bands which are gapped at all values of $k$. Bottom: both 
zero quasienergy topological states and non-topological states appear in the Floquet eigenvalue plot. Figure 2: $h_1 = 1,~\omega =2$. Top: the Floquet bulk bands are gapless at quasienergy 
$\epsilon = \omega/2$ and momenta $k = 0$ and $\pi$. Bottom: Edge
states are present at zero quasienergy, but non-topological edge states are not present. 
Figure 3: $h_1 = 1.2,~\omega =1$. Top: the Floquet bulk bands are gapless at zero quasienergy and momenta $k = 0$ and $\pi$. Bottom: non-topological edge states are present, but zero 
quasienergy states are not present due to the gap closing at $\epsilon = 0$. Figure 4: 
$h_1 = 1.53,~\omega =1.5$. Top: the Floquet bulk bands are gapless at quasienergy $\epsilon = 0$ and momentum $k = \pi/2$. Bottom: non-topological edge states are present, 
but due to the gap closing at $\epsilon =0$ edge states are not present at zero quasienergy. Figure (5): $h_1 = 1.9,~\omega =1.4$. Top: the Floquet bulk bands are 
gapped at all momenta. Bottom: both zero quasienergy topological
and non-topological edge states 
are present. Figure 6: $h_1 = 2,~\omega =1.15$. Top: the Floquet bulk bands are
gapless at $\epsilon = \omega/2$, but not at any special momenta like $k = 0, ~\pi$ or $\pi/2$, but rather at $k$ close to $\pi/3$. Bottom: Zero quasienergy
topological edge states are present but not non-topological edge states.} \label{fig2} \end{figure*}

In Fig.~\ref{fig2}, we plot the bulk bands and the Floquet eigenvalues which show edge states for six different values of $(h_1, \omega)$ for $J_x =1.8, ~J_y =0.2$ and
$h_0 = 0$. The isolated points in the Floquet eigenvalue plots indicate the edge states. These figures show that zero quasienergy topological
edge states can exist even if the gap closes at quasienergy $\epsilon = \omega/2$, and the non-topological edge states can exist even
if the band gap closes at quasienergy $\epsilon = 0$. We also see that the Floquet eigenvalues of the non-topological edge states vary continuously depending on the
values of $\omega$ and $h_1$.

To summarize this section, we have shown that sinusoidal driving of the
transverse field of the spin-1/2 $XY$ chain can generate edge states at 
quasienergies equal to 0 and $\om /2$ (topological edge states) and 
also at other values of the quasienergy (non-topological edge states), depending
on the driving parameters.

\subsection{Effects of edge states on OTOC}
\label{sec3c}

\subsubsection{The undriven model}
\label{sec3c1}

\begin{figure}[ht]
\centering
\includegraphics[width=0.8\textwidth]{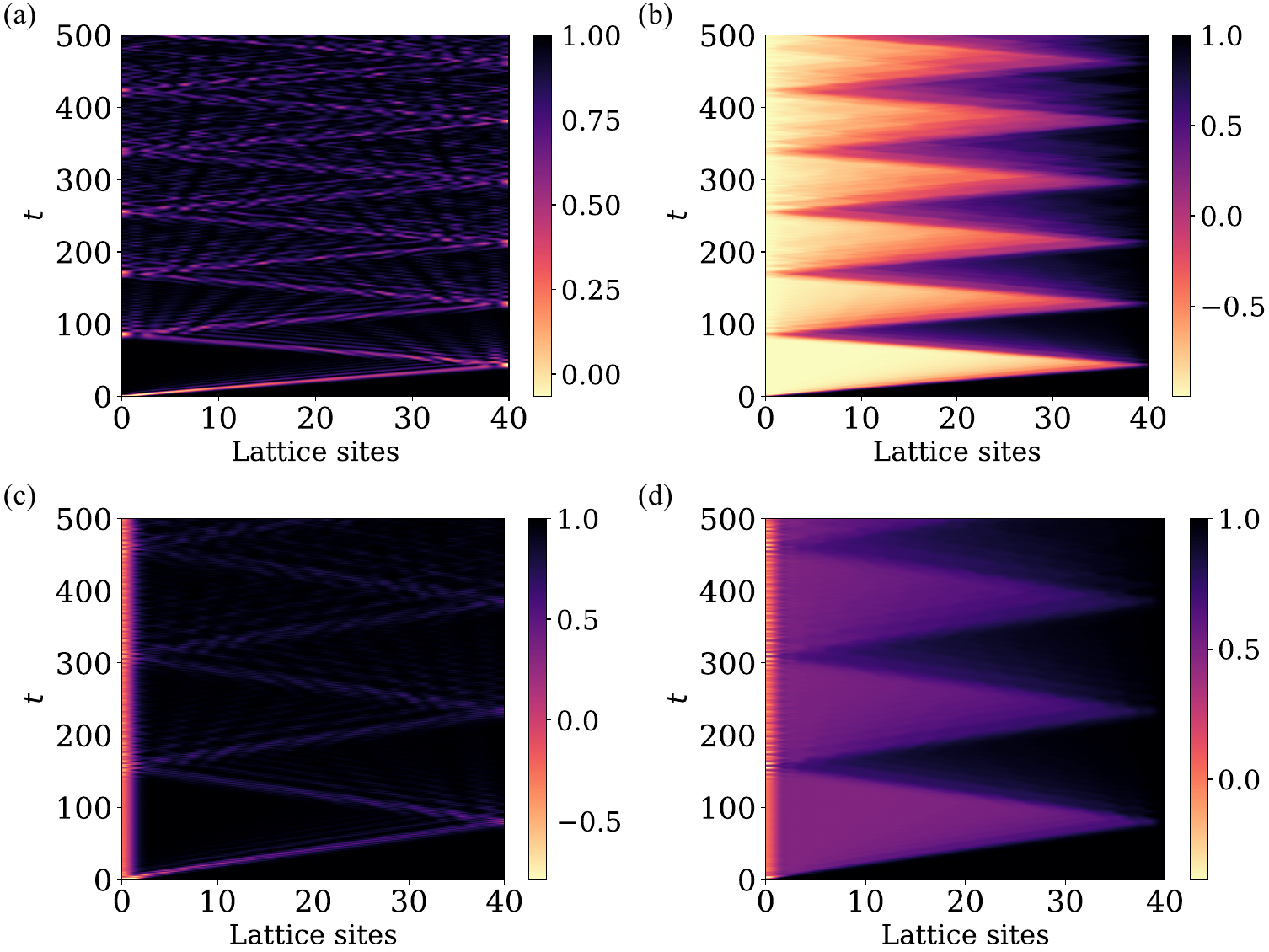}
\caption{Plots of $F^{zz}_{l}(t)$ (plots (a) and (c)) and $F^{xx}_{l}(t)$ (plots
(b) and (d)) for the undriven model as a 
function of position $l$ and time $t$. These functions are shown for $J_x =1.8$, $J_y 
= 0.2$, and $h_0 =2.5$ in (a) and (b) (top), and for $J_x =1.8$, $J_y = 0.2$, and $h_0
=0.5$ in (c) and (d) (bottom). In (c) and (d) we see the presence of a zero quasienergy topological
edge state as $h_0 < (J_x+J_y)/2 = 1$. The $\sigma^{x}$ OTOC plots in (b) and (d) show 
the scrambling of information and then unscrambling of information upon reflection 
from the ends of the system, while the $\sigma^z$ OTOC plots in (a) and (c) do not 
show any scrambling or unscrambling of information.} \label{fig3} \end{figure}

We first present our numerical results for an undriven system with a constant magnetic field 
term with coefficient $h_0$. We take a finite system with size $N=40$ and choose the couplings to be $J_x =1.8$ and $J_y =0.2 $. The plots of $F^{zz}_{l}(t)$ and $F^{xx}_{l}(t)$ as functions of position $l$ and time $t$ are shown in 
Fig.~\ref{fig3}. Figures 
\ref{fig3} (a) and (b) show plots of $F^{zz}_{l}(t)$ and $F^{xx}_{l}(t)$ plots for $h_0 = 2.5$, which ensures that the system is in the magnetically disordered phase (or 
non-topological phase in terms of Jordan-Wigner fermions) and therefore has no edge states. In both Figs.~\ref{fig3} (a) and (b), we see the propagation of information 
which is initiated by
the action of $\sigma^z$ and $\sigma^x$ operators on the left-most site ($l=0$) of the system. The dark regions, where the OTOC is close to 1, are the regions where the 
square of the commutator, $C_{i,l}(t)$, is close to zero, suggesting an absence of 
information propagation. On the other hand, the bright regions where the OTOC is much less than 1, imply that 
$C_{i,l}(t)$ is non-zero, and hence the wavefront of information of the local operator has reached that point. In particular, there are regions where OTOC is close 
to $-1$, which 
implies that the two local operators $W(t)$ and $V$ effectively {\it anticommute} 
with each other in these regions. 

We also see from the plot that there is a finite propagation speed of 
the wavefront, which gets reflected repeatedly upon reaching the edges of the system.
The velocity turns out to be exactly the maximum velocity $v_{max} = (\partial 
E_k / \partial k)_{max}$ as obtained from the dispersion of the bulk Hamiltonian,
\begin{equation} E_k = \sqrt{\left (\frac{J_x+J_y}{2} \cos k - h \right )^{2} 
+ \left (\frac{J_x-J_y}{2} \sin k \right )^2}. \end{equation}
{This indicates that the wave front of the propagating information of any 
local operator is governed by the bulk states with the maximum velocity possible 
in the system known as the Lieb-Robinson bound~\cite{lieb1972finite}.} 
Numerically, we find that $v_{max} \simeq 1$ for the parameter values given above.

However, there is a striking contrast between the OTOCs $F^{zz}_{l}(t)$ and $F^{xx}_{l}(t)$ which has been reported earlier for a system without 
boundaries~\cite{lin2018out}. For the $\sigma^z$, once the wavefront has passed 
through a lattice 
point, the OTOC again becomes approximately equal to 1, implying the absence of scrambling of information. Such a propagation of information starting from a single 
point may be visualized as having a `shell-like' structure in space-time. 
On the contrary, for the case of $\sigma^x$, we see that the OTOC remains the same (much less than 1), even 
after the wavefront of information has passed the lattice point. In fact, for a system where there are no edge states, the OTOC in the entire time-like region is close to 
$-1$, implying an approximate anticommutation between $\sigma^{x}_{l}(t)$ and $\sigma^{x}_{0}$. This propagation, starting from a single point, can be thought of as 
having a `solid ball-like' structure in space-time. The reason is that for an exactly solvable model, the OTOCs of spin operators which are local in terms of 
Jordan-Wigner fermions show no signature of scrambling or delocalization of information inside the light cone. On the other hand, the OTOC of spin operators that are non-local 
in Jordan-Wigner fermions show scrambling~\cite{lin2018out}. However, in this case, the OTOC is close to $-1$ instead of zero. 

The new effect that we report here is that the OTOC for an open ended system 
shows the effect of reflection from the edges for the wavefront of information propagation. 
If the perturbing operator is $\sigma^z$, the OTOC just propagates from one end and gets reflected upon reaching the other end. Even after repeated reflections from either end,
we see that there is very little decay in the value of the OTOC and almost negligible wavefront broadening for a long interval of time. However for the case of $\sigma^x$, we see an interesting effect upon every reflection from either end. The scrambled 
information inside the light cone starts unscrambling after reflection, thereby making the value of the OTOC approach 1 again. Such an unscrambling of information 
after reflection from the ends of the system has not been reported in the 
literature so far to the best of our knowledge. 

Next we turn to the OTOC plots shown in Fig.~\ref{fig3} (c) and (d) for the case when topological edge states are present in the system. We choose $h_0 = 0.5$ so that the 
undriven part of the spin model is in the ordered phase (namely, topological phase in terms of Jordan-Wigner 
fermions). For both the OTOCs, $F^{zz}_{l}(t)$ and $F^{xx}_{l}(t)$, a substantial portion of the information of the action of the operator stays localized at the edge and a small part is propagated into the system through the bulk states with a velocity 
$v_{max} \simeq 0.54$. The extent of localization of information is quantified by 
the deviation of the OTOC value from one, at the edge and inside the bulk of the system. We note that in absence of an edge state, $F^{zz}_{l}(t)$ at the edge ($l = 0$) is 
almost always equal to one (Fig.~\ref{fig3} (a)) except at the recurrence of wavefront. However, the presence of edge states contributes to a non-zero commutator and brings 
down the value of $F^{zz}_{l}(t)$ at the edge to a value much lesser than one (Fig.~\ref{fig3} (c)). 

On the other hand, for the $\sigma^x$ OTOC, $F^{xx}_{l}(t)$ at the edge stays close to $-1$ in the absence of edge states, implying an approximate anticommutation 
between $\sigma^{x}_{0}(t)$ and $\sigma^{x}_{0}$. The presence of edge state in this case contributes to a non-zero anticommutator and increases the OTOC value to become 
greater than $-1$ at the edge. In this way, the presence of zero quasienergy 
topological states at the edge of the system can modify the OTOC and influence the spreading of information from one end of the system.

To summarize this section, we have shown that in an undriven system, OTOCs 
can detect the presence of edge states. More precisely, the edge states trap part 
of the information at the end thereby reducing the values of the OTOCs 
in the bulk. Further, reflections from the ends of the system produces an
unscrambling effect for the $\si^x$ OTOC.

\subsubsection{Periodically driven model}
\label{sec3c2}

\begin{figure*}[htb]
\centering
\includegraphics[width=\textwidth]{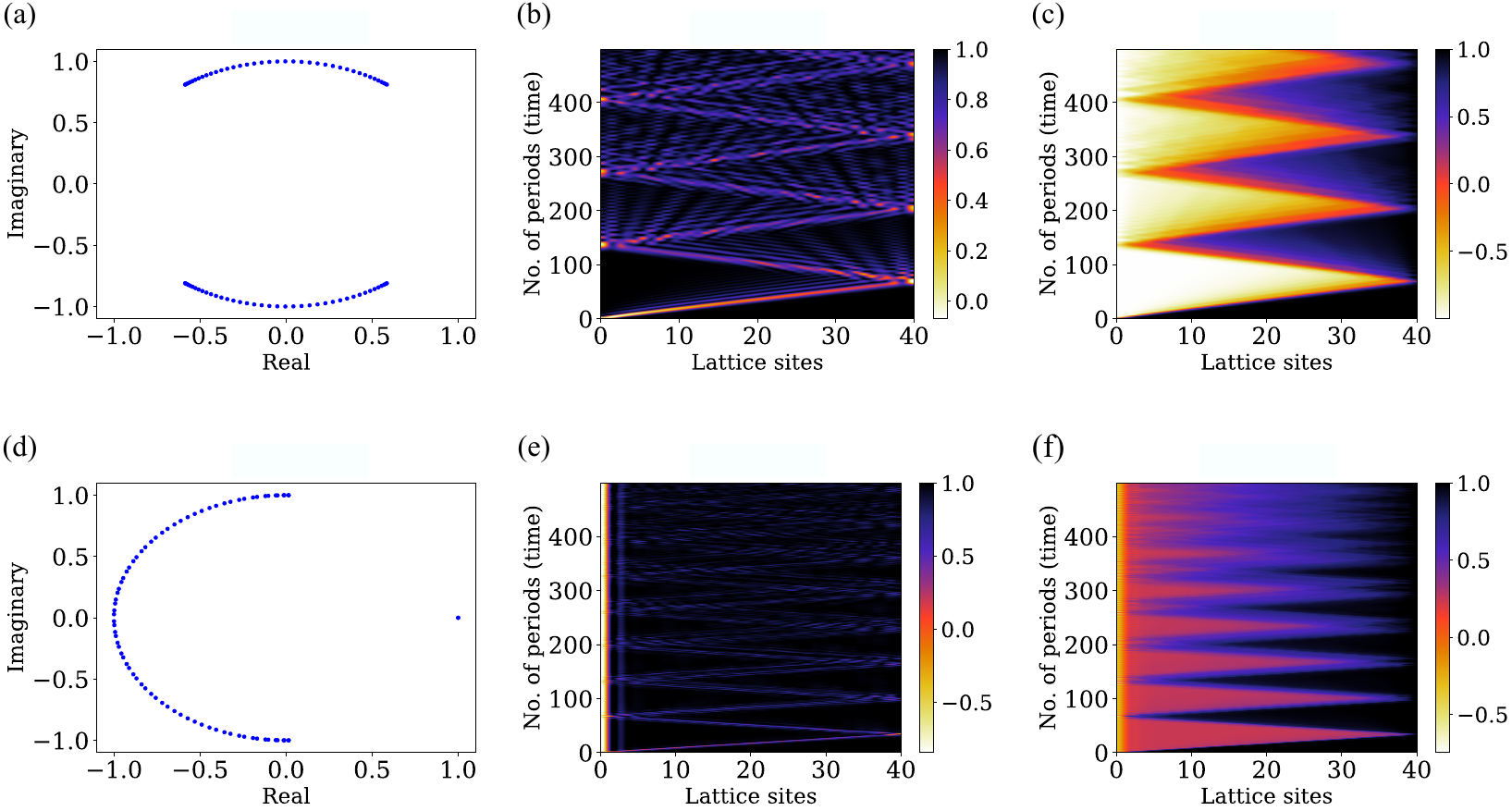}
\caption{Plots of (a) Floquet eigenvalues, (b) $F^{zz}_{l}(t)$ and (c) $F^{xx}_{l}(t)$ for $J_x = 1.8, ~J_y =0.2, ~h_0 = 2.5, ~h_1 = 0.6$ and $\omega = 10$. No edge states are present in the system for these values of the parameters. The same three
quantities are shown for $J_x = 1.8, ~J_y =0.2, ~h_0 = 0, ~h_1 = 1$ and $\omega = 2$ in plots (d), (e) and (f). In this case there are edge states with zero
quasienergy as can be seen in the plot of Floquet eigenvalues. These states 
localize a part of the information at the edges of the system which can be 
seen in plots (e) $F^{zz}_{l}(t)$ and (f) $F^{xx}_{l}(t)$. We also see the unscrambling effects upon reflection from the ends in the $\sigma^{x}$ OTOC plots 
(c) and (f).} \label{fig4} \end{figure*}

Having discussed some features of the undriven model, we present our numerical
results for the periodically driven model. 
If we turn on a driving with amplitude $h_1 = 0.6$ and frequency $\omega = 10$, a
static magnetic field $h_0 =2.5$, $J_x =1.8, ~J_y =0.2$ and system size $N=40$,
we obtain the plots shown in Fig.~\ref{fig4} (a), (b) and (c) for the Floquet 
eigenvalue, and the $\sigma^z$ and $\sigma^x$ OTOCs respectively. For a 
periodically driven system the OTOC 
is plotted as a function of the distance $l$ and number of periodic drives $p$. The 
time corresponding to $p$ drives is given by $2 \pi p/\omega $. 

For the chosen value of the static magnetic field $h_0$, there are no edge states present, as can be seen from the Floquet eigenvalue plot. The OTOC plots look 
qualitatively very similar to the case of the static model without edge states. Namely, $F^{zz}_{l}(t)$ shows no signs of scrambling whereas $F^{xx}_{l}(t)$ shows scrambling 
inside the light cone of information even in the case of a periodically driven system. At the edge, $F^{zz}_{l}(t)$ is almost always close to 1 and $F^{xx}_{l}(t)$ is always close to $-1$ for the time interval we have considered. The velocity of information propagation, as seen in the OTOC plot, can be obtained by looking at the dispersion
of the Floquet bulk band and numerically calculating the maximum velocity, which is found to be $v_{\text{max}} = 0.98$ in this case.

If we turn off the static part of the magnetic field $h_0$ and set $h_1 = 1, ~
\omega = 2$, we obtain a system with an edge state with quasienergy 
$\epsilon = 0$ as shown in Fig.~\ref{fig4} (d). 
Further, since $\omega = 2 = J_x+J_y$, the quasienergy spectrum is gapless at quasienergy $\epsilon = \omega/2$ (see Fig.~\ref{fig2} (2) for the bulk dispersion). The effect of this edge state is 
clearly captured in the $F^{zz}_{l}(t)$ plot in Fig.~\ref{fig4} (e). The local information at the end site $n=0$ stays localized at all times due to the presence of a zero quasienergy topological state. This can be explained by the non-zero commutator of the edge $\sigma^z$ operators at different times, just as
in the case of the static system. $F^{xx}_{l}(t)$ also shows localization of information at the edge, due to the non-zero anticommutation of edge $\sigma^x$ operators at different times. In this case, 
the information carried by the bulk states into the system is suppressed, which can be observed 
from the fainter lines across the bulk in the plots of both $F^{zz}_{l}(t)$ and $F^{xx}_{l}(t)$. The velocity of information propagation turns out to be $v_{max} = 0.4$, which is much less than the case when $h_0$ is present. 

\begin{figure*}[htb]
\centering
\includegraphics[width=\textwidth]{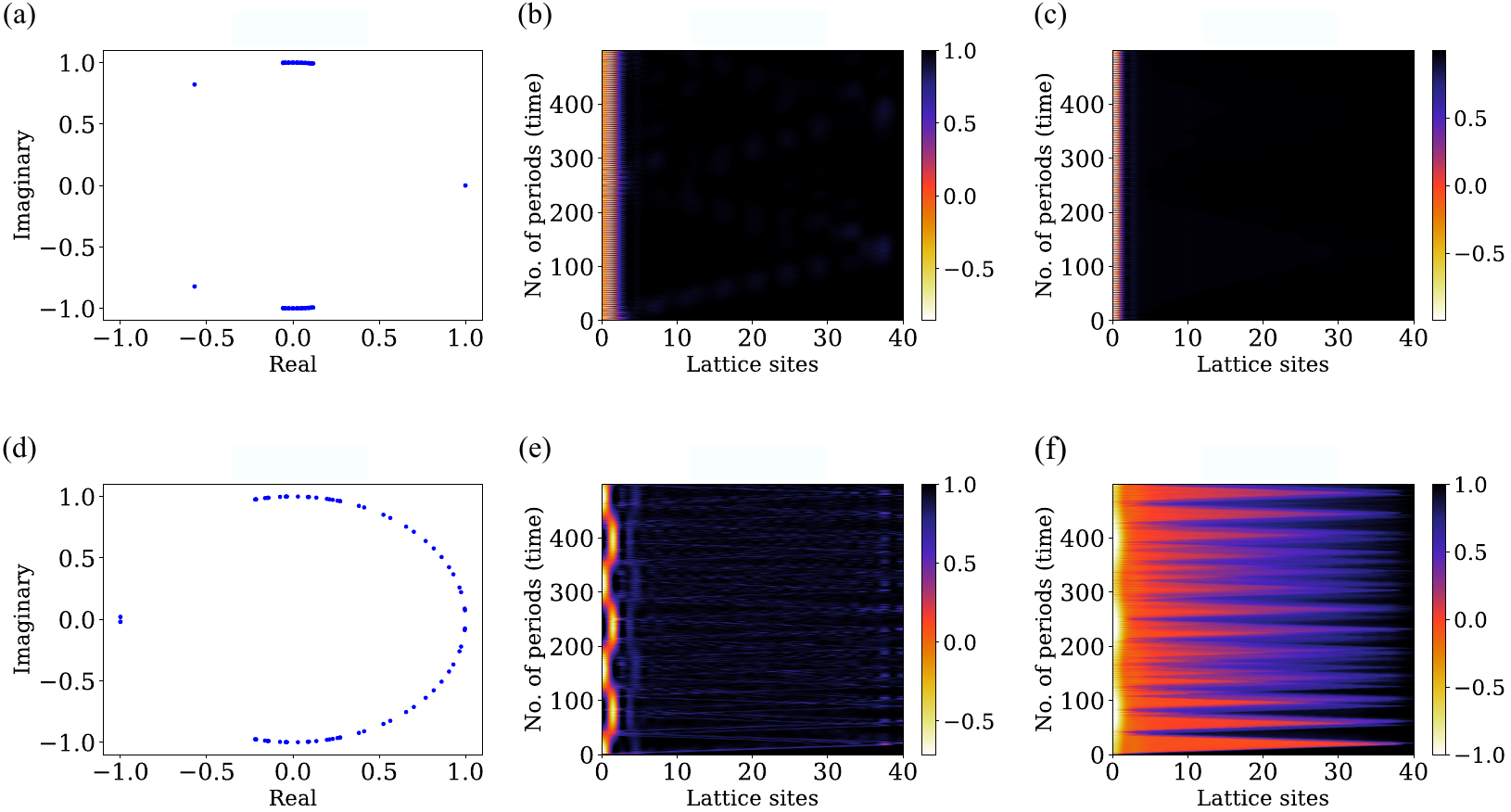}
\caption{Plots of Floquet eigenvalues, $F^{zz}_{l}(t)$ and $F^{xx}_{l}(t)$ for 
$J_x =1.8, ~J_y = 0.2, ~h_1 = 0.6$, and $\omega = 1.3$ in figures (a), (b) and (c) respectively. For these values of the parameters
both zero quasienergy topological edge states and non-topological edge states are present. The non-topological edge states give rise to oscillations at the edge in the OTOC plots. The same quantities are shown for $J_x =1.8, ~J_y = 0.2, ~h_1 = 1.2$ and $\omega = 1$ in figures (d), (e) and 
(f) respectively. The oscillations due to the non-topological edge states have a much longer period in this case. This is because the oscillation period is inversely proportional to the gap between the quasienergies of the non-topological edge states.} \label{fig5} \end{figure*}

We now consider the cases where there are non-topological edge states. We have 
considered two cases, one where both zero quasienergy topological 
edge states and non-topological edge states are present, and another where only the 
non-topological edge states are present. In Figs.~\ref{fig5} (a), (b) and (c) we show plots of the Floquet eigenvalues, $\sigma^z$ OTOC and $\sigma^x$ OTOC for the driving parameters $h_1 = 0.6$ and $\omega = 1.3$. The static part $h_0$ of the 
transverse field is set to zero. We observe that in addition to 
a pair of zero quasienergy topological states, we have two degenerate pairs of non-topological edge states, whose Floquet eigenvalues are well separated from the bulk states. The presence of non-topological edge states show a 
interesting effect at the edges of the OTOC plots. The information stays localized near the edge as 
in the cases with only zero quasienergy topological
edge states. In addition, we see oscillations of the OTOC in time, with 
a period $\tau \approx 15.6$ that depends inversely on the different between the quasienergies of the non-topological edge 
states. In this particular case, we also have a perfect localization of information near the edge 
with no information leaking into the bulk of the system. This is because of the nearly zero contribution of the bulk states at the edge site (see Fig.~\ref{fig2} (1)).

In the other case, the Floquet eigenvalues, $F^{zz}_{l}(t)$ and $F^{xx}_{l}(t)$ are shown for the driving parameters $h_1 = 1.2$ and $\omega = 1$ in Figs.~\ref{fig5} 
(d), (e) and (f). The parameters are chosen in such a way that the bulk dispersion is gapless at quasienergy $\epsilon = 0$ (see Fig.~\ref{fig2} (3)), and only two pairs 
of non-topological edge states are present. The OTOC at the edge shows an oscillatory behavior as in the previous case. The oscillation period in this case is much longer, 
with $\tau \approx 1005$, than in the previous case. This is because the difference between the quasienergies of the non-topological edge states is much smaller here. 
We will provide a more quantitative explanation of these oscillations of OTOC at the 
edge is given in Sec.~\ref{sec3d} in terms of the Floquet eigenfunctions and eigenvalues.
The values of the OTOC inside the bulk of the system indicate that information moving into the bulk is partially suppressed but not entirely. 

The $\si^x$ OTOC plots in Figs.~\ref{fig4} (c), \ref{fig4} (f) and 
\ref{fig5} (f) clearly show the unscrambling of quantum information.
We would like to emphasize here that the unscrambling effect is entirely due 
to reflections from the ends of the system. It does {\it not} occur in a system with
periodic boundary conditions~\cite{zamani2022out}. Further, unscrambling is not 
related to the phenomenon of Loschmidt revivals. A system with open ends and one with
periodic boundary conditions both have finite recurrence times; however only
the former system shows unscrambling.

We also note from Figs.~\ref{fig4} (c), \ref{fig4} (f) and \ref{fig5} (f) that the
unscrambling effect is very prominent following the first few reflections but gradually disappears after a large number of reflections. We have studied a larger
system with 80 sites and the same parameter values as in Figs.~\ref{fig4} (a-c),
and have found that the unscrambling still occurs after a few reflections and then gradually disappears (we have not shown this plot here). Thus the degree of 
unscrambling seems to depend only on the number of reflections and not the system
size.

\subsection{Exact expressions for OTOCs and the effects of edge states}
\label{sec3d}

We now provide some analytical expressions for the $\sigma^z$ OTOC and $\sigma^x$ OTOC 
using the Floquet eigenvalues and eigenstates. For $F_{l}^{zz}(t)$, we have relatively
simple expressions for any lattice site $l$. But for $F_{l}^{xx}(t)$ we have a simple
expression only at the edge site $l =0$ since, in the fermionic language
(Eq.~\eqref{jw}), the $\sigma^x_l$ operator at any site $l$ contains a string of
operators $e^{i \pi {\hat n}_j}$ from site 0 to site $l-1$ which becomes increasingly complicated as $l$ increases.

{\it (a) Expression for $F_{l}^{zz}(t)$:} We begin with the relation between 
the OTOC and the squared commutator given by
\begin{eqnarray} F_{l}^{zz}(t) &=& 1 - \frac{1}{2} \braket{ [\sigma^{z}_{l}(t), \sigma^{z}_{0}]^{\dagger} [\sigma^{z}_{l}(t),\sigma^{z}_{0}]} \non \\
&=& 1 + \frac{1}{2}\braket{[\sigma^{z}_{l}(t),\sigma^{z}_{0}][\sigma^{z}_{l}(t),
\sigma^{z}_{0}]}, \label{OTOC_com} \end{eqnarray}
where $\braket{O} = (1/2^{N}) \text{Tr}(O)$ denotes the infinite-temperature 
expectation 
value of an operator $O$. The commutator $[\sigma^{z}_{l}(t),\sigma^{z}_{0}]$ can be written in terms of Majorana operators as $-[a_{2l}(t)a_{2l+1}(t), a_0 a_1]$. We now use the mode expansions for the Majorana operators given by Eq.~\eqref{am}. In our analysis we will use the terms corresponding to the driven (Floquet) system. For an undriven system, 
we can replace $e^{-i p \theta_j}$ by $e^{-i E_j t}$, $a_m(pT)$ by $a_m(t)$ and Floquet eigenvectors by energy eigenvectors. For a driven system, $\alpha_j$ and $\alpha_j^{\dagger}$ in the mode expansion can be expressed as Majorana operators at time $t=0$ using Eq.~\eqref{alphaop}. Substituting these expressions in the commutator, we arrive at
\begin{equation} [\sigma^{z}_{l}(t=pT),\sigma^{z}_{0}] ~=~ -\sum_{m,n =0}^{2N-1} \mathcal{C}^{(m,n)}_{2l,2l+1}(t=pT) [a_m a_n, a_0 a_1], \end{equation}
where 
\bea \mathcal{C}^{(m,n)}_{2l,2l+1}(t=pT) &=& \frac{1}{2} \sum_{j,j'=0}^{N-1} ~
\big((\mathbb{A}^{-1})_{2l,2j} (\mathbf{v}^{+}_{j})_m e^{~-i p \theta_j} + (\mathbb{A}^{-1})_{2l,2j+1} (\mathbf{v}^{-}_{j})_m e^{~i p \theta_j} \big) \non \\
&& ~~~~~~~~~~~~~\times ~\big((\mathbb{A}^{-1})_{2l+1,2j'} (\mathbf{v}^{+}_{j'})_n 
e^{~-i p \theta_{j'}} + (\mathbb{A}^{-1})_{2l+1,2j'+1} (\mathbf{v}^{-}_{j'})_n
e^{~i p \theta_{j'}} \big). \label{cmn} \eea

Recalling that $(\mathbf{v}^{\pm}_{j})_m$ corresponds to the $m$-th component of the eigenvector with eigenvalue $e^{\pm i \theta_j}$, we rewrite $(\mathbf{v}^{+}_{j})_m$ ($(\mathbf{v}^{-}_{j})_m$) as $\psi_j(m)$ ($\psi_j^{\ast}(m)$) which are the wave functions in position space. Now, the wave function corresponding to the eigenvalue $e^{-i \theta_j}$ is the complex conjugate of the wave function corresponding to the eigenvalue $e^{i \theta_j}$ because the Floquet unitary matrix $U$ is purely real in the Majorana picture. Hence the orthogonality condition gives $\sum_{m} \psi_j(m) \psi_j(m) = \sum_{m} [\psi_j(m)]^{2} = 0$. The matrix $\mathbb{A}$ is made up of 
the orthonormal Floquet eigenvectors $\frac{1}{\sqrt{2}}\mathbf{v}^{+}_j$ and $\frac{1}{\sqrt{2}}\mathbf{v}^{-}_j$ along its rows. Hence the matrix elements of $\mathbb{A}^{-1}$ are given by $(\mathbb{A}^{-1})_{2l,2j} = \sqrt{2} \psi^{\ast}_j(2l)$ and $(\mathbb{A}^{-1})_{2l,2j+1} = \sqrt{2} \psi_j(2l)$.
Eq.~\eqref{cmn} then becomes
\bea \mathcal{C}^{(m,n)}_{2l,2l+1}(t=pT) &=& \sum_{j,j'=0}^{N-1} \big( \psi^{\ast}_j(2l) \psi_j(m) e^{~-i p \theta_j} + \psi_j(2l) \psi^{\ast}_j(m) 
e^{~i p \theta_j} \big) \non \\
&& ~~~~~~~~~~~\times ~\big(\psi^{\ast}_{j'}(2l+1) \psi_{j'}(m) e^{~-i p \theta_{j'}} + \psi_{j'}(2l+1) \psi^{\ast}_{j'}(m) e^{~i p \theta_{j'}} \big).
\eea

The commutator $[a_m a_n, a_0 a_1]$ is non-zero only when $m = 0$ or 1 but $n \neq 0,1$, or when $n=0$ or 1 but $m \neq 0,1$. Hence the $\sigma^z$-commutator is given by
\beq [\sigma^{z}_{l}(pT),\sigma^{z}_{0}] ~=~ \sum_{n \neq 0,1}^{2N-1} 2\Big[\big(\mathcal{C}^{(0,n)}_{2l,2l+1}(pT) - \mathcal{C}^{(n,0)}_{2l,2l+1}(pT)\big) a_0 a_n ~+~ \big(\mathcal{C}^{(n,1)}_{2l,2l+1}(pT)) - \mathcal{C}^{(1,n)}_{2l,2l+1}(pT)\big) a_1 a_n \Big]. \label{zcom} \eeq

Squaring Eq.~\eqref{zcom} and taking the trace in the many-body basis, we obtain the final expression for the infinite-temperature OTOC,
\begin{widetext}
\begin{align}
F_{l}^{zz}(pT) = 1 - 2\sum_{n=2}^{2N-1}\Big[ \big( \mathcal{\xi}^{(0)}_{2l}(pT) \mathcal{\xi}^{(n)}_{2l+1}(pT) - \mathcal{\xi}^{(n)}_{2l}(pT) \mathcal{\xi}^{(0)}_{2l+1}(pT) \big) ^{2} 
+ \big( \mathcal{\xi}^{(n)}_{2l}(pT) \mathcal{\xi}^{(1)}_{2l+1}(pT) - \mathcal{\xi}^{(1)}_{2l}(pT) \mathcal{\xi}^{(n)}_{2l+1}(pT) \big) ^{2} \Big],
\label{Fzz}
\end{align}
where
\begin{align}
\mathcal{\xi}^{(m)}_{2l}(pT) &= \sum_{j =0}^{N-1} ~\big(\psi^{\ast}_j(2l) \psi_j(m) e^{~-i p \theta_{j}} + \psi_j(2l)\psi^{\ast}_j(m) e^{~i p \theta_{j}} \big) \non \\
&= \sum_{j =0}^{N-1} ~2 ~\text{Re} \big(\psi^{\ast}_j(2l) \psi_j(m)\big) \cos
(p \theta_{j} + K^{m}_{j,2l}), \label{xi2l}
\end{align}
and $K^{m}_{j,2l}$ is a constant which depends on $\psi^{\ast}_j(2l)$ and $\psi_j(m)$.
\end{widetext}

Although the expressions in Eqs.~\eqref{Fzz} and \eqref{xi2l} look complicated, we can read out the effect of the edge states easily. Since $\psi_j(m)$ is the wave function 
in the Majorana basis, $m =0$ and 1 correspond to the sites at the left end in terms of Dirac fermions. If edge states are not present, we will have only bulk states whose wave functions are almost zero at the edge.
We now consider the OTOC only at the edge, i.e., $F_{l=0}^{zz}(t)$. In the absence of edge states, $\mathcal{\xi}^{(m)}_{2l = 0}(pT)$, $\mathcal{\xi}^{(m)}_{2l+1 = 1}(pT)$, $\mathcal{\xi}^{(m)}_{2l =0}(pT)$ and $\mathcal{\xi}^{(m)}_{2l+1 = 1}(pT)$ are all 
very close to zero; hence the term inside the summation on the right hand side of Eq.~\eqref{Fzz} almost vanishes. Thus, $F_{l=0}^{zz}(t) \approx 1$ at all times in the absence of edge states. We see this in the OTOC plots without edge states (see 
Figs.~\ref{fig3} (a) and \ref{fig4} (b)). On the contrary, if there are 
zero quasienergy topological edge states, the second term on the right hand side of Eq.~\eqref{Fzz} becomes 
non-zero. Also, since $e^{\pm i p\theta_j} =1$ in the case of zero quasienergy topological edge states (i.e., $\theta_j = 0$), $F_{l=0}^{zz}(t)$ becomes almost $p$-independent 
(time independent) and significantly less than $1$. This manifests as the bright lines at the edge of the system in Figs.~\ref{fig3} (c) and \ref{fig4} (e). 

If two almost degenerate pairs of non-topological edge states are present, the
two states have Floquet eigenvalues $e^{\pm i \theta_{1}}$ and $e^{\pm i 
\theta_{2}}$, where $\theta_1 \approx \theta_2$. Therefore $\xi^{m}_0$ and 
$\xi^{m}_1$ terms have a time-dependence which is
dominated by sums of terms involving $\cos (p\theta_1)$ 
and $\cos (p\theta_2)$ with almost equal factors since both the edge states have almost the same contribution at the edge. Putting these terms in 
Eq.~\eqref{Fzz}, we find 
that the squared terms in $F_{l=0}^{zz}(t)$ have a sinusoidal dependence on the sums and differences of the Floquet eigenvalues of the edge states. Quantitatively, 
$F_{l=0}^{zz}(t)$ becomes oscillatory, with the dominant frequencies being dependent 
on $\theta_1 +\theta_2$ and $\theta_1 -\theta_2$. Since the Floquet eigenvalues are almost equal, the oscillation with the shorter period is due to the frequency 
$\theta_1 +\theta_2 \approx 2 \theta_1 $. However if $2 \theta_1$ is close to $2 \pi$, we can take the frequency to be $|2\pi - 2\theta_1|$ since there is no difference between $\cos (2 \theta_1 p)$ and $\cos (|2 \pi - 2 
\theta_1| p)$ for integer values of $p$. We can clearly see these oscillations at 
the edge of the system in the OTOC plots in Figs.~\ref{fig5} (b) and (e). 

\begin{figure}[H]
\centering
\includegraphics[width=0.5\textwidth]{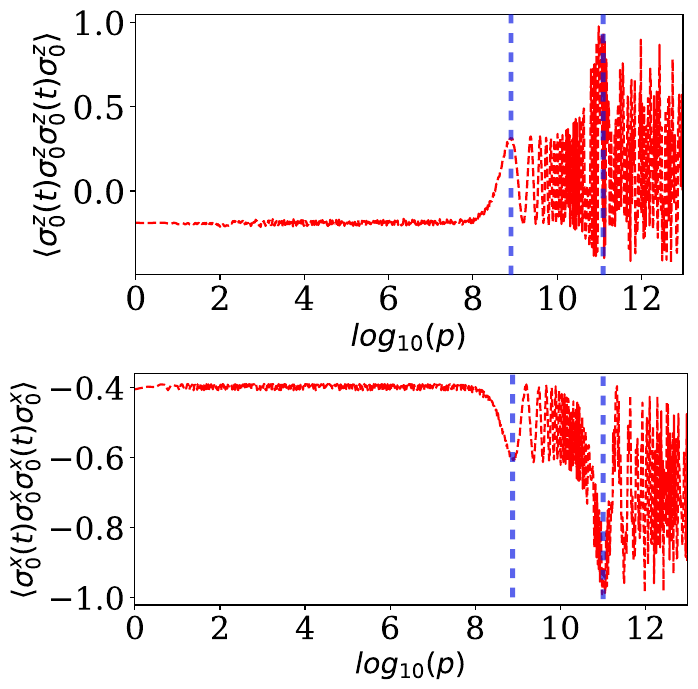}
\caption{Oscillations in OTOC with extremely large time periods due to the gaps between almost degenerate edge states for $J_x = 1.8, ~J_y = 0.2, ~h_1 = 0.8$,
$\omega = 1.4$, and $N=40$. Both zero quasienergy topological edge states and non-topological edge 
states exist for these parameter values with a degeneracy gap of $\Delta \epsilon \simeq 10^{-11}$ and
$~ 10^{-9}$ respectively. The oscillations are shown in a semi-log plot where a small-time average ($pT \sim 100$) has been performed to highlight only the long-time 
oscillations. The two long time periods of $10^9$ and $10^{11}$ have been marked 
using blue dotted lines.} \label{fig6} \end{figure}

We also have oscillations with a much larger period due to the small gaps $\theta_1 -
\theta_2$ between the almost degenerate non-topological edge states and also between the
almost degenerate zero quasienergy topological edge
states. Both these gaps are due to 
the finite sizes of the systems used for the calculation and they vanish
as we approach the thermodynamic limit. We demonstrate the long-period oscillations of $F_{l=0}^{zz}(pT)$ due to these gaps as a function of $\log_{10}(p)$ in 
Fig.~\ref{fig6} (top) for $J_x = 1.8$, $J_y =0.2$, $h_1 = 0.8$, $\omega =1.4$ and
$N=40$. To 
highlight the long periods we have done a time average over a smaller time scale $pT \sim 100$ to smoothen out the plots at that time scale. The dashed lines mark the periods on the log scale, and their inverses are the quasienergy gaps between the almost degenerate edge states. In this case the 
quasienergy gaps are $\Delta \epsilon \approx 10^{-9}$ for the non-topological edge states and 
$\Delta \epsilon \approx 10^{-11}$ for the zero quasienergy topological states. (We recall that the quasienergy $\epsilon_j$ is related to $\theta_j$ as $\theta_j = \epsilon_j T$).

The expressions in Eqs.~\eqref{Fzz} and \eqref{xi2l} also give us an understanding of the velocity of 
information propagation for the OTOC $F_l^{zz} (pT)$ in Eq.~\eqref{Fzz}
as follows. We note that the wave functions $\psi_j (2l)$ and 
$\psi_j (2l+1)$ correspond to a momentum $k= 2\pi j/N$,
and their dependence on $l$ therefore has the plane wave form $e^{ikl}$. The
factors $e^{\pm i p \theta_j}$ are equal to $e^{\pm i \epsilon_k t}$ where
$t=pT$. We therefore have a superposition of wave functions of the form $e^{i (kl \pm \epsilon_k t)}$ in Eq.~\eqref{xi2l}. A wave packet with this form will travel with 
a maximum velocity $dl/dt$ given by $v_{max} = (\partial \epsilon_k /\partial k)_{max}$. This explains the velocity with which the information is carried 
by the OTOC in Eq.~\eqref{Fzz}.

{\it (b) Expression for $F_{l=0}^{xx}(t)$:} The OTOC can also be written using the squared {\it anticommutator}
\begin{equation} F^{xx}_l(t) = -1 + \frac{1}{2} \langle \{\sigma^{x}_{l}(t),\sigma^{x}_{0}\}^{2} \rangle. \end{equation}
For $l =0$, the anticommutator $\{\sigma^{x}_{0}(t),\sigma^{x}_{0}\}$ is equal to 
$\{ a_0(t), a_0\}$ in terms of Majorana operators. We use the mode expansions of 
the Majorana operator $a_0(t)$ to obtain 
\begin{align}
\{a_0(t), a_0\} &= 2 \sum_{j=0}^{N-1} [\psi^{\ast}_{j}(0) \psi_{j}(0) e^{-ip\theta_j} + \psi_{j}(0) \psi^{\ast}_{j}(0) e^{ip\theta_j}] \non \\
&= 4 \sum_{j=0}^{N-1} |\psi_j(0)|^{2} \cos (p\theta_j). \label{anti}
\end{align}

In the absence of edge states of any type, $\psi_j(0) \approx 0$ for any $j$;
hence the contribution from the anticommutator is almost zero. Therefore the $\sigma^{x}$ OTOC is close to 
$-1$ at the edge as we can see in Figs.~\ref{fig3} (b) and \ref{fig4} (c). If zero quasienergy topological edge states are present, the coefficients $|\psi_j(0) |^{2}$ are only 
dominated by edge states and the time-dependence goes away. Therefore the OTOC at the edge, i.e., $F^{xx}_{l=0}(t)$ has a constant value greater than $-1$ which can be 
seen in Figs.~\ref{fig3} (d) and \ref{fig4} (f).

The presence of two pairs of non-topological edge states contributes to a term like $\cos (p \theta_1) + \cos (p \theta_2)$ in the anticommutator in 
Eq.~\eqref{anti}, where $\theta_1 \approx 
\theta_2$. The square of the anticommutator and hence the OTOC contain terms like $\cos (2p \theta_1)$, $\cos (2p \theta_2)$, $\cos (p(\theta_1 +\theta_2))$, and 
$\cos (p(\theta_1- \theta_2))$. Due to the almost degeneracy, the first three terms give rise to oscillations in $F^{xx}_{l=0}(t)$ with a frequency $2\theta_1$ or 
$|2\pi-2\theta_1|$,
whichever is smaller. These can be seen at the edges of the OTOC plots in 
Figs.~\ref{fig4} (c) and (f).

Similar to $F^{zz}_{l=0}(t)$, $F^{xx}_{l=0}(t)$ also has oscillations with a very long period due to the presence of a $\cos (p(\theta_1- \theta_2))$ term. Since $\theta_1 
\approx \theta_2$, this frequency is small and we have oscillations with a large period, which becomes infinity in the limit of infinite system size. These oscillations as well as the oscillations due to the quasienergy gap between
zero quasienergy topological edge states are shown in Fig.~\ref{fig6} (bottom) for $J_x = 1.8$, $J_y =0.2$, $h_1 = 0.8$ and $\omega =1.4$. Here again we have performed a time average over $pT \sim 100$ to highlight the long-period oscillations. The two long periods of $10^9$ and $10^{11}$ are marked with blue dashed lines.

To conclude, the presence of edge modes implies that the OTOCs at the end 
site, $l=0$, will remain non-zero (with or without oscillations) for an infinite 
amount of time; this follows from the discussion after Eq.~\eqref{anti}. This is
understandable since our model is integrable as mentioned at the end of Sec.~\ref{sec2a}.

To summarize this section, we have used mode expansions of $\si_l^z$ at any
site $l$ and $\si_l^x$ at $l=0$ in terms of Majorana
operators to quantitatively derive the effects of edge states on the OTOCs. In 
particular, both short-time and long-time oscillations in the OTOCS at the edge
of the system can be explained in terms of the differences in quasienergies of the
edge states.

\subsection{Behavior of $\sigma^{x}$-OTOC over very long times}
\label{sec3e}

\begin{figure}[H]
\centering
\includegraphics[width=0.95\textwidth]{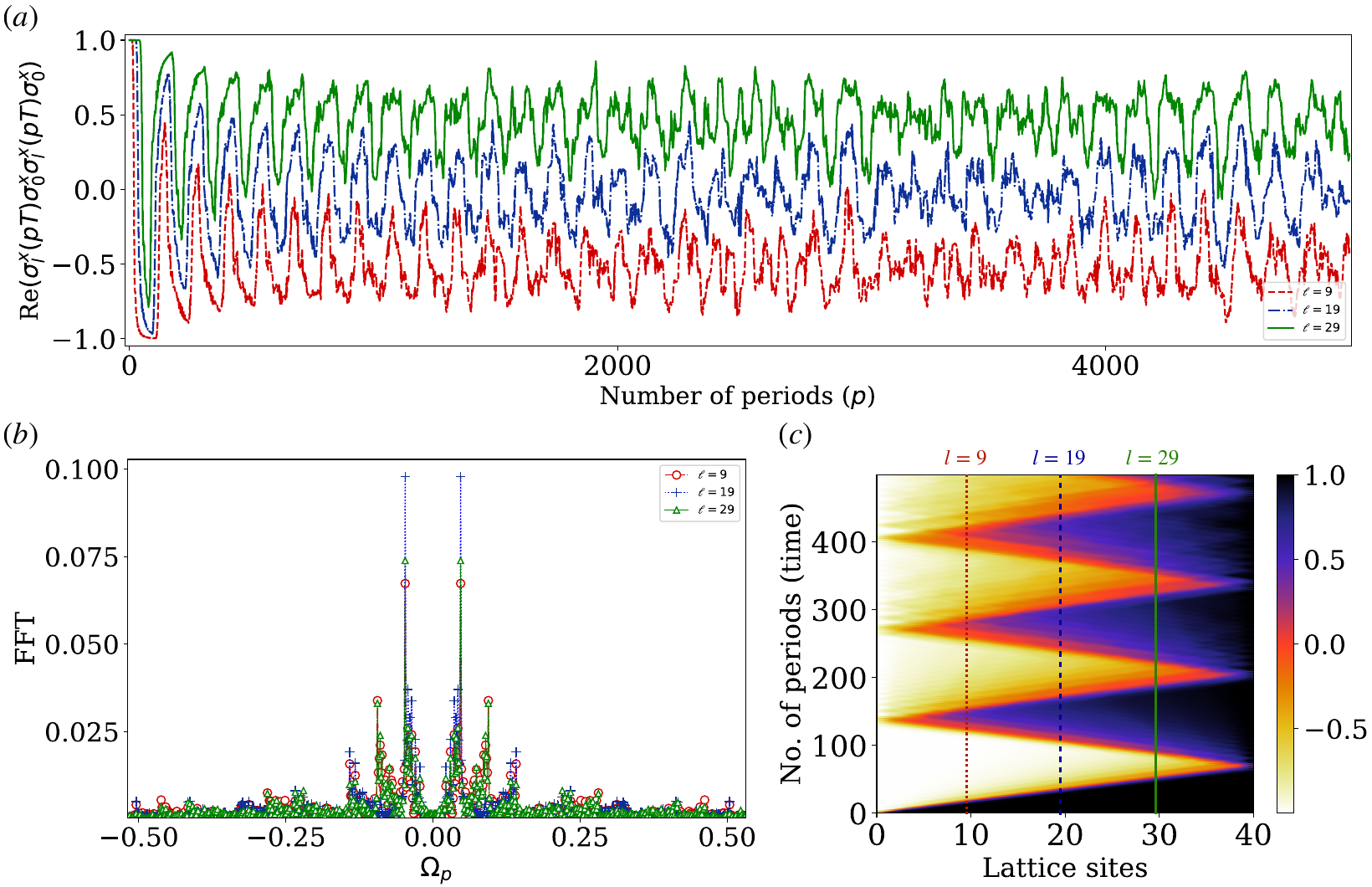}
\caption{{(a) $\sigma^{x}$-OTOC shown for a time $t=pT$,
where $p$ goes from 0 to 5000, for the driven system with parameters 
$J_x = 1.8, ~J_y =0.2, ~h_0 = 2.5, ~h_1 = 0.6$, $\omega = 10$ 
at the lattice sites $l = 9, ~19$ and $29$ for a 40-site
system. (b) The Fourier transforms of the $\sigma^{x}$-OTOC for the time interval $p=[1000,5000]$ show pronounced peaks at 
$\Omega_p \simeq 2 \pi \times 0.0075$. The smaller peaks correspond to harmonics with double and triple the fundamental frequency. (c) The lattice sites where the 
long-time behavior has been shown in plots (a) and (b) are marked on the 
space-time plot of the $\sigma^{x}$-OTOC.}} \label{fig7} \end{figure}

It is interesting to look at the scrambling and unscrambling given by the 
$\sigma^x$-OTOC over very long times at different sites. This is shown in 
Fig.~\ref{fig7} for a periodically driven system with parameters 
$J_x = 1.8, ~J_y =0.2, ~h_0 = 2.5, ~h_1 = 0.6$, $\omega = 10$ 
at three different lattice sites $l = 9, ~19$ and $29$ for a 40-site
system; we note that there are no end modes for these parameter values.
(Since the lattice site label goes from 0 to 39, the three chosen
sites are located at distances equal to $L/4$, $L/2$ and $3L/4$ from
the left end of the system). Remarkably, we see in Fig.~\ref{fig7} (a)
that the oscillations in the OTOC denoting successive scrambling and 
unscrambling persist for a very long time, and they do not show an appreciable
decay with time; we do not have an analytical understanding of this
phenomenon. Fig.~\ref{fig7} (b) shows the Fourier transform of the 
$\sigma^{x}$-OTOC for the time interval $p=[1000,5000]$; we see pronounced peaks 
at $\Omega_p \simeq 2 \pi \times 0.0075$. (To calculate the Fourier transform we
have omitted the first 1000 values of $p$ to avoid the initial
transient effects. Also,
we have not shown the Fourier transforms at zero frequency since those are
much larger than the peaks at $\pm \Omega_p$; their values are given by
the long-time averages of the OTOCs in Fig.~\ref{fig7} (a) which are non-zero).
This value of $\Omega_p$ corresponds to the periodic structure of the OTOC 
arising from successive scrambling and unscrambling with a time period of 
$\Delta t = (\Delta p) T \simeq 133$; this is the recurrence time given by 
$2L/v_{max}$ for one back-and-forth motion across the system,
where $L=40$ and $v_{max}$ is the maximum velocity. Hence $\Omega_p =
2 \pi/\Delta t$. The smaller peaks correspond to higher harmonics 
lying at $\pm 2 \Omega_p$ and $\pm 3 \Omega_p$. Fig.~\ref{fig7} (c) indicates
the three lattice sites where the long-time behavior of the OTOC
has been shown in Figs.~\ref{fig7} (a) and (b). Interestingly, we find that
the long-time average values of the $\sigma^x$-OTOC at the three sites 
$l=9, 19$ and $29$ are given by $-0.513,~ - 0.025$ and $0.462$ respectively.
These values seem to be increasing linearly with $l$ which suggests that the long-time average of the OTOC at site $l$, which goes from
0 to $L-1$ in an $L$-site system, is 
given approximately by the simple expression $(2l/(L-1)) - 1$.

\subsection{Velocity of information propagation for different drive frequencies}
\label{sec3f}

\begin{widetext}
\begin{figure*}[htb]
\centering
\includegraphics[width=0.9\textwidth,height=0.5\textwidth]{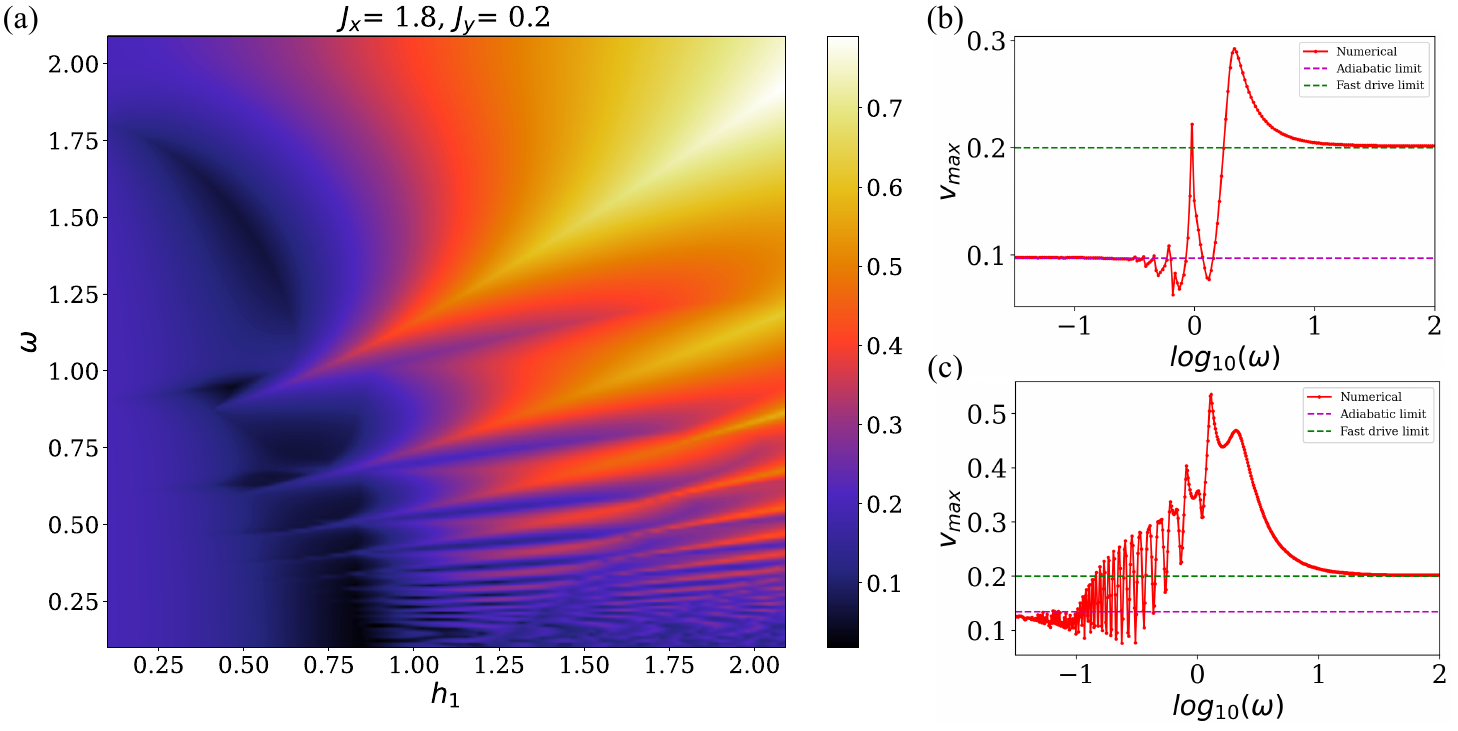}
\caption{(a) Variation of the velocity of information propagation with the driving amplitude $h_1$ and frequency $\omega$ for $J_x =1.8$ and $J_y =0.2$. (b) and (c) 
shows the same quantities for fixed values of $h_1 = 0.6$ and $1.2$ respectively 
for a larger range of $\omega$. In either case the velocities match the limits found 
analytically for very fast driving ($\omega \gg J_x, ~J_y, ~h_1$ ) and very slow 
driving ($\omega \ll J_x, ~J_y, ~h_1$).} \label{fig9} \end{figure*}
\end{widetext}

We have demonstrated that the propagation of local information through a driven spin chain can be better understood in terms of the Jordan-Wigner fermions, and it is carried by 
the bulk states having the maximum velocity, i.e., $v_{\text{information}} = \bigl( \partial E_k / \partial k \bigr)_\text{max} = v_{\text{max}}$. Further, the propagation from one end of the system is seen to be suppressed if edge states are present since such states tend to partially trap some information at the end. 
We find numerically that the velocity depends non-trivially on the amplitude and 
frequency of the periodic drive. The variation of the velocity with the driving amplitude $h_1$ and frequency $\omega$ is shown in the Fig.~\ref{fig9} (a). The 
regions in the plot where the velocity changes abruptly has the values of both 
$h_1$ and $\omega$ close 
to 1, and therefore, we cannot provide a perturbative understanding of these
regions. However we observe that the velocity generally increases with $h_1$.
We have also shown the dependence on $\omega$ of the velocity for $h_1 = 0.6$ and $1.2$ in Figs.~\ref{fig9} (b) and (c). For a particular $h_1$ we can understand 
the fast 
driving limit ($\omega \gg J_x, J_y, h_1$) and the adiabatic limit $\omega \ll J_x, J_y, h_1$). For very large $\omega$ the effective bulk Hamiltonian in $k-$ space can 
be obtained from Eq.~\eqref{ham-k2} as
\bea h^{\text{fast}}_{k} &=& \frac{1}{T}~ \int_{0}^{T} dt ~\Tilde{h}_{k}(t) 
\non \\
&=& - ~\frac{(J_x + J_y)}{2} \cos k ~\tau^{z} ~+~ \frac{(J_x -J_y)}{2} \sin k ~\tau^{y}. \eea
Note that $h_k^{\rm fast}$ is simply the time-independent part of the Hamiltonian in Eq.~\eqref{ham-k2}; this is expected in the high-frequency limit~\cite{bukov2015,mikami2016}.
The velocity is obtained by differentiating the dispersion $E^{\text{fast}}_{k}$ with respect to $k$ and then finding the maximum group velocity. In this case the dispersion looks like,
\begin{equation}
E^{\text{fast}}_{k} ~=~ \pm ~\sqrt{\left (\frac{J_x+J_y}{2} \cos k \right)^{2} + \left (\frac{J_x-J_y}{2} \sin k \right )^2}. \end{equation}
We note that for very fast driving the dispersion and hence the velocity are independent of $h_1$. Indeed we observe that for $J_x = 1.8$ and $J_y =0.2$, 
$v_{\text{max}}$ turns out to be $0.2$ irrespective of the value of $h_1$, as shown by the green dashed lines in Figs.~\ref{fig9} (b) and (c).

On the other hand for very slow driving frequency $\omega$, we can make an adiabatic approximation of the Hamiltonian. In this case the variable $\omega t$ is considered 
to be a slowly varying parameter. The Floquet quasienergy bands are then obtained by integrating the instantaneous eigenvalues $E_k(t)$ of $\Tilde{h}_{k}(t)$ over a 
period $T$~\cite{sen2021}. Namely,
\beq E^{\text{slow}}_{k} ~=~ \pm ~\frac{1}{T} ~\int_{0}^{T} dt ~\sqrt{ \left( \frac{J_x+J_y}{2} ~\cos k - h_1 \sin (\omega t) \right)^{2} ~+~
\left(\frac{J_x-J_y}{2} ~\sin k \right)^2 }. \eeq

We find that $v_{\text{max}} = 0.097$ and $0.134$ for $h_1 = 0.6$ and $1.2$ respectively, which are marked with purple dashed lines in Figs.~\ref{fig8} 
(b) and (c). For moderate values of $\omega$, the maximum velocity fluctuates rapidly and attains a maximum at some value of $\omega$ before decreasing steadily to the
value corresponding to fast-drive limit. The maximum value depends on the parameters $J_x$, $J_y$ and the driving amplitude $h_1$.

\subsection{Infinite-temperature OTOCs in the presence of interactions}
\label{sec3g}

In this section we will briefly study the effects of interactions which
break integrability on the infinite-temperature OTOCs. For convenience, we will 
consider an undriven $XY$ chain with a constant $\sigma^z$ field and 
interactions. More precisely, the Hamiltonian of the model is given by
\begin{equation}
\mathcal{H} ~=~ \frac{J_x}{4} ~\sum_{j =0}^{N-2} \sigma^{x}_{j}
\sigma^{x}_{j+1} ~+~ \frac{J_y}{4} ~\sum_{j =0}^{N-2} \sigma^{y}_{j}
\sigma^{y}_{j+1} ~+~ \frac{J_z}{4} ~\sum_{j =0}^{N-2} \sigma^{z}_{j}
\sigma^{z}_{j+1} ~+~ \frac{h}{2} ~\sum_{j=0}^{N-1} \sigma^{z}_{j},
\label{xyzham} \end{equation}
where $J_z$ denotes the strength of density-density interactions 
between the Jordan-Wigner fermions ($\sigma_j^z$ is related to the
density
of fermions according to Eq.~\eqref{jw}). The model is 
non-integrable if $J_x \ne J_y$, $J_z \ne 0$ and $h \ne 0$.
Note that the Hamiltonian
is symmetric under the $Z_2$ transformation $\sigma_j^x \to - 
\sigma_j^x$, $\sigma_j^y \to - \sigma_j^y$ and $\sigma_j^z \to \sigma_j^z$. If we choose $J_x$ to be much larger than the
other parameters, the system will be in a $\sigma^x$ spin-ordered phase or a topological phase in the fermionic language.

\begin{widetext}
\begin{figure*}[htb]
\centering
\includegraphics[width=0.9\textwidth,height=0.5\textwidth]{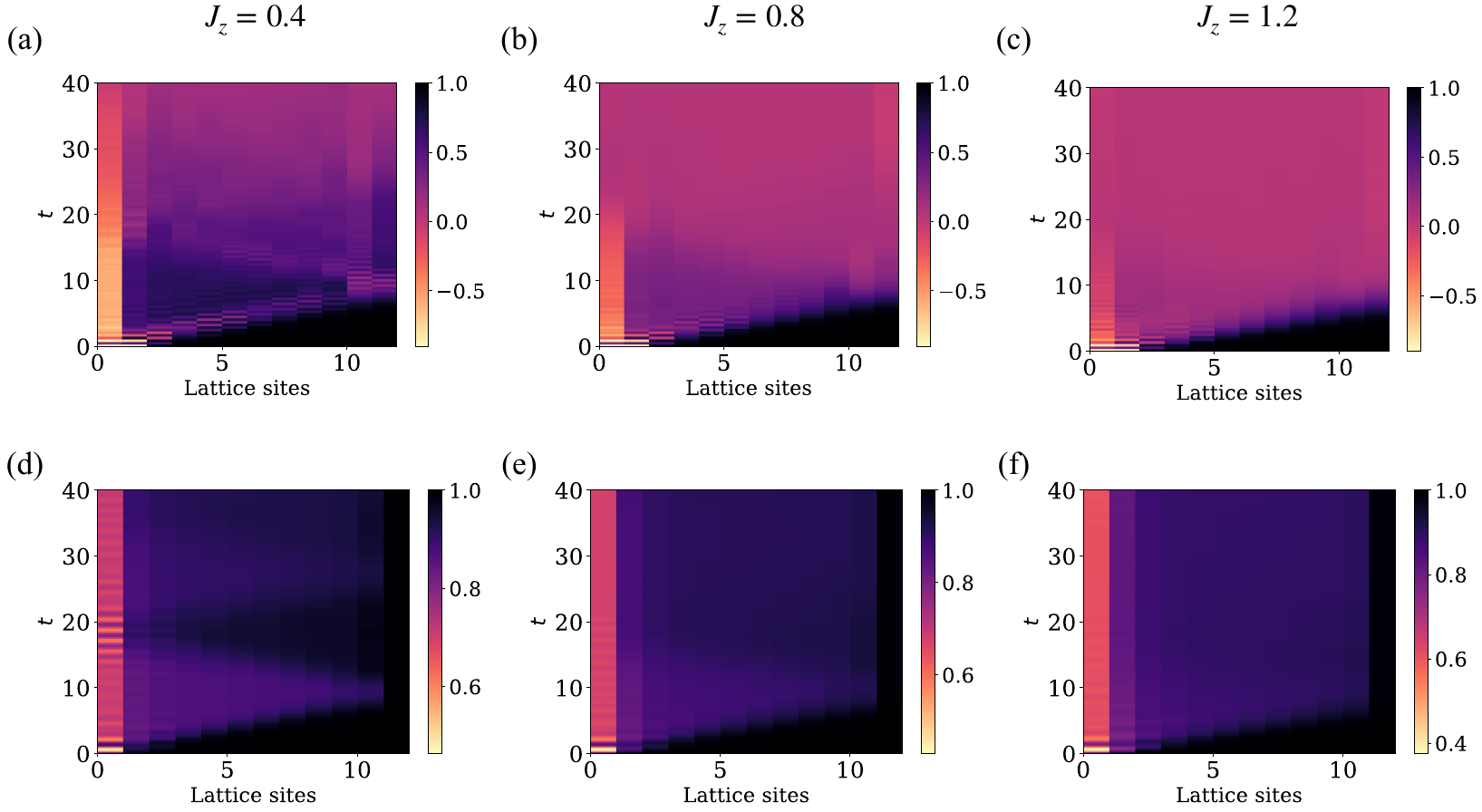}
\caption{Plots of $F^{zz}(l,t)$ for the model given in Eq.~\eqref{xyzham} for 
$J_x = 7.2$, $J_y = 0.8$, and $h = 1$, while $J_z$ takes the values 
(a) $0.4$, (b) $0.8$, and (c) $1.2$, for a system with 12 sites.
The local quantum information gets completely scrambled throughout the system as the interaction is increased. The plots of $F^{xx}(l,t)$ for different values of $J_z$, (d) 
$0.4$, (e) $0.8$, and (f) $1.2$, show partial localization of the OTOC at the edge due to the presence of an almost strong zero mode. However, the unscrambling effect of 
$F^{xx}(l,t)$ goes away with increasing interaction.} \label{fig8} 
\end{figure*}
\end{widetext}

Since we cannot use Wick's theorem to evaluate the OTOCs in this model with
interactions, we evaluate them numerically using
exact diagonalization for a system of size $N=12$. We will plot $F^{zz}(l,t)$ for 
different values of $J_z$, while the other parameters are chosen to be $J_x = 
7.2$, $J_y = 0.8$, and $h = 1$; this choice of parameters makes the model lie in
the topological phase in terms of Jordan-Wigner 
fermions. The plots for $F^{zz}(l,t)$ for $J_z = 0.4, ~0.8$ and 
$1.2$ are shown in Figs.~\ref{fig8} (a), (b) and (c) respectively, for a 
12-site system. While the 
$\sigma^z$ OTOC for a non-interacting system only showed lines 
of information propagation and no scrambling, we find that even with a small 
interaction, the local information gets scrambled over the entire 
range of $l$ and $t$. Moreover, the localization of the $\sigma^z$
OTOC at the edge gradually goes away with increasing interaction 
strength after a finite 
amount of time. Hence the $\sigma^{z}$ OTOC does not show localization due to 
topological end modes when the interaction becomes moderately strong.


The plots for $F^{xx}(l,t)$ in the presence of interactions, shown in Figs.~\ref{fig8} (d), (e) and (f), 
correspond to the same parameter values as above. In striking contrast to $F^{zz}(l,t)$, we see here that the localization 
of the OTOC at the edge persists even in the presence of interactions. This remarkable observation can be attributed 
to the presence of an almost strong zero edge mode which is described below.

The ordered phase of the non-interacting 
model on an open-ended chain without a $J_z$ term is known to host a strong zero 
mode operator localized at the edge. The strong zero mode commutes with the
Hamiltonian in the thermodynamic limit and switches 
between the different fermion parity states in the ordered phase~\cite{fendley2012,fendley2015}. 
Numerically, a local operator at the edge ($\sigma^{x}_0$ in this case) having 
significant overlap with the strong zero mode operator 
exhibits a long coherence time under unitary evolution by the 
Hamiltonian~\cite{kemp2017long,yates2020}. 
This time scale can be shown to be exponential in the system size. In the 
presence of the $J_z$ (interaction) term, the strong zero mode becomes 
what is known as an almost strong zero mode~\cite{kemp2017long}, which 
also exhibits a long coherence time for 
the same edge operators($\sigma^{x}_0$), although it is not exponential in system 
size anymore. This gets reflected in the 
value of $F^{xx}(l,t)$ at the edge ($l=0$), which does not fade away within 
the time scale considered even when the interaction strength is increased.
To conclude, in the topological phase, the model has a strong zero mode in
the absence of interactions and an almost strong zero mode in the presence
of interactions. The two kinds of modes have almost the same effect near
the end of the system till we go to very large times which have not been 
considered in this paper.

Finally, we observe that (i) the unscrambling effect in the $\si^x$ OTOC in 
the bulk of the system goes away with increasing interaction strength, and 
(ii) only part of the local operator $\sigma_0^x$ stays at the
edge while the rest of it gets scrambled over the entire system.

\section{Discussion}
\label{sec4}

In this paper, we have used OTOCs to study the effects of edge states on
information propagation across a periodically driven spin-1/2 chain with open
boundary conditions. The spin chain is taken to have nearest-neighbor $xy$ couplings
and a transverse magnetic field; we have studied what happens when this field is driven
sinusoidally in time. The spin-1/2 operators can be mapped to 
Majorana operators using the Jordan-Wigner transformation. The Majorana
formalism simplifies the analytical and numerical calculations because the
Hamiltonian turns out to be quadratic in terms of Majorana operators, and
correlations functions like OTOCs can be computed using Wick's theorem in some cases. 
The OTOCs are given by expectation values of products of four spin operators of the 
form $\si_l^\al (t) \si_0^\al \si_l^al (t) \si_0^\al$ in the infinite-temperature 
ensemble (i.e., the contributions of all Floquet eigenstates are added 
up with equal weights). Here the subscripts 0 and $l$ refer to the lattice
site; we have chosen two of
the lattice sites to be at 0 since we are specifically interested in the 
effects of edge states on the OTOCs. We allow the other site index, $l$, 
to vary over all the sites of the lattice and
the time $t$ (taken to be $pT$, where $p$ is an integer and $T$ is
the driving time period) to vary over a range which is much larger than $T$ 
in order to obtain
a detailed space-time picture of the OTOCs. We have studied OTOCs for $\al = x, z$, 
i.e., $\si^z$ and $\si^x$ spin operators. These show different behaviors due to the 
fact that the former is local while the latter is non-local in terms of the Majorana
operators. We find that a sinusoidal driving of the transverse field can give rise 
to both topological edge states and non-topological edge states at the
ends of a long chain. (The topological states have quasienergies lying at the
centre and edge of the Floquet Brillouin zone (quasienergies equal to zero and
$\om/2$ respectively), while the non-topological states have quasienergies lying
away from both zero and $\om/2$). We find that the two kinds of edge 
states, topological and non-topological, have strikingly different effects on the 
information propagation from the ends of the system.

We have studied a number of different values of the parameters of the driven system 
which give rise to different 
edge state structures: there may be no edge states, only zero quasienergy 
topological or 
only non-topological edge states, or both zero quasienergy topological
and non-topological edge states.
We find that the behavior of both $\si^z$ and $\si^x$ OTOCs in the bulk (i.e., 
when the site index 
$l$ is not too close to either end of the system) does not depend qualitatively 
on the presence of edge states; however, the values of the OTOCs deviate less
from 1 if edge states are present because edge states trap some of the information thereby reducing the contribution from the bulk states. 
The $\si^z$ OTOCs are close to 1 (i.e., $\si_0^z (0)$ and $\si_l^z (t)$
almost commute) in most of the $l-t$ plane
except on certain straight lines which traverse the system
from one end to the other. The slopes of these lines are given by $\pm 
v_{max}$, where $v_{max}$ is the maximum value of the velocity 
$v = \partial \epsilon_k /\partial k$, with $\epsilon_k$
being the quasienergy of the bulk state with momentum $k$. 

The $\si^x$ OTOCs show a more complicated behavior in the bulk. As $l$ increases 
linearly with $t$ on going from the initial space-time point $(l,t) = (0,0)$ 
lying at the left end of the system to
the right end $(l=N,t=N/v_{max})$, we find that the OTOC in the region 
above that line differs significantly from 1. This region 
corresponds to the scrambling of information ($\si_0^x (0)$ and $\si_l^x (t)$ 
no longer commute). However, after there is a 
reflection from the right end and $(l,t)$
returns from $(l=N,t=N/v_{max})$ back to the left end $(l=0,t=2N/v_{max})$, 
the OTOC in the region above the second line is again 
close to 1, corresponding to unscrambling of information. This pattern
repeats: the $\si^x$ OTOC in the regions between the lines $(l=0,t=2pN/v_{max})$ 
and $(l=N,t=(2p+1)N/v_{max})$ (where $p=0,1,2,\cdots$)
differs significantly from 1 (scrambling of information), while in all the 
other regions it is close to 1. The repeated scrambling and unscrambling of 
information as $(l,t)$ moves back and forth across the system is clear
from the plots of the $\si^x$ OTOCs, but a quantitative explanation of 
these phenomena is currently lacking~\cite{lin2018out}. This may be an 
interesting problem for future studies.
{Further, we have studied the $\si^x$ OTOC over a very long period of
time at different sites in the bulk of a periodically driven system. We 
find that this appears to oscillate forever, indicating a persistent
scrambling and unscrambling of information.}

The maximum velocity $v_{max}$ depends on the system parameters, and its
value can be obtained from the slopes of the OTOCs in the $(l,t)$ plane.
It can also be found analytically in the limits of very large or very small driving
frequencies, and we find that these limiting values agree with the numerically
obtained results.

In contrast to the OTOCs in the bulk, the behaviors of both $\si^z$ and $\si^x$ 
OTOCs near the left end ($l=0$) depend strongly on the presence and type of edge 
states. If there 
are no edge states, the OTOCs do not show any special features at $l=0$.
If there are only zero quasienergy topological
edge states, then we find that the OTOCs
at $l=0$ differ considerably from 1; however, their values do not vary
much with the time $t$. (This shows up in the plots as a vertical line at 
$l=0$ which has a different color from the OTOCs in the bulk). This is 
because zero quasienergy states have Floquet
eigenvalues equal to $e^{i\theta_j} = 1$ and therefore $e^{i p \theta_j} =1$
for all $p$. However, if there are pairs of non-topological edge states with Floquet 
eigenvalues $e^{\pm i \theta_j}$ which are not equal to 1 
or $-1$. Then there are cross-terms arising from the product of these states;
these terms take the form $\cos (2p \theta_j)$ and therefore oscillate with $p$. 
This leads to oscillations of the OTOCs as $t=pT$ increases. These oscillations 
clearly show up in the plots of the OTOCs close to $l=0$. The oscillations
are not localized exactly at $l=0$ and are instead visible in a finite but small region 
around $l=0$; this occurs because the wave functions of the non-topological edge 
states have a finite decay length.

For comparison, we have also looked at the OTOCs of the undriven model where the 
transverse field
is time-independent. This model either has no edge states or only has zero energy 
edge states, depending on whether it is in the non-topological phase or topological
phase. The behaviors of the $\si^z$ and $\si^x$ OTOCs is similar to those of the
driven model for the cases of no edge states or zero quasienergy 
topological edge states.
Note that the undriven model has no analog of non-topological edge states, hence the
OTOC at $l=0$ never shows any oscillations with time.

For the driven system, the edge states (either topological
or non-topological) have
a two-fold degeneracy in their quasienergies if the system is infinitely long,
since the states at the two edges will have the same quasienergy in that case.
For a system with a finite length $N$, the states at the two edges hybridize with
each other, and this breaks the degeneracy by an amount which is exponentially 
small in the system size (the quasienergy splitting is of the order of 
$e^{-N/\lambda}$, where $\lambda$ is the decay length of the edge states).
The cross-terms arising from pairs of such states
give rise to oscillations with very long periods, since the oscillation period
is inversely proportional to the quasienergy splitting. We have numerically
verified the existence of these oscillations with extremely long time periods
arising from either the zero quasienergy topological or the non-topological edge states.

The model we have studied in this paper is integrable because
the Hamiltonian at any time is quadratic in terms of Majorana operators. 
We have briefly studied the effects of the breaking of integrability
on the OTOCs by looking at an undriven model with four-fermion 
interactions and choosing the parameters in such a way that the
system lies in a topological phase. We find that the localization of
the $\sigma^x$ OTOC at the end site persists for a long time in the presence of
interactions due to the presence of an almost strong zero mode. 
However the unscrambling effect in the $\sigma^x$ OTOC disappears 
and a scrambling effect appears in the $\sigma^z$ OTOC as the interaction
strength is increased.

To summarize, our main conclusions are as follows. First, the OTOC of 
$\si^x$ (which is non-local in terms of fermions) in the bulk of a system with 
open boundary conditions shows both scrambling and unscrambling effects as one of the
coordinates of the OTOC goes between the two ends of the system. The 
unscrambling effect has not been reported earlier; it arises due to reflections
from the ends of the system. Second, the presence of edge states (both topological 
and non-topological) gives rise to clear signatures in the OTOCs 
near the ends of the system. Namely, the OTOC has a very different value at the 
ends compared to the bulk due to the trapping of some of the information by edge 
states, and non-topological edge states give rise to prominent oscillations in the OTOCs.

Finally, we turn to experimental measurements and proposals for studying
OTOCs in different systems~\cite{Swingle,Zhu,Garttner2017,Li,Lewis,Nie,Braumuller2022,zhao2022}.
OTOCs have been found to be useful for measuring entanglement entropy and the
velocity of correlation propagation~\cite{Li} and detecting dynamical quantum phase 
transitions~\cite{Nie} in the transverse field Ising chain using an NMR quantum
simulator with 4 qubits. OTOCs have been used to study the buildup of many-body correlations in an Ising spin quantum simulator based on more than 100 trapped ions~\cite{Garttner2017}, and to study the propagation of quantum information
in a superconducting circuit which simulates a hard-core Bose-Hubbard model~\cite{Braumuller2022}. OTOCs have been used to study operator spreading 
in a periodically driven superconducting chain with 10 qubits~\cite{zhao2022}.
There are proposals to use OTOCs for studying spin models for a system
of cold atoms in an optical cavity~\cite{Swingle} and a cavity quantum electrodynamics setup~\cite{Zhu}. There is also a proposal to study OTOCs in the
Dicke model (a single large spin coupled to a simple harmonic oscillator) to
understand the relations between scrambling, entanglement and thermalization~\cite{Lewis}. Given the wide variety of systems in which
OTOCs can be measured, we believe that it would be possible to test our
results on the effects of edge states on the OTOCs of a one-dimensional
system. Such a study would of course require a large enough system so that 
the effects of edge and bulk states can be distinguished from each other.

\vspace{.8cm}
\centerline{\bf Acknowledgments}
\vspace{.4cm}

S.S. thanks MHRD, India for financial support through the PMRF.
D.S. acknowledges funding from SERB, India (JBR/2020/000043).

\bibliography{references}

\begin{thebibliography}{115}%
\makeatletter
\providecommand \@ifxundefined [1]{%
 \@ifx{#1\undefined}
}%
\providecommand \@ifnum [1]{%
 \ifnum #1\expandafter \@firstoftwo
 \else \expandafter \@secondoftwo
 \fi
}%
\providecommand \@ifx [1]{%
 \ifx #1\expandafter \@firstoftwo
 \else \expandafter \@secondoftwo
 \fi
}%
\providecommand \natexlab [1]{#1}%
\providecommand \enquote  [1]{``#1''}%
\providecommand \bibnamefont  [1]{#1}%
\providecommand \bibfnamefont [1]{#1}%
\providecommand \citenamefont [1]{#1}%
\providecommand \href@noop [0]{\@secondoftwo}%
\providecommand \href [0]{\begingroup \@sanitize@url \@href}%
\providecommand \@href[1]{\@@startlink{#1}\@@href}%
\providecommand \@@href[1]{\endgroup#1\@@endlink}%
\providecommand \@sanitize@url [0]{\catcode `\\12\catcode `\$12\catcode
  `\&12\catcode `\#12\catcode `\^12\catcode `\_12\catcode `\%12\relax}%
\providecommand \@@startlink[1]{}%
\providecommand \@@endlink[0]{}%
\providecommand \url  [0]{\begingroup\@sanitize@url \@url }%
\providecommand \@url [1]{\endgroup\@href {#1}{\urlprefix }}%
\providecommand \urlprefix  [0]{URL }%
\providecommand \Eprint [0]{\href }%
\providecommand \doibase [0]{http://dx.doi.org/}%
\providecommand \selectlanguage [0]{\@gobble}%
\providecommand \bibinfo  [0]{\@secondoftwo}%
\providecommand \bibfield  [0]{\@secondoftwo}%
\providecommand \translation [1]{[#1]}%
\providecommand \BibitemOpen [0]{}%
\providecommand \bibitemStop [0]{}%
\providecommand \bibitemNoStop [0]{.\EOS\space}%
\providecommand \EOS [0]{\spacefactor3000\relax}%
\providecommand \BibitemShut  [1]{\csname bibitem#1\endcsname}%
\let\auto@bib@innerbib\@empty
\bibitem [{\citenamefont {Roberts}\ and\ \citenamefont
  {Swingle}(2016)}]{Roberts}%
  \BibitemOpen
  \bibfield  {author} {\bibinfo {author} {\bibfnamefont {D.~A.}\ \bibnamefont
  {Roberts}}\ and\ \bibinfo {author} {\bibfnamefont {B.}~\bibnamefont
  {Swingle}},\ }\href {\doibase 10.1103/PhysRevLett.117.091602} {\bibfield
  {journal} {\bibinfo  {journal} {Phys. Rev. Lett.}\ }\textbf {\bibinfo
  {volume} {117}},\ \bibinfo {pages} {091602} (\bibinfo {year}
  {2016})}\BibitemShut {NoStop}%
\bibitem [{\citenamefont {Luitz}\ and\ \citenamefont {Bar~Lev}(2017)}]{Luitz}%
  \BibitemOpen
  \bibfield  {author} {\bibinfo {author} {\bibfnamefont {D.~J.}\ \bibnamefont
  {Luitz}}\ and\ \bibinfo {author} {\bibfnamefont {Y.}~\bibnamefont
  {Bar~Lev}},\ }\href {\doibase 10.1103/PhysRevB.96.020406} {\bibfield
  {journal} {\bibinfo  {journal} {Phys. Rev. B}\ }\textbf {\bibinfo {volume}
  {96}},\ \bibinfo {pages} {020406} (\bibinfo {year} {2017})}\BibitemShut
  {NoStop}%
\bibitem [{\citenamefont {Wei}\ \emph {et~al.}(2018)\citenamefont {Wei},
  \citenamefont {Ramanathan},\ and\ \citenamefont {Cappellaro}}]{Wei}%
  \BibitemOpen
  \bibfield  {author} {\bibinfo {author} {\bibfnamefont {K.~X.}\ \bibnamefont
  {Wei}}, \bibinfo {author} {\bibfnamefont {C.}~\bibnamefont {Ramanathan}}, \
  and\ \bibinfo {author} {\bibfnamefont {P.}~\bibnamefont {Cappellaro}},\
  }\href {\doibase 10.1103/PhysRevLett.120.070501} {\bibfield  {journal}
  {\bibinfo  {journal} {Phys. Rev. Lett.}\ }\textbf {\bibinfo {volume} {120}},\
  \bibinfo {pages} {070501} (\bibinfo {year} {2018})}\BibitemShut {NoStop}%
\bibitem [{\citenamefont {Iyoda}\ and\ \citenamefont
  {Sagawa}(2018)}]{iyoda2018}%
  \BibitemOpen
  \bibfield  {author} {\bibinfo {author} {\bibfnamefont {E.}~\bibnamefont
  {Iyoda}}\ and\ \bibinfo {author} {\bibfnamefont {T.}~\bibnamefont {Sagawa}},\
  }\href {\doibase 10.1103/PhysRevA.97.042330} {\bibfield  {journal} {\bibinfo
  {journal} {Phys. Rev. A}\ }\textbf {\bibinfo {volume} {97}},\ \bibinfo
  {pages} {042330} (\bibinfo {year} {2018})}\BibitemShut {NoStop}%
\bibitem [{\citenamefont {Von~Keyserlingk}\ \emph {et~al.}(2018)\citenamefont
  {Von~Keyserlingk}, \citenamefont {Rakovszky}, \citenamefont {Pollmann},\ and\
  \citenamefont {Sondhi}}]{von2018}%
  \BibitemOpen
  \bibfield  {author} {\bibinfo {author} {\bibfnamefont {C.}~\bibnamefont
  {Von~Keyserlingk}}, \bibinfo {author} {\bibfnamefont {T.}~\bibnamefont
  {Rakovszky}}, \bibinfo {author} {\bibfnamefont {F.}~\bibnamefont {Pollmann}},
  \ and\ \bibinfo {author} {\bibfnamefont {S.~L.}\ \bibnamefont {Sondhi}},\
  }\href {\doibase 10.1103/PhysRevX.8.021013} {\bibfield  {journal} {\bibinfo
  {journal} {Phys. Rev. X}\ }\textbf {\bibinfo {volume} {8}},\ \bibinfo {pages}
  {021013} (\bibinfo {year} {2018})}\BibitemShut {NoStop}%
\bibitem [{\citenamefont {Niknam}\ \emph {et~al.}(2020)\citenamefont {Niknam},
  \citenamefont {Santos},\ and\ \citenamefont {Cory}}]{niknam2020}%
  \BibitemOpen
  \bibfield  {author} {\bibinfo {author} {\bibfnamefont {M.}~\bibnamefont
  {Niknam}}, \bibinfo {author} {\bibfnamefont {L.~F.}\ \bibnamefont {Santos}},
  \ and\ \bibinfo {author} {\bibfnamefont {D.~G.}\ \bibnamefont {Cory}},\
  }\href {\doibase 10.1103/PhysRevResearch.2.013200} {\bibfield  {journal}
  {\bibinfo  {journal} {Phys. Rev. Research}\ }\textbf {\bibinfo {volume}
  {2}},\ \bibinfo {pages} {013200} (\bibinfo {year} {2020})}\BibitemShut
  {NoStop}%
\bibitem [{\citenamefont {Shukla}\ \emph {et~al.}(2022)\citenamefont {Shukla},
  \citenamefont {Lakshminarayan},\ and\ \citenamefont {Mishra}}]{shukla2022}%
  \BibitemOpen
  \bibfield  {author} {\bibinfo {author} {\bibfnamefont {R.~K.}\ \bibnamefont
  {Shukla}}, \bibinfo {author} {\bibfnamefont {A.}~\bibnamefont
  {Lakshminarayan}}, \ and\ \bibinfo {author} {\bibfnamefont {S.~K.}\
  \bibnamefont {Mishra}},\ }\href {\doibase 10.1103/PhysRevB.105.224307}
  {\bibfield  {journal} {\bibinfo  {journal} {Phys. Rev. B}\ }\textbf {\bibinfo
  {volume} {105}},\ \bibinfo {pages} {224307} (\bibinfo {year}
  {2022})}\BibitemShut {NoStop}%
\bibitem [{\citenamefont {Alavirad}\ and\ \citenamefont
  {Lavasani}(2019)}]{alavirad2019}%
  \BibitemOpen
  \bibfield  {author} {\bibinfo {author} {\bibfnamefont {Y.}~\bibnamefont
  {Alavirad}}\ and\ \bibinfo {author} {\bibfnamefont {A.}~\bibnamefont
  {Lavasani}},\ }\href {\doibase 10.1103/PhysRevA.99.043602} {\bibfield
  {journal} {\bibinfo  {journal} {Phys. Rev. A}\ }\textbf {\bibinfo {volume}
  {99}},\ \bibinfo {pages} {043602} (\bibinfo {year} {2019})}\BibitemShut
  {NoStop}%
\bibitem [{\citenamefont {Hosur}\ \emph {et~al.}(2016)\citenamefont {Hosur},
  \citenamefont {Qi}, \citenamefont {Roberts},\ and\ \citenamefont
  {Yoshida}}]{hosur2016}%
  \BibitemOpen
  \bibfield  {author} {\bibinfo {author} {\bibfnamefont {P.}~\bibnamefont
  {Hosur}}, \bibinfo {author} {\bibfnamefont {X.-L.}\ \bibnamefont {Qi}},
  \bibinfo {author} {\bibfnamefont {D.~A.}\ \bibnamefont {Roberts}}, \ and\
  \bibinfo {author} {\bibfnamefont {B.}~\bibnamefont {Yoshida}},\ }\href
  {\doibase https://doi.org/10.1007/JHEP02(2016)004} {\bibfield  {journal}
  {\bibinfo  {journal} {Journal of High Energy Physics}\ }\textbf {\bibinfo
  {volume} {2016}},\ \bibinfo {pages} {1} (\bibinfo {year} {2016})}\BibitemShut
  {NoStop}%
\bibitem [{\citenamefont {Swingle}\ and\ \citenamefont
  {Chowdhury}(2017)}]{swingle2017}%
  \BibitemOpen
  \bibfield  {author} {\bibinfo {author} {\bibfnamefont {B.}~\bibnamefont
  {Swingle}}\ and\ \bibinfo {author} {\bibfnamefont {D.}~\bibnamefont
  {Chowdhury}},\ }\href {\doibase 10.1103/PhysRevB.95.060201} {\bibfield
  {journal} {\bibinfo  {journal} {Phys. Rev. B}\ }\textbf {\bibinfo {volume}
  {95}},\ \bibinfo {pages} {060201} (\bibinfo {year} {2017})}\BibitemShut
  {NoStop}%
\bibitem [{\citenamefont {Schleier-Smith}(2017)}]{schleier2017}%
  \BibitemOpen
  \bibfield  {author} {\bibinfo {author} {\bibfnamefont {M.}~\bibnamefont
  {Schleier-Smith}},\ }\href {\doibase https://doi.org/10.1038/nphys4165}
  {\bibfield  {journal} {\bibinfo  {journal} {Nature Physics}\ }\textbf
  {\bibinfo {volume} {13}},\ \bibinfo {pages} {724} (\bibinfo {year}
  {2017})}\BibitemShut {NoStop}%
\bibitem [{\citenamefont {Pappalardi}\ \emph {et~al.}(2018)\citenamefont
  {Pappalardi}, \citenamefont {Russomanno}, \citenamefont
  {{\v{Z}}unkovi{\v{c}}}, \citenamefont {Iemini}, \citenamefont {Silva},\ and\
  \citenamefont {Fazio}}]{pappalardi2018}%
  \BibitemOpen
  \bibfield  {author} {\bibinfo {author} {\bibfnamefont {S.}~\bibnamefont
  {Pappalardi}}, \bibinfo {author} {\bibfnamefont {A.}~\bibnamefont
  {Russomanno}}, \bibinfo {author} {\bibfnamefont {B.}~\bibnamefont
  {{\v{Z}}unkovi{\v{c}}}}, \bibinfo {author} {\bibfnamefont {F.}~\bibnamefont
  {Iemini}}, \bibinfo {author} {\bibfnamefont {A.}~\bibnamefont {Silva}}, \
  and\ \bibinfo {author} {\bibfnamefont {R.}~\bibnamefont {Fazio}},\ }\href
  {\doibase 10.1103/PhysRevB.98.134303} {\bibfield  {journal} {\bibinfo
  {journal} {Phys. Rev. B}\ }\textbf {\bibinfo {volume} {98}},\ \bibinfo
  {pages} {134303} (\bibinfo {year} {2018})}\BibitemShut {NoStop}%
\bibitem [{\citenamefont {Klug}\ \emph {et~al.}(2018)\citenamefont {Klug},
  \citenamefont {Scheurer},\ and\ \citenamefont {Schmalian}}]{klug2018}%
  \BibitemOpen
  \bibfield  {author} {\bibinfo {author} {\bibfnamefont {M.~J.}\ \bibnamefont
  {Klug}}, \bibinfo {author} {\bibfnamefont {M.~S.}\ \bibnamefont {Scheurer}},
  \ and\ \bibinfo {author} {\bibfnamefont {J.}~\bibnamefont {Schmalian}},\
  }\href {\doibase 10.1103/PhysRevB.98.045102} {\bibfield  {journal} {\bibinfo
  {journal} {Phys. Rev. B}\ }\textbf {\bibinfo {volume} {98}},\ \bibinfo
  {pages} {045102} (\bibinfo {year} {2018})}\BibitemShut {NoStop}%
\bibitem [{\citenamefont {Khemani}\ \emph {et~al.}(2018)\citenamefont
  {Khemani}, \citenamefont {Vishwanath},\ and\ \citenamefont
  {Huse}}]{khemani2018}%
  \BibitemOpen
  \bibfield  {author} {\bibinfo {author} {\bibfnamefont {V.}~\bibnamefont
  {Khemani}}, \bibinfo {author} {\bibfnamefont {A.}~\bibnamefont {Vishwanath}},
  \ and\ \bibinfo {author} {\bibfnamefont {D.~A.}\ \bibnamefont {Huse}},\
  }\href {\doibase 10.1103/PhysRevX.8.031057} {\bibfield  {journal} {\bibinfo
  {journal} {Phys. Rev. X}\ }\textbf {\bibinfo {volume} {8}},\ \bibinfo {pages}
  {031057} (\bibinfo {year} {2018})}\BibitemShut {NoStop}%
\bibitem [{\citenamefont {Maldacena}\ \emph {et~al.}(2016)\citenamefont
  {Maldacena}, \citenamefont {Shenker},\ and\ \citenamefont
  {Stanford}}]{maldacena2016}%
  \BibitemOpen
  \bibfield  {author} {\bibinfo {author} {\bibfnamefont {J.}~\bibnamefont
  {Maldacena}}, \bibinfo {author} {\bibfnamefont {S.~H.}\ \bibnamefont
  {Shenker}}, \ and\ \bibinfo {author} {\bibfnamefont {D.}~\bibnamefont
  {Stanford}},\ }\href {\doibase https://doi.org/10.1007/JHEP08(2016)106}
  {\bibfield  {journal} {\bibinfo  {journal} {Journal of High Energy Physics}\
  }\textbf {\bibinfo {volume} {2016}},\ \bibinfo {pages} {1} (\bibinfo {year}
  {2016})}\BibitemShut {NoStop}%
\bibitem [{\citenamefont {Shenker}\ and\ \citenamefont
  {Stanford}(2014)}]{shenker2014}%
  \BibitemOpen
  \bibfield  {author} {\bibinfo {author} {\bibfnamefont {S.~H.}\ \bibnamefont
  {Shenker}}\ and\ \bibinfo {author} {\bibfnamefont {D.}~\bibnamefont
  {Stanford}},\ }\href {\doibase https://doi.org/10.1007/JHEP03(2014)067}
  {\bibfield  {journal} {\bibinfo  {journal} {Journal of High Energy Physics}\
  }\textbf {\bibinfo {volume} {2014}},\ \bibinfo {pages} {1} (\bibinfo {year}
  {2014})}\BibitemShut {NoStop}%
\bibitem [{\citenamefont {Kitaev}\ and\ \citenamefont
  {Suh}(2018)}]{kitaev2018}%
  \BibitemOpen
  \bibfield  {author} {\bibinfo {author} {\bibfnamefont {A.}~\bibnamefont
  {Kitaev}}\ and\ \bibinfo {author} {\bibfnamefont {S.~J.}\ \bibnamefont
  {Suh}},\ }\href {\doibase https://doi.org/10.1007/JHEP05(2018)183} {\bibfield
   {journal} {\bibinfo  {journal} {Journal of High Energy Physics}\ }\textbf
  {\bibinfo {volume} {2018}},\ \bibinfo {pages} {1} (\bibinfo {year}
  {2018})}\BibitemShut {NoStop}%
\bibitem [{\citenamefont {Fan}\ \emph {et~al.}(2017)\citenamefont {Fan},
  \citenamefont {Zhang}, \citenamefont {Shen},\ and\ \citenamefont
  {Zhai}}]{fan2017}%
  \BibitemOpen
  \bibfield  {author} {\bibinfo {author} {\bibfnamefont {R.}~\bibnamefont
  {Fan}}, \bibinfo {author} {\bibfnamefont {P.}~\bibnamefont {Zhang}}, \bibinfo
  {author} {\bibfnamefont {H.}~\bibnamefont {Shen}}, \ and\ \bibinfo {author}
  {\bibfnamefont {H.}~\bibnamefont {Zhai}},\ }\href {\doibase
  https://doi.org/10.1016/j.scib.2017.04.011} {\bibfield  {journal} {\bibinfo
  {journal} {Science Bulletin}\ }\textbf {\bibinfo {volume} {62}},\ \bibinfo
  {pages} {707} (\bibinfo {year} {2017})}\BibitemShut {NoStop}%
\bibitem [{\citenamefont {Huang}\ \emph {et~al.}(2017)\citenamefont {Huang},
  \citenamefont {Zhang},\ and\ \citenamefont {Chen}}]{huang2017out}%
  \BibitemOpen
  \bibfield  {author} {\bibinfo {author} {\bibfnamefont {Y.}~\bibnamefont
  {Huang}}, \bibinfo {author} {\bibfnamefont {Y.-L.}\ \bibnamefont {Zhang}}, \
  and\ \bibinfo {author} {\bibfnamefont {X.}~\bibnamefont {Chen}},\ }\href
  {\doibase 10.1002/andp.201600318} {\bibfield  {journal} {\bibinfo  {journal}
  {Annalen der Physik}\ }\textbf {\bibinfo {volume} {529}},\ \bibinfo {pages}
  {1600318} (\bibinfo {year} {2017})}\BibitemShut {NoStop}%
\bibitem [{\citenamefont {Swingle}\ \emph {et~al.}(2016)\citenamefont
  {Swingle}, \citenamefont {Bentsen}, \citenamefont {Schleier-Smith},\ and\
  \citenamefont {Hayden}}]{Swingle}%
  \BibitemOpen
  \bibfield  {author} {\bibinfo {author} {\bibfnamefont {B.}~\bibnamefont
  {Swingle}}, \bibinfo {author} {\bibfnamefont {G.}~\bibnamefont {Bentsen}},
  \bibinfo {author} {\bibfnamefont {M.}~\bibnamefont {Schleier-Smith}}, \ and\
  \bibinfo {author} {\bibfnamefont {P.}~\bibnamefont {Hayden}},\ }\href
  {\doibase 10.1103/PhysRevA.94.040302} {\bibfield  {journal} {\bibinfo
  {journal} {Phys. Rev. A}\ }\textbf {\bibinfo {volume} {94}},\ \bibinfo
  {pages} {040302} (\bibinfo {year} {2016})}\BibitemShut {NoStop}%
\bibitem [{\citenamefont {Zhu}\ \emph {et~al.}(2016)\citenamefont {Zhu},
  \citenamefont {Hafezi},\ and\ \citenamefont {Grover}}]{Zhu}%
  \BibitemOpen
  \bibfield  {author} {\bibinfo {author} {\bibfnamefont {G.}~\bibnamefont
  {Zhu}}, \bibinfo {author} {\bibfnamefont {M.}~\bibnamefont {Hafezi}}, \ and\
  \bibinfo {author} {\bibfnamefont {T.}~\bibnamefont {Grover}},\ }\href
  {\doibase 10.1103/PhysRevA.94.062329} {\bibfield  {journal} {\bibinfo
  {journal} {Phys. Rev. A}\ }\textbf {\bibinfo {volume} {94}},\ \bibinfo
  {pages} {062329} (\bibinfo {year} {2016})}\BibitemShut {NoStop}%
\bibitem [{\citenamefont {G{\"a}rttner}\ \emph {et~al.}(2017)\citenamefont
  {G{\"a}rttner}, \citenamefont {Bohnet}, \citenamefont {Safavi-Naini},
  \citenamefont {Wall}, \citenamefont {Bollinger},\ and\ \citenamefont
  {Rey}}]{Garttner2017}%
  \BibitemOpen
  \bibfield  {author} {\bibinfo {author} {\bibfnamefont {M.}~\bibnamefont
  {G{\"a}rttner}}, \bibinfo {author} {\bibfnamefont {J.~G.}\ \bibnamefont
  {Bohnet}}, \bibinfo {author} {\bibfnamefont {A.}~\bibnamefont
  {Safavi-Naini}}, \bibinfo {author} {\bibfnamefont {M.~L.}\ \bibnamefont
  {Wall}}, \bibinfo {author} {\bibfnamefont {J.~J.}\ \bibnamefont {Bollinger}},
  \ and\ \bibinfo {author} {\bibfnamefont {A.~M.}\ \bibnamefont {Rey}},\ }\href
  {\doibase 10.1038/nphys4119} {\bibfield  {journal} {\bibinfo  {journal}
  {Nature Physics}\ }\textbf {\bibinfo {volume} {13}},\ \bibinfo {pages} {781}
  (\bibinfo {year} {2017})}\BibitemShut {NoStop}%
\bibitem [{\citenamefont {Li}\ \emph {et~al.}(2017)\citenamefont {Li},
  \citenamefont {Fan}, \citenamefont {Wang}, \citenamefont {Ye}, \citenamefont
  {Zeng}, \citenamefont {Zhai}, \citenamefont {Peng},\ and\ \citenamefont
  {Du}}]{Li}%
  \BibitemOpen
  \bibfield  {author} {\bibinfo {author} {\bibfnamefont {J.}~\bibnamefont
  {Li}}, \bibinfo {author} {\bibfnamefont {R.}~\bibnamefont {Fan}}, \bibinfo
  {author} {\bibfnamefont {H.}~\bibnamefont {Wang}}, \bibinfo {author}
  {\bibfnamefont {B.}~\bibnamefont {Ye}}, \bibinfo {author} {\bibfnamefont
  {B.}~\bibnamefont {Zeng}}, \bibinfo {author} {\bibfnamefont {H.}~\bibnamefont
  {Zhai}}, \bibinfo {author} {\bibfnamefont {X.}~\bibnamefont {Peng}}, \ and\
  \bibinfo {author} {\bibfnamefont {J.}~\bibnamefont {Du}},\ }\href {\doibase
  10.1103/PhysRevX.7.031011} {\bibfield  {journal} {\bibinfo  {journal} {Phys.
  Rev. X}\ }\textbf {\bibinfo {volume} {7}},\ \bibinfo {pages} {031011}
  (\bibinfo {year} {2017})}\BibitemShut {NoStop}%
\bibitem [{\citenamefont {Lewis-Swan}\ \emph {et~al.}(2019)\citenamefont
  {Lewis-Swan}, \citenamefont {Safavi-Naini}, \citenamefont {Bollinger},\ and\
  \citenamefont {Rey}}]{Lewis}%
  \BibitemOpen
  \bibfield  {author} {\bibinfo {author} {\bibfnamefont {R.~J.}\ \bibnamefont
  {Lewis-Swan}}, \bibinfo {author} {\bibfnamefont {A.}~\bibnamefont
  {Safavi-Naini}}, \bibinfo {author} {\bibfnamefont {J.~J.}\ \bibnamefont
  {Bollinger}}, \ and\ \bibinfo {author} {\bibfnamefont {A.~M.}\ \bibnamefont
  {Rey}},\ }\href {\doibase 10.1038/s41467-019-09436-y} {\bibfield  {journal}
  {\bibinfo  {journal} {Nature Communications}\ }\textbf {\bibinfo {volume}
  {10}},\ \bibinfo {pages} {1581} (\bibinfo {year} {2019})}\BibitemShut
  {NoStop}%
\bibitem [{\citenamefont {Nie}\ \emph {et~al.}(2020)\citenamefont {Nie},
  \citenamefont {Wei}, \citenamefont {Chen}, \citenamefont {Zhang},
  \citenamefont {Zhao}, \citenamefont {Qiu}, \citenamefont {Tian},
  \citenamefont {Ji}, \citenamefont {Xin}, \citenamefont {Lu},\ and\
  \citenamefont {Li}}]{Nie}%
  \BibitemOpen
  \bibfield  {author} {\bibinfo {author} {\bibfnamefont {X.}~\bibnamefont
  {Nie}}, \bibinfo {author} {\bibfnamefont {B.-B.}\ \bibnamefont {Wei}},
  \bibinfo {author} {\bibfnamefont {X.}~\bibnamefont {Chen}}, \bibinfo {author}
  {\bibfnamefont {Z.}~\bibnamefont {Zhang}}, \bibinfo {author} {\bibfnamefont
  {X.}~\bibnamefont {Zhao}}, \bibinfo {author} {\bibfnamefont {C.}~\bibnamefont
  {Qiu}}, \bibinfo {author} {\bibfnamefont {Y.}~\bibnamefont {Tian}}, \bibinfo
  {author} {\bibfnamefont {Y.}~\bibnamefont {Ji}}, \bibinfo {author}
  {\bibfnamefont {T.}~\bibnamefont {Xin}}, \bibinfo {author} {\bibfnamefont
  {D.}~\bibnamefont {Lu}}, \ and\ \bibinfo {author} {\bibfnamefont
  {J.}~\bibnamefont {Li}},\ }\href {\doibase 10.1103/PhysRevLett.124.250601}
  {\bibfield  {journal} {\bibinfo  {journal} {Phys. Rev. Lett.}\ }\textbf
  {\bibinfo {volume} {124}},\ \bibinfo {pages} {250601} (\bibinfo {year}
  {2020})}\BibitemShut {NoStop}%
\bibitem [{\citenamefont {Braum{\"u}ller}\ \emph {et~al.}(2022)\citenamefont
  {Braum{\"u}ller}, \citenamefont {Karamlou}, \citenamefont {Yanay},
  \citenamefont {Kannan}, \citenamefont {Kim}, \citenamefont {Kjaergaard},
  \citenamefont {Melville}, \citenamefont {Niedzielski}, \citenamefont {Sung},
  \citenamefont {Veps{\"a}l{\"a}inen}, \citenamefont {Winik}, \citenamefont
  {Yoder}, \citenamefont {Orlando}, \citenamefont {Gustavsson}, \citenamefont
  {Tahan},\ and\ \citenamefont {Oliver}}]{Braumuller2022}%
  \BibitemOpen
  \bibfield  {author} {\bibinfo {author} {\bibfnamefont {J.}~\bibnamefont
  {Braum{\"u}ller}}, \bibinfo {author} {\bibfnamefont {A.~H.}\ \bibnamefont
  {Karamlou}}, \bibinfo {author} {\bibfnamefont {Y.}~\bibnamefont {Yanay}},
  \bibinfo {author} {\bibfnamefont {B.}~\bibnamefont {Kannan}}, \bibinfo
  {author} {\bibfnamefont {D.}~\bibnamefont {Kim}}, \bibinfo {author}
  {\bibfnamefont {M.}~\bibnamefont {Kjaergaard}}, \bibinfo {author}
  {\bibfnamefont {A.}~\bibnamefont {Melville}}, \bibinfo {author}
  {\bibfnamefont {B.~M.}\ \bibnamefont {Niedzielski}}, \bibinfo {author}
  {\bibfnamefont {Y.}~\bibnamefont {Sung}}, \bibinfo {author} {\bibfnamefont
  {A.}~\bibnamefont {Veps{\"a}l{\"a}inen}}, \bibinfo {author} {\bibfnamefont
  {R.}~\bibnamefont {Winik}}, \bibinfo {author} {\bibfnamefont {J.~L.}\
  \bibnamefont {Yoder}}, \bibinfo {author} {\bibfnamefont {T.~P.}\ \bibnamefont
  {Orlando}}, \bibinfo {author} {\bibfnamefont {S.}~\bibnamefont {Gustavsson}},
  \bibinfo {author} {\bibfnamefont {C.}~\bibnamefont {Tahan}}, \ and\ \bibinfo
  {author} {\bibfnamefont {W.~D.}\ \bibnamefont {Oliver}},\ }\href {\doibase
  10.1038/s41567-021-01430-w} {\bibfield  {journal} {\bibinfo  {journal}
  {Nature Physics}\ }\textbf {\bibinfo {volume} {18}},\ \bibinfo {pages} {172}
  (\bibinfo {year} {2022})}\BibitemShut {NoStop}%
\bibitem [{\citenamefont {Zhao}\ \emph {et~al.}(2022)\citenamefont {Zhao},
  \citenamefont {Ge}, \citenamefont {Xiang}, \citenamefont {Xue}, \citenamefont
  {Yan}, \citenamefont {Wang}, \citenamefont {Wang}, \citenamefont {Xu},
  \citenamefont {Su}, \citenamefont {Yang}, \citenamefont {Zhang},
  \citenamefont {Zhang}, \citenamefont {Guo}, \citenamefont {Xu}, \citenamefont
  {Tian}, \citenamefont {Yu}, \citenamefont {Zheng}, \citenamefont {Fan},\ and\
  \citenamefont {Zhao}}]{zhao2022}%
  \BibitemOpen
  \bibfield  {author} {\bibinfo {author} {\bibfnamefont {S.~K.}\ \bibnamefont
  {Zhao}}, \bibinfo {author} {\bibfnamefont {Z.-Y.}\ \bibnamefont {Ge}},
  \bibinfo {author} {\bibfnamefont {Z.}~\bibnamefont {Xiang}}, \bibinfo
  {author} {\bibfnamefont {G.~M.}\ \bibnamefont {Xue}}, \bibinfo {author}
  {\bibfnamefont {H.~S.}\ \bibnamefont {Yan}}, \bibinfo {author} {\bibfnamefont
  {Z.~T.}\ \bibnamefont {Wang}}, \bibinfo {author} {\bibfnamefont
  {Z.}~\bibnamefont {Wang}}, \bibinfo {author} {\bibfnamefont {H.~K.}\
  \bibnamefont {Xu}}, \bibinfo {author} {\bibfnamefont {F.~F.}\ \bibnamefont
  {Su}}, \bibinfo {author} {\bibfnamefont {Z.~H.}\ \bibnamefont {Yang}},
  \bibinfo {author} {\bibfnamefont {H.}~\bibnamefont {Zhang}}, \bibinfo
  {author} {\bibfnamefont {Y.-R.}\ \bibnamefont {Zhang}}, \bibinfo {author}
  {\bibfnamefont {X.-Y.}\ \bibnamefont {Guo}}, \bibinfo {author} {\bibfnamefont
  {K.}~\bibnamefont {Xu}}, \bibinfo {author} {\bibfnamefont {Y.}~\bibnamefont
  {Tian}}, \bibinfo {author} {\bibfnamefont {H.~F.}\ \bibnamefont {Yu}},
  \bibinfo {author} {\bibfnamefont {D.~N.}\ \bibnamefont {Zheng}}, \bibinfo
  {author} {\bibfnamefont {H.}~\bibnamefont {Fan}}, \ and\ \bibinfo {author}
  {\bibfnamefont {S.~P.}\ \bibnamefont {Zhao}},\ }\href {\doibase
  10.1103/PhysRevLett.129.160602} {\bibfield  {journal} {\bibinfo  {journal}
  {Phys. Rev. Lett.}\ }\textbf {\bibinfo {volume} {129}},\ \bibinfo {pages}
  {160602} (\bibinfo {year} {2022})}\BibitemShut {NoStop}%
\bibitem [{\citenamefont {Larkin}\ and\ \citenamefont
  {Ovchinnikov}(1969)}]{larkin1969}%
  \BibitemOpen
  \bibfield  {author} {\bibinfo {author} {\bibfnamefont {A.}~\bibnamefont
  {Larkin}}\ and\ \bibinfo {author} {\bibfnamefont {Y.~N.}\ \bibnamefont
  {Ovchinnikov}},\ }\href {\doibase
  https://ui.adsabs.harvard.edu/abs/1969JETP...28.1200L} {\bibfield  {journal}
  {\bibinfo  {journal} {JETP}\ }\textbf {\bibinfo {volume} {28}},\ \bibinfo
  {pages} {1200} (\bibinfo {year} {1969})}\BibitemShut {NoStop}%
\bibitem [{\citenamefont {Maldacena}\ \emph {et~al.}(2017)\citenamefont
  {Maldacena}, \citenamefont {Stanford},\ and\ \citenamefont
  {Yang}}]{maldacena2017}%
  \BibitemOpen
  \bibfield  {author} {\bibinfo {author} {\bibfnamefont {J.}~\bibnamefont
  {Maldacena}}, \bibinfo {author} {\bibfnamefont {D.}~\bibnamefont {Stanford}},
  \ and\ \bibinfo {author} {\bibfnamefont {Z.}~\bibnamefont {Yang}},\ }\href
  {\doibase 10.1002/prop.201700034} {\bibfield  {journal} {\bibinfo  {journal}
  {Fortschritte der Physik}\ }\textbf {\bibinfo {volume} {65}},\ \bibinfo
  {pages} {1700034} (\bibinfo {year} {2017})}\BibitemShut {NoStop}%
\bibitem [{\citenamefont {Maldacena}\ and\ \citenamefont
  {Stanford}(2016)}]{maldacena2016remarks}%
  \BibitemOpen
  \bibfield  {author} {\bibinfo {author} {\bibfnamefont {J.}~\bibnamefont
  {Maldacena}}\ and\ \bibinfo {author} {\bibfnamefont {D.}~\bibnamefont
  {Stanford}},\ }\href {\doibase 10.1103/PhysRevD.94.106002} {\bibfield
  {journal} {\bibinfo  {journal} {Phys. Rev. D}\ }\textbf {\bibinfo {volume}
  {94}},\ \bibinfo {pages} {106002} (\bibinfo {year} {2016})}\BibitemShut
  {NoStop}%
\bibitem [{\citenamefont {Sachdev}\ and\ \citenamefont
  {Ye}(1993)}]{sachdev1993gapless}%
  \BibitemOpen
  \bibfield  {author} {\bibinfo {author} {\bibfnamefont {S.}~\bibnamefont
  {Sachdev}}\ and\ \bibinfo {author} {\bibfnamefont {J.}~\bibnamefont {Ye}},\
  }\href {\doibase 10.1103/PhysRevLett.70.3339} {\bibfield  {journal} {\bibinfo
   {journal} {Phys. Rev. Lett.}\ }\textbf {\bibinfo {volume} {70}},\ \bibinfo
  {pages} {3339} (\bibinfo {year} {1993})}\BibitemShut {NoStop}%
\bibitem [{\citenamefont {Patel}\ \emph {et~al.}(2017)\citenamefont {Patel},
  \citenamefont {Chowdhury}, \citenamefont {Sachdev},\ and\ \citenamefont
  {Swingle}}]{patel2017quantum1}%
  \BibitemOpen
  \bibfield  {author} {\bibinfo {author} {\bibfnamefont {A.~A.}\ \bibnamefont
  {Patel}}, \bibinfo {author} {\bibfnamefont {D.}~\bibnamefont {Chowdhury}},
  \bibinfo {author} {\bibfnamefont {S.}~\bibnamefont {Sachdev}}, \ and\
  \bibinfo {author} {\bibfnamefont {B.}~\bibnamefont {Swingle}},\ }\href
  {\doibase 10.1103/PhysRevX.7.031047} {\bibfield  {journal} {\bibinfo
  {journal} {Phys. Rev. X}\ }\textbf {\bibinfo {volume} {7}},\ \bibinfo {pages}
  {031047} (\bibinfo {year} {2017})}\BibitemShut {NoStop}%
\bibitem [{\citenamefont {Patel}\ and\ \citenamefont
  {Sachdev}(2017)}]{patel2017quantum}%
  \BibitemOpen
  \bibfield  {author} {\bibinfo {author} {\bibfnamefont {A.~A.}\ \bibnamefont
  {Patel}}\ and\ \bibinfo {author} {\bibfnamefont {S.}~\bibnamefont
  {Sachdev}},\ }\href {\doibase 10.1073/pnas.1618185114} {\bibfield  {journal}
  {\bibinfo  {journal} {Proceedings of the National Academy of Sciences}\
  }\textbf {\bibinfo {volume} {114}},\ \bibinfo {pages} {1844} (\bibinfo {year}
  {2017})}\BibitemShut {NoStop}%
\bibitem [{\citenamefont {D{\'o}ra}\ and\ \citenamefont
  {Moessner}(2017)}]{dora2017out}%
  \BibitemOpen
  \bibfield  {author} {\bibinfo {author} {\bibfnamefont {B.}~\bibnamefont
  {D{\'o}ra}}\ and\ \bibinfo {author} {\bibfnamefont {R.}~\bibnamefont
  {Moessner}},\ }\href {\doibase 10.1103/PhysRevLett.119.026802} {\bibfield
  {journal} {\bibinfo  {journal} {Phys. Rev. Lett.}\ }\textbf {\bibinfo
  {volume} {119}},\ \bibinfo {pages} {026802} (\bibinfo {year}
  {2017})}\BibitemShut {NoStop}%
\bibitem [{\citenamefont {Shen}\ \emph {et~al.}(2017)\citenamefont {Shen},
  \citenamefont {Zhang}, \citenamefont {Fan},\ and\ \citenamefont
  {Zhai}}]{shen2017out}%
  \BibitemOpen
  \bibfield  {author} {\bibinfo {author} {\bibfnamefont {H.}~\bibnamefont
  {Shen}}, \bibinfo {author} {\bibfnamefont {P.}~\bibnamefont {Zhang}},
  \bibinfo {author} {\bibfnamefont {R.}~\bibnamefont {Fan}}, \ and\ \bibinfo
  {author} {\bibfnamefont {H.}~\bibnamefont {Zhai}},\ }\href {\doibase
  10.1103/PhysRevB.96.054503} {\bibfield  {journal} {\bibinfo  {journal} {Phys.
  Rev. B}\ }\textbf {\bibinfo {volume} {96}},\ \bibinfo {pages} {054503}
  (\bibinfo {year} {2017})}\BibitemShut {NoStop}%
\bibitem [{\citenamefont {Heyl}\ \emph {et~al.}(2018)\citenamefont {Heyl},
  \citenamefont {Pollmann},\ and\ \citenamefont
  {D{\'o}ra}}]{heyl2018detecting}%
  \BibitemOpen
  \bibfield  {author} {\bibinfo {author} {\bibfnamefont {M.}~\bibnamefont
  {Heyl}}, \bibinfo {author} {\bibfnamefont {F.}~\bibnamefont {Pollmann}}, \
  and\ \bibinfo {author} {\bibfnamefont {B.}~\bibnamefont {D{\'o}ra}},\ }\href
  {\doibase 10.1103/PhysRevLett.121.016801} {\bibfield  {journal} {\bibinfo
  {journal} {Phys. Rev. Lett.}\ }\textbf {\bibinfo {volume} {121}},\ \bibinfo
  {pages} {016801} (\bibinfo {year} {2018})}\BibitemShut {NoStop}%
\bibitem [{\citenamefont {Sahu}\ \emph {et~al.}(2019)\citenamefont {Sahu},
  \citenamefont {Xu},\ and\ \citenamefont {Swingle}}]{sahu2019scrambling}%
  \BibitemOpen
  \bibfield  {author} {\bibinfo {author} {\bibfnamefont {S.}~\bibnamefont
  {Sahu}}, \bibinfo {author} {\bibfnamefont {S.}~\bibnamefont {Xu}}, \ and\
  \bibinfo {author} {\bibfnamefont {B.}~\bibnamefont {Swingle}},\ }\href
  {\doibase 10.1103/PhysRevLett.123.165902} {\bibfield  {journal} {\bibinfo
  {journal} {Phys. Rev. Lett.}\ }\textbf {\bibinfo {volume} {123}},\ \bibinfo
  {pages} {165902} (\bibinfo {year} {2019})}\BibitemShut {NoStop}%
\bibitem [{\citenamefont {Lin}\ and\ \citenamefont
  {Motrunich}(2018)}]{lin2018out}%
  \BibitemOpen
  \bibfield  {author} {\bibinfo {author} {\bibfnamefont {C.-J.}\ \bibnamefont
  {Lin}}\ and\ \bibinfo {author} {\bibfnamefont {O.~I.}\ \bibnamefont
  {Motrunich}},\ }\href {\doibase 10.1103/PhysRevB.97.144304} {\bibfield
  {journal} {\bibinfo  {journal} {Phys. Rev. B}\ }\textbf {\bibinfo {volume}
  {97}},\ \bibinfo {pages} {144304} (\bibinfo {year} {2018})}\BibitemShut
  {NoStop}%
\bibitem [{\citenamefont {McGinley}\ \emph {et~al.}(2019)\citenamefont
  {McGinley}, \citenamefont {Nunnenkamp},\ and\ \citenamefont
  {Knolle}}]{mcginley2019slow}%
  \BibitemOpen
  \bibfield  {author} {\bibinfo {author} {\bibfnamefont {M.}~\bibnamefont
  {McGinley}}, \bibinfo {author} {\bibfnamefont {A.}~\bibnamefont
  {Nunnenkamp}}, \ and\ \bibinfo {author} {\bibfnamefont {J.}~\bibnamefont
  {Knolle}},\ }\href {\doibase 10.1103/PhysRevLett.122.020603} {\bibfield
  {journal} {\bibinfo  {journal} {Phys. Rev. Lett.}\ }\textbf {\bibinfo
  {volume} {122}},\ \bibinfo {pages} {020603} (\bibinfo {year}
  {2019})}\BibitemShut {NoStop}%
\bibitem [{\citenamefont {Da{\u{g}}}\ \emph {et~al.}(2020)\citenamefont
  {Da{\u{g}}}, \citenamefont {Duan},\ and\ \citenamefont
  {Sun}}]{daug2020topologically}%
  \BibitemOpen
  \bibfield  {author} {\bibinfo {author} {\bibfnamefont {C.~B.}\ \bibnamefont
  {Da{\u{g}}}}, \bibinfo {author} {\bibfnamefont {L.-M.}\ \bibnamefont {Duan}},
  \ and\ \bibinfo {author} {\bibfnamefont {K.}~\bibnamefont {Sun}},\ }\href
  {\doibase 10.1103/PhysRevB.101.104415} {\bibfield  {journal} {\bibinfo
  {journal} {Phys. Rev. B}\ }\textbf {\bibinfo {volume} {101}},\ \bibinfo
  {pages} {104415} (\bibinfo {year} {2020})}\BibitemShut {NoStop}%
\bibitem [{\citenamefont {Campisi}\ and\ \citenamefont
  {Goold}(2017)}]{campisi2017thermodynamics}%
  \BibitemOpen
  \bibfield  {author} {\bibinfo {author} {\bibfnamefont {M.}~\bibnamefont
  {Campisi}}\ and\ \bibinfo {author} {\bibfnamefont {J.}~\bibnamefont
  {Goold}},\ }\href {\doibase 10.1103/PhysRevE.95.062127} {\bibfield  {journal}
  {\bibinfo  {journal} {Phys. Rev. E}\ }\textbf {\bibinfo {volume} {95}},\
  \bibinfo {pages} {062127} (\bibinfo {year} {2017})}\BibitemShut {NoStop}%
\bibitem [{\citenamefont {Chenu}\ \emph {et~al.}(2018)\citenamefont {Chenu},
  \citenamefont {Egusquiza}, \citenamefont {Molina-Vilaplana},\ and\
  \citenamefont {del Campo}}]{chenu2018quantum}%
  \BibitemOpen
  \bibfield  {author} {\bibinfo {author} {\bibfnamefont {A.}~\bibnamefont
  {Chenu}}, \bibinfo {author} {\bibfnamefont {I.~L.}\ \bibnamefont
  {Egusquiza}}, \bibinfo {author} {\bibfnamefont {J.}~\bibnamefont
  {Molina-Vilaplana}}, \ and\ \bibinfo {author} {\bibfnamefont
  {A.}~\bibnamefont {del Campo}},\ }\href {\doibase 10.1038/s41598-018-30982-w}
  {\bibfield  {journal} {\bibinfo  {journal} {Scientific Reports}\ }\textbf
  {\bibinfo {volume} {8}},\ \bibinfo {pages} {1} (\bibinfo {year}
  {2018})}\BibitemShut {NoStop}%
\bibitem [{\citenamefont {Bao}\ and\ \citenamefont {Zhang}(2020)}]{bao2020out}%
  \BibitemOpen
  \bibfield  {author} {\bibinfo {author} {\bibfnamefont {J.-H.}\ \bibnamefont
  {Bao}}\ and\ \bibinfo {author} {\bibfnamefont {C.-Y.}\ \bibnamefont
  {Zhang}},\ }\href {\doibase 10.1088/1572-9494/ab8a28} {\bibfield  {journal}
  {\bibinfo  {journal} {Communications in Theoretical Physics}\ }\textbf
  {\bibinfo {volume} {72}},\ \bibinfo {pages} {085103} (\bibinfo {year}
  {2020})}\BibitemShut {NoStop}%
\bibitem [{\citenamefont {Kirkby}\ \emph {et~al.}(2019)\citenamefont {Kirkby},
  \citenamefont {Mumford},\ and\ \citenamefont {O'Dell}}]{odell2019}%
  \BibitemOpen
  \bibfield  {author} {\bibinfo {author} {\bibfnamefont {W.}~\bibnamefont
  {Kirkby}}, \bibinfo {author} {\bibfnamefont {J.}~\bibnamefont {Mumford}}, \
  and\ \bibinfo {author} {\bibfnamefont {D.~H.~J.}\ \bibnamefont {O'Dell}},\
  }\href {\doibase 10.1103/PhysRevResearch.1.033135} {\bibfield  {journal}
  {\bibinfo  {journal} {Phys. Rev. Res.}\ }\textbf {\bibinfo {volume} {1}},\
  \bibinfo {pages} {033135} (\bibinfo {year} {2019})}\BibitemShut {NoStop}%
\bibitem [{\citenamefont {Riddell}\ and\ \citenamefont
  {S{\o}rensen}(2019)}]{riddell2019out}%
  \BibitemOpen
  \bibfield  {author} {\bibinfo {author} {\bibfnamefont {J.}~\bibnamefont
  {Riddell}}\ and\ \bibinfo {author} {\bibfnamefont {E.~S.}\ \bibnamefont
  {S{\o}rensen}},\ }\href {\doibase 10.1103/PhysRevB.99.054205} {\bibfield
  {journal} {\bibinfo  {journal} {Phys. Rev. B}\ }\textbf {\bibinfo {volume}
  {99}},\ \bibinfo {pages} {054205} (\bibinfo {year} {2019})}\BibitemShut
  {NoStop}%
\bibitem [{\citenamefont {Lee}\ \emph {et~al.}(2019)\citenamefont {Lee},
  \citenamefont {Kim},\ and\ \citenamefont {Kim}}]{lee2019typical}%
  \BibitemOpen
  \bibfield  {author} {\bibinfo {author} {\bibfnamefont {J.}~\bibnamefont
  {Lee}}, \bibinfo {author} {\bibfnamefont {D.}~\bibnamefont {Kim}}, \ and\
  \bibinfo {author} {\bibfnamefont {D.-H.}\ \bibnamefont {Kim}},\ }\href
  {\doibase 10.1103/PhysRevB.99.184202} {\bibfield  {journal} {\bibinfo
  {journal} {Phys. Rev. B}\ }\textbf {\bibinfo {volume} {99}},\ \bibinfo
  {pages} {184202} (\bibinfo {year} {2019})}\BibitemShut {NoStop}%
\bibitem [{\citenamefont {Da{\u{g}}}\ \emph {et~al.}(2019)\citenamefont
  {Da{\u{g}}}, \citenamefont {Sun},\ and\ \citenamefont
  {Duan}}]{dag2019detection}%
  \BibitemOpen
  \bibfield  {author} {\bibinfo {author} {\bibfnamefont {C.~B.}\ \bibnamefont
  {Da{\u{g}}}}, \bibinfo {author} {\bibfnamefont {K.}~\bibnamefont {Sun}}, \
  and\ \bibinfo {author} {\bibfnamefont {L.-M.}\ \bibnamefont {Duan}},\ }\href
  {\doibase 10.1103/PhysRevLett.123.140602} {\bibfield  {journal} {\bibinfo
  {journal} {Phys. Rev. Lett.}\ }\textbf {\bibinfo {volume} {123}},\ \bibinfo
  {pages} {140602} (\bibinfo {year} {2019})}\BibitemShut {NoStop}%
\bibitem [{\citenamefont {Wei}\ \emph {et~al.}(2019)\citenamefont {Wei},
  \citenamefont {Sun},\ and\ \citenamefont {Hwang}}]{wei2019dynamical}%
  \BibitemOpen
  \bibfield  {author} {\bibinfo {author} {\bibfnamefont {B.-B.}\ \bibnamefont
  {Wei}}, \bibinfo {author} {\bibfnamefont {G.}~\bibnamefont {Sun}}, \ and\
  \bibinfo {author} {\bibfnamefont {M.-J.}\ \bibnamefont {Hwang}},\ }\href
  {\doibase 10.1103/PhysRevB.100.195107} {\bibfield  {journal} {\bibinfo
  {journal} {Phys. Rev. B}\ }\textbf {\bibinfo {volume} {100}},\ \bibinfo
  {pages} {195107} (\bibinfo {year} {2019})}\BibitemShut {NoStop}%
\bibitem [{\citenamefont {Shukla}\ \emph {et~al.}(2021)\citenamefont {Shukla},
  \citenamefont {Naik},\ and\ \citenamefont {Mishra}}]{shukla2021out}%
  \BibitemOpen
  \bibfield  {author} {\bibinfo {author} {\bibfnamefont {R.~K.}\ \bibnamefont
  {Shukla}}, \bibinfo {author} {\bibfnamefont {G.~K.}\ \bibnamefont {Naik}}, \
  and\ \bibinfo {author} {\bibfnamefont {S.~K.}\ \bibnamefont {Mishra}},\
  }\href {\doibase 10.1209/0295-5075/132/47003} {\bibfield  {journal} {\bibinfo
   {journal} {EPL}\ }\textbf {\bibinfo {volume} {132}},\ \bibinfo {pages}
  {47003} (\bibinfo {year} {2021})}\BibitemShut {NoStop}%
\bibitem [{\citenamefont {Shukla}\ and\ \citenamefont
  {Mishra}(2022)}]{shukla2022characteristic}%
  \BibitemOpen
  \bibfield  {author} {\bibinfo {author} {\bibfnamefont {R.~K.}\ \bibnamefont
  {Shukla}}\ and\ \bibinfo {author} {\bibfnamefont {S.~K.}\ \bibnamefont
  {Mishra}},\ }\href {\doibase 10.1103/PhysRevA.106.022403} {\bibfield
  {journal} {\bibinfo  {journal} {Phys. Rev. A}\ }\textbf {\bibinfo {volume}
  {106}},\ \bibinfo {pages} {022403} (\bibinfo {year} {2022})}\BibitemShut
  {NoStop}%
\bibitem [{\citenamefont {Zamani}\ \emph {et~al.}(2022)\citenamefont {Zamani},
  \citenamefont {Jafari},\ and\ \citenamefont {Langari}}]{zamani2022out}%
  \BibitemOpen
  \bibfield  {author} {\bibinfo {author} {\bibfnamefont {S.}~\bibnamefont
  {Zamani}}, \bibinfo {author} {\bibfnamefont {R.}~\bibnamefont {Jafari}}, \
  and\ \bibinfo {author} {\bibfnamefont {A.}~\bibnamefont {Langari}},\ }\href
  {\doibase 10.1103/PhysRevB.105.094304} {\bibfield  {journal} {\bibinfo
  {journal} {Phys. Rev. B}\ }\textbf {\bibinfo {volume} {105}},\ \bibinfo
  {pages} {094304} (\bibinfo {year} {2022})}\BibitemShut {NoStop}%
\bibitem [{\citenamefont {Nizami}(2020)}]{nizami2020quantum}%
  \BibitemOpen
  \bibfield  {author} {\bibinfo {author} {\bibfnamefont {A.~A.}\ \bibnamefont
  {Nizami}},\ }\href@noop {} {\bibfield  {journal} {\bibinfo  {journal}
  {arXiv:2007.07283v2}\ } (\bibinfo {year} {2020})}\BibitemShut {NoStop}%
\bibitem [{\citenamefont {Kitamura}\ \emph {et~al.}(2017)\citenamefont
  {Kitamura}, \citenamefont {Oka},\ and\ \citenamefont
  {Aoki}}]{kitamura2017probing}%
  \BibitemOpen
  \bibfield  {author} {\bibinfo {author} {\bibfnamefont {S.}~\bibnamefont
  {Kitamura}}, \bibinfo {author} {\bibfnamefont {T.}~\bibnamefont {Oka}}, \
  and\ \bibinfo {author} {\bibfnamefont {H.}~\bibnamefont {Aoki}},\ }\href
  {\doibase 10.1103/PhysRevB.96.014406} {\bibfield  {journal} {\bibinfo
  {journal} {Phys. Rev. B}\ }\textbf {\bibinfo {volume} {96}},\ \bibinfo
  {pages} {014406} (\bibinfo {year} {2017})}\BibitemShut {NoStop}%
\bibitem [{\citenamefont {Claassen}\ \emph {et~al.}(2017)\citenamefont
  {Claassen}, \citenamefont {Jiang}, \citenamefont {Moritz},\ and\
  \citenamefont {Devereaux}}]{claassen2017dynamical}%
  \BibitemOpen
  \bibfield  {author} {\bibinfo {author} {\bibfnamefont {M.}~\bibnamefont
  {Claassen}}, \bibinfo {author} {\bibfnamefont {H.-C.}\ \bibnamefont {Jiang}},
  \bibinfo {author} {\bibfnamefont {B.}~\bibnamefont {Moritz}}, \ and\ \bibinfo
  {author} {\bibfnamefont {T.~P.}\ \bibnamefont {Devereaux}},\ }\href {\doibase
  10.1038/s41467-017-00876-y} {\bibfield  {journal} {\bibinfo  {journal}
  {Nature Communications}\ }\textbf {\bibinfo {volume} {8}},\ \bibinfo {pages}
  {1} (\bibinfo {year} {2017})}\BibitemShut {NoStop}%
\bibitem [{\citenamefont {Sriram}\ and\ \citenamefont
  {Claassen}(2022)}]{sriram2022light}%
  \BibitemOpen
  \bibfield  {author} {\bibinfo {author} {\bibfnamefont {A.}~\bibnamefont
  {Sriram}}\ and\ \bibinfo {author} {\bibfnamefont {M.}~\bibnamefont
  {Claassen}},\ }\href {\doibase 10.1103/PhysRevResearch.4.L032036} {\bibfield
  {journal} {\bibinfo  {journal} {Phys. Rev. Research}\ }\textbf {\bibinfo
  {volume} {4}},\ \bibinfo {pages} {L032036} (\bibinfo {year}
  {2022})}\BibitemShut {NoStop}%
\bibitem [{\citenamefont {Sur}\ \emph {et~al.}(2022)\citenamefont {Sur},
  \citenamefont {Udupa},\ and\ \citenamefont {Sen}}]{sur2022driven}%
  \BibitemOpen
  \bibfield  {author} {\bibinfo {author} {\bibfnamefont {S.}~\bibnamefont
  {Sur}}, \bibinfo {author} {\bibfnamefont {A.}~\bibnamefont {Udupa}}, \ and\
  \bibinfo {author} {\bibfnamefont {D.}~\bibnamefont {Sen}},\ }\href {\doibase
  10.1103/PhysRevB.105.054423} {\bibfield  {journal} {\bibinfo  {journal}
  {Phys. Rev. B}\ }\textbf {\bibinfo {volume} {105}},\ \bibinfo {pages}
  {054423} (\bibinfo {year} {2022})}\BibitemShut {NoStop}%
\bibitem [{\citenamefont {Decker}\ \emph {et~al.}(2020)\citenamefont {Decker},
  \citenamefont {Karrasch}, \citenamefont {Eisert},\ and\ \citenamefont
  {Kennes}}]{decker2020floquet}%
  \BibitemOpen
  \bibfield  {author} {\bibinfo {author} {\bibfnamefont {K.~S.}\ \bibnamefont
  {Decker}}, \bibinfo {author} {\bibfnamefont {C.}~\bibnamefont {Karrasch}},
  \bibinfo {author} {\bibfnamefont {J.}~\bibnamefont {Eisert}}, \ and\ \bibinfo
  {author} {\bibfnamefont {D.~M.}\ \bibnamefont {Kennes}},\ }\href {\doibase
  10.1103/PhysRevLett.124.190601} {\bibfield  {journal} {\bibinfo  {journal}
  {Phys. Rev. Lett.}\ }\textbf {\bibinfo {volume} {124}},\ \bibinfo {pages}
  {190601} (\bibinfo {year} {2020})}\BibitemShut {NoStop}%
\bibitem [{\citenamefont {Rudner}\ \emph {et~al.}(2013)\citenamefont {Rudner},
  \citenamefont {Lindner}, \citenamefont {Berg},\ and\ \citenamefont
  {Levin}}]{rudner2013anomalous}%
  \BibitemOpen
  \bibfield  {author} {\bibinfo {author} {\bibfnamefont {M.~S.}\ \bibnamefont
  {Rudner}}, \bibinfo {author} {\bibfnamefont {N.~H.}\ \bibnamefont {Lindner}},
  \bibinfo {author} {\bibfnamefont {E.}~\bibnamefont {Berg}}, \ and\ \bibinfo
  {author} {\bibfnamefont {M.}~\bibnamefont {Levin}},\ }\href {\doibase
  10.1103/PhysRevX.3.031005} {\bibfield  {journal} {\bibinfo  {journal} {Phys.
  Rev. X}\ }\textbf {\bibinfo {volume} {3}},\ \bibinfo {pages} {031005}
  (\bibinfo {year} {2013})}\BibitemShut {NoStop}%
\bibitem [{\citenamefont {Nathan}\ and\ \citenamefont
  {Rudner}(2015)}]{nathan2015topological}%
  \BibitemOpen
  \bibfield  {author} {\bibinfo {author} {\bibfnamefont {F.}~\bibnamefont
  {Nathan}}\ and\ \bibinfo {author} {\bibfnamefont {M.~S.}\ \bibnamefont
  {Rudner}},\ }\href {\doibase 10.1088/1367-2630/17/12/125014} {\bibfield
  {journal} {\bibinfo  {journal} {New J. Phys.}\ }\textbf {\bibinfo {volume}
  {17}},\ \bibinfo {pages} {125014} (\bibinfo {year} {2015})}\BibitemShut
  {NoStop}%
\bibitem [{\citenamefont {Kitagawa}\ \emph {et~al.}(2010)\citenamefont
  {Kitagawa}, \citenamefont {Berg}, \citenamefont {Rudner},\ and\ \citenamefont
  {Demler}}]{kitagawa2010topological}%
  \BibitemOpen
  \bibfield  {author} {\bibinfo {author} {\bibfnamefont {T.}~\bibnamefont
  {Kitagawa}}, \bibinfo {author} {\bibfnamefont {E.}~\bibnamefont {Berg}},
  \bibinfo {author} {\bibfnamefont {M.}~\bibnamefont {Rudner}}, \ and\ \bibinfo
  {author} {\bibfnamefont {E.}~\bibnamefont {Demler}},\ }\href {\doibase
  10.1103/PhysRevB.82.235114} {\bibfield  {journal} {\bibinfo  {journal} {Phys.
  Rev. B}\ }\textbf {\bibinfo {volume} {82}},\ \bibinfo {pages} {235114}
  (\bibinfo {year} {2010})}\BibitemShut {NoStop}%
\bibitem [{\citenamefont {Kitagawa}\ \emph {et~al.}(2011)\citenamefont
  {Kitagawa}, \citenamefont {Oka}, \citenamefont {Brataas}, \citenamefont
  {Fu},\ and\ \citenamefont {Demler}}]{kitagawa2011transport}%
  \BibitemOpen
  \bibfield  {author} {\bibinfo {author} {\bibfnamefont {T.}~\bibnamefont
  {Kitagawa}}, \bibinfo {author} {\bibfnamefont {T.}~\bibnamefont {Oka}},
  \bibinfo {author} {\bibfnamefont {A.}~\bibnamefont {Brataas}}, \bibinfo
  {author} {\bibfnamefont {L.}~\bibnamefont {Fu}}, \ and\ \bibinfo {author}
  {\bibfnamefont {E.}~\bibnamefont {Demler}},\ }\href {\doibase
  10.1103/PhysRevB.84.235108} {\bibfield  {journal} {\bibinfo  {journal} {Phys.
  Rev. B}\ }\textbf {\bibinfo {volume} {84}},\ \bibinfo {pages} {235108}
  (\bibinfo {year} {2011})}\BibitemShut {NoStop}%
\bibitem [{\citenamefont {Lindner}\ \emph {et~al.}(2011)\citenamefont
  {Lindner}, \citenamefont {Refael},\ and\ \citenamefont
  {Galitski}}]{lindner2011floquet}%
  \BibitemOpen
  \bibfield  {author} {\bibinfo {author} {\bibfnamefont {N.~H.}\ \bibnamefont
  {Lindner}}, \bibinfo {author} {\bibfnamefont {G.}~\bibnamefont {Refael}}, \
  and\ \bibinfo {author} {\bibfnamefont {V.}~\bibnamefont {Galitski}},\ }\href
  {\doibase 10.1038/nphys1926} {\bibfield  {journal} {\bibinfo  {journal}
  {Nature Physics}\ }\textbf {\bibinfo {volume} {7}},\ \bibinfo {pages} {490}
  (\bibinfo {year} {2011})}\BibitemShut {NoStop}%
\bibitem [{\citenamefont {Liu}\ \emph {et~al.}(2013)\citenamefont {Liu},
  \citenamefont {Levchenko},\ and\ \citenamefont {Baranger}}]{liu}%
  \BibitemOpen
  \bibfield  {author} {\bibinfo {author} {\bibfnamefont {D.~E.}\ \bibnamefont
  {Liu}}, \bibinfo {author} {\bibfnamefont {A.}~\bibnamefont {Levchenko}}, \
  and\ \bibinfo {author} {\bibfnamefont {H.~U.}\ \bibnamefont {Baranger}},\
  }\href {\doibase 10.1103/PhysRevLett.111.047002} {\bibfield  {journal}
  {\bibinfo  {journal} {Phys. Rev. Lett.}\ }\textbf {\bibinfo {volume} {111}},\
  \bibinfo {pages} {047002} (\bibinfo {year} {2013})}\BibitemShut {NoStop}%
\bibitem [{\citenamefont {Kundu}\ and\ \citenamefont
  {Seradjeh}(2013)}]{kundu2013transport}%
  \BibitemOpen
  \bibfield  {author} {\bibinfo {author} {\bibfnamefont {A.}~\bibnamefont
  {Kundu}}\ and\ \bibinfo {author} {\bibfnamefont {B.}~\bibnamefont
  {Seradjeh}},\ }\href {\doibase 10.1103/PhysRevLett.111.136402} {\bibfield
  {journal} {\bibinfo  {journal} {Phys. Rev. Lett.}\ }\textbf {\bibinfo
  {volume} {111}},\ \bibinfo {pages} {136402} (\bibinfo {year}
  {2013})}\BibitemShut {NoStop}%
\bibitem [{\citenamefont {Kundu}\ \emph {et~al.}(2014)\citenamefont {Kundu},
  \citenamefont {Fertig},\ and\ \citenamefont {Seradjeh}}]{kundu2014effective}%
  \BibitemOpen
  \bibfield  {author} {\bibinfo {author} {\bibfnamefont {A.}~\bibnamefont
  {Kundu}}, \bibinfo {author} {\bibfnamefont {H.}~\bibnamefont {Fertig}}, \
  and\ \bibinfo {author} {\bibfnamefont {B.}~\bibnamefont {Seradjeh}},\ }\href
  {\doibase 10.1103/PhysRevLett.113.236803} {\bibfield  {journal} {\bibinfo
  {journal} {Phys. Rev. Lett.}\ }\textbf {\bibinfo {volume} {113}},\ \bibinfo
  {pages} {236803} (\bibinfo {year} {2014})}\BibitemShut {NoStop}%
\bibitem [{\citenamefont {Thakurathi}\ \emph {et~al.}(2013)\citenamefont
  {Thakurathi}, \citenamefont {Patel}, \citenamefont {Sen},\ and\ \citenamefont
  {Dutta}}]{thakurathi2013floquet}%
  \BibitemOpen
  \bibfield  {author} {\bibinfo {author} {\bibfnamefont {M.}~\bibnamefont
  {Thakurathi}}, \bibinfo {author} {\bibfnamefont {A.~A.}\ \bibnamefont
  {Patel}}, \bibinfo {author} {\bibfnamefont {D.}~\bibnamefont {Sen}}, \ and\
  \bibinfo {author} {\bibfnamefont {A.}~\bibnamefont {Dutta}},\ }\href
  {\doibase 10.1103/PhysRevB.88.155133} {\bibfield  {journal} {\bibinfo
  {journal} {Phys. Rev. B}\ }\textbf {\bibinfo {volume} {88}},\ \bibinfo
  {pages} {155133} (\bibinfo {year} {2013})}\BibitemShut {NoStop}%
\bibitem [{\citenamefont {Thakurathi}\ \emph {et~al.}(2014)\citenamefont
  {Thakurathi}, \citenamefont {Sengupta},\ and\ \citenamefont
  {Sen}}]{thakurathi2014majorana}%
  \BibitemOpen
  \bibfield  {author} {\bibinfo {author} {\bibfnamefont {M.}~\bibnamefont
  {Thakurathi}}, \bibinfo {author} {\bibfnamefont {K.}~\bibnamefont
  {Sengupta}}, \ and\ \bibinfo {author} {\bibfnamefont {D.}~\bibnamefont
  {Sen}},\ }\href {\doibase 10.1103/PhysRevB.89.235434} {\bibfield  {journal}
  {\bibinfo  {journal} {Phys. Rev. B}\ }\textbf {\bibinfo {volume} {89}},\
  \bibinfo {pages} {235434} (\bibinfo {year} {2014})}\BibitemShut {NoStop}%
\bibitem [{\citenamefont {Deb}\ and\ \citenamefont
  {Sen}(2017)}]{deb2017generating}%
  \BibitemOpen
  \bibfield  {author} {\bibinfo {author} {\bibfnamefont {O.}~\bibnamefont
  {Deb}}\ and\ \bibinfo {author} {\bibfnamefont {D.}~\bibnamefont {Sen}},\
  }\href {\doibase 10.1103/PhysRevB.95.144311} {\bibfield  {journal} {\bibinfo
  {journal} {Phys. Rev. B}\ }\textbf {\bibinfo {volume} {95}},\ \bibinfo
  {pages} {144311} (\bibinfo {year} {2017})}\BibitemShut {NoStop}%
\bibitem [{\citenamefont {Saha}\ \emph {et~al.}(2017)\citenamefont {Saha},
  \citenamefont {Sivarajan},\ and\ \citenamefont {Sen}}]{saha2017generating}%
  \BibitemOpen
  \bibfield  {author} {\bibinfo {author} {\bibfnamefont {S.}~\bibnamefont
  {Saha}}, \bibinfo {author} {\bibfnamefont {S.~N.}\ \bibnamefont {Sivarajan}},
  \ and\ \bibinfo {author} {\bibfnamefont {D.}~\bibnamefont {Sen}},\ }\href
  {\doibase 10.1103/PhysRevB.95.174306} {\bibfield  {journal} {\bibinfo
  {journal} {Phys. Rev. B}\ }\textbf {\bibinfo {volume} {95}},\ \bibinfo
  {pages} {174306} (\bibinfo {year} {2017})}\BibitemShut {NoStop}%
\bibitem [{\citenamefont {Balabanov}\ and\ \citenamefont
  {Johannesson}(2017)}]{balabanov}%
  \BibitemOpen
  \bibfield  {author} {\bibinfo {author} {\bibfnamefont {O.}~\bibnamefont
  {Balabanov}}\ and\ \bibinfo {author} {\bibfnamefont {H.}~\bibnamefont
  {Johannesson}},\ }\href {\doibase 10.1103/PhysRevB.96.035149} {\bibfield
  {journal} {\bibinfo  {journal} {Phys. Rev. B}\ }\textbf {\bibinfo {volume}
  {96}},\ \bibinfo {pages} {035149} (\bibinfo {year} {2017})}\BibitemShut
  {NoStop}%
\bibitem [{\citenamefont {Yates}\ and\ \citenamefont
  {Mitra}(2017)}]{yates2017}%
  \BibitemOpen
  \bibfield  {author} {\bibinfo {author} {\bibfnamefont {D.~J.}\ \bibnamefont
  {Yates}}\ and\ \bibinfo {author} {\bibfnamefont {A.}~\bibnamefont {Mitra}},\
  }\href {\doibase 10.1103/PhysRevB.96.115108} {\bibfield  {journal} {\bibinfo
  {journal} {Phys. Rev. B}\ }\textbf {\bibinfo {volume} {96}},\ \bibinfo
  {pages} {115108} (\bibinfo {year} {2017})}\BibitemShut {NoStop}%
\bibitem [{\citenamefont {Molignini}\ \emph {et~al.}(2018)\citenamefont
  {Molignini}, \citenamefont {Chen},\ and\ \citenamefont {Chitra}}]{molignini}%
  \BibitemOpen
  \bibfield  {author} {\bibinfo {author} {\bibfnamefont {P.}~\bibnamefont
  {Molignini}}, \bibinfo {author} {\bibfnamefont {W.}~\bibnamefont {Chen}}, \
  and\ \bibinfo {author} {\bibfnamefont {R.}~\bibnamefont {Chitra}},\ }\href
  {\doibase 10.1103/PhysRevB.98.125129} {\bibfield  {journal} {\bibinfo
  {journal} {Phys. Rev. B}\ }\textbf {\bibinfo {volume} {98}},\ \bibinfo
  {pages} {125129} (\bibinfo {year} {2018})}\BibitemShut {NoStop}%
\bibitem [{\citenamefont {Li}\ \emph {et~al.}(2019)\citenamefont {Li},
  \citenamefont {Li}, \citenamefont {Wang},\ and\ \citenamefont {Zhou}}]{lili}%
  \BibitemOpen
  \bibfield  {author} {\bibinfo {author} {\bibfnamefont {X.~P.}\ \bibnamefont
  {Li}}, \bibinfo {author} {\bibfnamefont {C.~F.}\ \bibnamefont {Li}}, \bibinfo
  {author} {\bibfnamefont {L.~C.}\ \bibnamefont {Wang}}, \ and\ \bibinfo
  {author} {\bibfnamefont {L.}~\bibnamefont {Zhou}},\ }\href {\doibase
  10.1007/s10773-019-04054-2} {\bibfield  {journal} {\bibinfo  {journal} {Int.
  J. Theor. Phys}\ }\textbf {\bibinfo {volume} {58}},\ \bibinfo {pages} {1590}
  (\bibinfo {year} {2019})}\BibitemShut {NoStop}%
\bibitem [{\citenamefont {M\"uller}\ \emph {et~al.}(2020)\citenamefont
  {M\"uller}, \citenamefont {Kennes}, \citenamefont {Klinovaja}, \citenamefont
  {Loss},\ and\ \citenamefont {Schoeller}}]{muller}%
  \BibitemOpen
  \bibfield  {author} {\bibinfo {author} {\bibfnamefont {N.}~\bibnamefont
  {M\"uller}}, \bibinfo {author} {\bibfnamefont {D.~M.}\ \bibnamefont
  {Kennes}}, \bibinfo {author} {\bibfnamefont {J.}~\bibnamefont {Klinovaja}},
  \bibinfo {author} {\bibfnamefont {D.}~\bibnamefont {Loss}}, \ and\ \bibinfo
  {author} {\bibfnamefont {H.}~\bibnamefont {Schoeller}},\ }\href {\doibase
  10.1103/PhysRevB.101.155417} {\bibfield  {journal} {\bibinfo  {journal}
  {Phys. Rev. B}\ }\textbf {\bibinfo {volume} {101}},\ \bibinfo {pages}
  {155417} (\bibinfo {year} {2020})}\BibitemShut {NoStop}%
\bibitem [{\citenamefont {Seshadri}\ \emph {et~al.}(2019)\citenamefont
  {Seshadri}, \citenamefont {Dutta},\ and\ \citenamefont
  {Sen}}]{seshadri2019generating}%
  \BibitemOpen
  \bibfield  {author} {\bibinfo {author} {\bibfnamefont {R.}~\bibnamefont
  {Seshadri}}, \bibinfo {author} {\bibfnamefont {A.}~\bibnamefont {Dutta}}, \
  and\ \bibinfo {author} {\bibfnamefont {D.}~\bibnamefont {Sen}},\ }\href
  {\doibase 10.1103/PhysRevB.100.115403} {\bibfield  {journal} {\bibinfo
  {journal} {Phys. Rev. B}\ }\textbf {\bibinfo {volume} {100}},\ \bibinfo
  {pages} {115403} (\bibinfo {year} {2019})}\BibitemShut {NoStop}%
\bibitem [{\citenamefont {Seshadri}\ and\ \citenamefont
  {Sen}(2022)}]{seshadri2022engineering}%
  \BibitemOpen
  \bibfield  {author} {\bibinfo {author} {\bibfnamefont {R.}~\bibnamefont
  {Seshadri}}\ and\ \bibinfo {author} {\bibfnamefont {D.}~\bibnamefont {Sen}},\
  }\href {\doibase 10.1103/PhysRevB.106.245401} {\bibfield  {journal} {\bibinfo
   {journal} {Phys. Rev. B}\ }\textbf {\bibinfo {volume} {106}},\ \bibinfo
  {pages} {245401} (\bibinfo {year} {2022})}\BibitemShut {NoStop}%
\bibitem [{\citenamefont {Sur}\ and\ \citenamefont
  {Sen}(2021)}]{sur2021floquet}%
  \BibitemOpen
  \bibfield  {author} {\bibinfo {author} {\bibfnamefont {S.}~\bibnamefont
  {Sur}}\ and\ \bibinfo {author} {\bibfnamefont {D.}~\bibnamefont {Sen}},\
  }\href {\doibase 10.1103/PhysRevB.103.085417} {\bibfield  {journal} {\bibinfo
   {journal} {Phys. Rev. B}\ }\textbf {\bibinfo {volume} {103}},\ \bibinfo
  {pages} {085417} (\bibinfo {year} {2021})}\BibitemShut {NoStop}%
\bibitem [{\citenamefont {Else}\ \emph {et~al.}(2016)\citenamefont {Else},
  \citenamefont {Bauer},\ and\ \citenamefont {Nayak}}]{else2016floquet}%
  \BibitemOpen
  \bibfield  {author} {\bibinfo {author} {\bibfnamefont {D.~V.}\ \bibnamefont
  {Else}}, \bibinfo {author} {\bibfnamefont {B.}~\bibnamefont {Bauer}}, \ and\
  \bibinfo {author} {\bibfnamefont {C.}~\bibnamefont {Nayak}},\ }\href
  {\doibase 10.1103/PhysRevLett.117.090402} {\bibfield  {journal} {\bibinfo
  {journal} {Phys. Rev. Lett.}\ }\textbf {\bibinfo {volume} {117}},\ \bibinfo
  {pages} {090402} (\bibinfo {year} {2016})}\BibitemShut {NoStop}%
\bibitem [{\citenamefont {Zeng}\ and\ \citenamefont
  {Sheng}(2017)}]{zeng2017prethermal}%
  \BibitemOpen
  \bibfield  {author} {\bibinfo {author} {\bibfnamefont {T.-S.}\ \bibnamefont
  {Zeng}}\ and\ \bibinfo {author} {\bibfnamefont {D.}~\bibnamefont {Sheng}},\
  }\href {\doibase 10.1103/PhysRevB.96.094202} {\bibfield  {journal} {\bibinfo
  {journal} {Phys. Rev. B}\ }\textbf {\bibinfo {volume} {96}},\ \bibinfo
  {pages} {094202} (\bibinfo {year} {2017})}\BibitemShut {NoStop}%
\bibitem [{\citenamefont {Zhang}\ \emph {et~al.}(2017)\citenamefont {Zhang},
  \citenamefont {Hess}, \citenamefont {Kyprianidis}, \citenamefont {Becker},
  \citenamefont {Lee}, \citenamefont {Smith}, \citenamefont {Pagano},
  \citenamefont {Potirniche}, \citenamefont {Potter}, \citenamefont
  {Vishwanath}, \citenamefont {Yao},\ and\ \citenamefont
  {Monroe}}]{zhang2017observation}%
  \BibitemOpen
  \bibfield  {author} {\bibinfo {author} {\bibfnamefont {J.}~\bibnamefont
  {Zhang}}, \bibinfo {author} {\bibfnamefont {P.~W.}\ \bibnamefont {Hess}},
  \bibinfo {author} {\bibfnamefont {A.}~\bibnamefont {Kyprianidis}}, \bibinfo
  {author} {\bibfnamefont {P.}~\bibnamefont {Becker}}, \bibinfo {author}
  {\bibfnamefont {A.}~\bibnamefont {Lee}}, \bibinfo {author} {\bibfnamefont
  {J.}~\bibnamefont {Smith}}, \bibinfo {author} {\bibfnamefont
  {G.}~\bibnamefont {Pagano}}, \bibinfo {author} {\bibfnamefont {I.-D.}\
  \bibnamefont {Potirniche}}, \bibinfo {author} {\bibfnamefont {A.~C.}\
  \bibnamefont {Potter}}, \bibinfo {author} {\bibfnamefont {A.}~\bibnamefont
  {Vishwanath}}, \bibinfo {author} {\bibfnamefont {N.~Y.}\ \bibnamefont {Yao}},
  \ and\ \bibinfo {author} {\bibfnamefont {C.}~\bibnamefont {Monroe}},\ }\href
  {\doibase 10.1038/nature21413} {\bibfield  {journal} {\bibinfo  {journal}
  {Nature}\ }\textbf {\bibinfo {volume} {543}},\ \bibinfo {pages} {217}
  (\bibinfo {year} {2017})}\BibitemShut {NoStop}%
\bibitem [{\citenamefont {Russomanno}\ \emph {et~al.}(2017)\citenamefont
  {Russomanno}, \citenamefont {Iemini}, \citenamefont {Dalmonte},\ and\
  \citenamefont {Fazio}}]{russomanno2017floquet}%
  \BibitemOpen
  \bibfield  {author} {\bibinfo {author} {\bibfnamefont {A.}~\bibnamefont
  {Russomanno}}, \bibinfo {author} {\bibfnamefont {F.}~\bibnamefont {Iemini}},
  \bibinfo {author} {\bibfnamefont {M.}~\bibnamefont {Dalmonte}}, \ and\
  \bibinfo {author} {\bibfnamefont {R.}~\bibnamefont {Fazio}},\ }\href
  {\doibase 10.1103/PhysRevB.95.214307} {\bibfield  {journal} {\bibinfo
  {journal} {Phys. Rev. B}\ }\textbf {\bibinfo {volume} {95}},\ \bibinfo
  {pages} {214307} (\bibinfo {year} {2017})}\BibitemShut {NoStop}%
\bibitem [{\citenamefont {Surace}\ \emph {et~al.}(2019)\citenamefont {Surace},
  \citenamefont {Russomanno}, \citenamefont {Dalmonte}, \citenamefont {Silva},
  \citenamefont {Fazio},\ and\ \citenamefont {Iemini}}]{surace2019floquet}%
  \BibitemOpen
  \bibfield  {author} {\bibinfo {author} {\bibfnamefont {F.~M.}\ \bibnamefont
  {Surace}}, \bibinfo {author} {\bibfnamefont {A.}~\bibnamefont {Russomanno}},
  \bibinfo {author} {\bibfnamefont {M.}~\bibnamefont {Dalmonte}}, \bibinfo
  {author} {\bibfnamefont {A.}~\bibnamefont {Silva}}, \bibinfo {author}
  {\bibfnamefont {R.}~\bibnamefont {Fazio}}, \ and\ \bibinfo {author}
  {\bibfnamefont {F.}~\bibnamefont {Iemini}},\ }\href {\doibase
  10.1103/PhysRevB.99.104303} {\bibfield  {journal} {\bibinfo  {journal} {Phys.
  Rev. B}\ }\textbf {\bibinfo {volume} {99}},\ \bibinfo {pages} {104303}
  (\bibinfo {year} {2019})}\BibitemShut {NoStop}%
\bibitem [{\citenamefont {Yarloo}\ \emph {et~al.}(2020)\citenamefont {Yarloo},
  \citenamefont {Kopaei},\ and\ \citenamefont
  {Langari}}]{yarloo2020homogeneous}%
  \BibitemOpen
  \bibfield  {author} {\bibinfo {author} {\bibfnamefont {H.}~\bibnamefont
  {Yarloo}}, \bibinfo {author} {\bibfnamefont {A.~E.}\ \bibnamefont {Kopaei}},
  \ and\ \bibinfo {author} {\bibfnamefont {A.}~\bibnamefont {Langari}},\ }\href
  {\doibase 10.1103/PhysRevB.102.224309} {\bibfield  {journal} {\bibinfo
  {journal} {Phys. Rev. B}\ }\textbf {\bibinfo {volume} {102}},\ \bibinfo
  {pages} {224309} (\bibinfo {year} {2020})}\BibitemShut {NoStop}%
\bibitem [{\citenamefont {Lazarides}\ \emph {et~al.}(2020)\citenamefont
  {Lazarides}, \citenamefont {Roy}, \citenamefont {Piazza},\ and\ \citenamefont
  {Moessner}}]{lazarides2020time}%
  \BibitemOpen
  \bibfield  {author} {\bibinfo {author} {\bibfnamefont {A.}~\bibnamefont
  {Lazarides}}, \bibinfo {author} {\bibfnamefont {S.}~\bibnamefont {Roy}},
  \bibinfo {author} {\bibfnamefont {F.}~\bibnamefont {Piazza}}, \ and\ \bibinfo
  {author} {\bibfnamefont {R.}~\bibnamefont {Moessner}},\ }\href {\doibase
  10.1103/PhysRevResearch.2.022002} {\bibfield  {journal} {\bibinfo  {journal}
  {Phys. Rev. Research}\ }\textbf {\bibinfo {volume} {2}},\ \bibinfo {pages}
  {022002} (\bibinfo {year} {2020})}\BibitemShut {NoStop}%
\bibitem [{\citenamefont {Natsheh}\ \emph {et~al.}(2021)\citenamefont
  {Natsheh}, \citenamefont {Gambassi},\ and\ \citenamefont
  {Mitra}}]{natsheh2021critical}%
  \BibitemOpen
  \bibfield  {author} {\bibinfo {author} {\bibfnamefont {M.}~\bibnamefont
  {Natsheh}}, \bibinfo {author} {\bibfnamefont {A.}~\bibnamefont {Gambassi}}, \
  and\ \bibinfo {author} {\bibfnamefont {A.}~\bibnamefont {Mitra}},\ }\href
  {\doibase 10.1103/PhysRevB.103.014305} {\bibfield  {journal} {\bibinfo
  {journal} {Phys. Rev. B}\ }\textbf {\bibinfo {volume} {103}},\ \bibinfo
  {pages} {014305} (\bibinfo {year} {2021})}\BibitemShut {NoStop}%
\bibitem [{\citenamefont {Yang}\ \emph {et~al.}(2019)\citenamefont {Yang},
  \citenamefont {Zhou}, \citenamefont {Ma}, \citenamefont {Kong}, \citenamefont
  {Wang}, \citenamefont {Qin}, \citenamefont {Rong}, \citenamefont {Wang},
  \citenamefont {Shi}, \citenamefont {Gong},\ and\ \citenamefont
  {Du}}]{yang2019floquet}%
  \BibitemOpen
  \bibfield  {author} {\bibinfo {author} {\bibfnamefont {K.}~\bibnamefont
  {Yang}}, \bibinfo {author} {\bibfnamefont {L.}~\bibnamefont {Zhou}}, \bibinfo
  {author} {\bibfnamefont {W.}~\bibnamefont {Ma}}, \bibinfo {author}
  {\bibfnamefont {X.}~\bibnamefont {Kong}}, \bibinfo {author} {\bibfnamefont
  {P.}~\bibnamefont {Wang}}, \bibinfo {author} {\bibfnamefont {X.}~\bibnamefont
  {Qin}}, \bibinfo {author} {\bibfnamefont {X.}~\bibnamefont {Rong}}, \bibinfo
  {author} {\bibfnamefont {Y.}~\bibnamefont {Wang}}, \bibinfo {author}
  {\bibfnamefont {F.}~\bibnamefont {Shi}}, \bibinfo {author} {\bibfnamefont
  {J.}~\bibnamefont {Gong}}, \ and\ \bibinfo {author} {\bibfnamefont
  {J.}~\bibnamefont {Du}},\ }\href {\doibase 10.1103/PhysRevB.100.085308}
  {\bibfield  {journal} {\bibinfo  {journal} {Phys. Rev. B}\ }\textbf {\bibinfo
  {volume} {100}},\ \bibinfo {pages} {085308} (\bibinfo {year}
  {2019})}\BibitemShut {NoStop}%
\bibitem [{\citenamefont {Zamani}\ \emph {et~al.}(2020)\citenamefont {Zamani},
  \citenamefont {Jafari},\ and\ \citenamefont {Langari}}]{zamani2020floquet}%
  \BibitemOpen
  \bibfield  {author} {\bibinfo {author} {\bibfnamefont {S.}~\bibnamefont
  {Zamani}}, \bibinfo {author} {\bibfnamefont {R.}~\bibnamefont {Jafari}}, \
  and\ \bibinfo {author} {\bibfnamefont {A.}~\bibnamefont {Langari}},\ }\href
  {\doibase 10.1103/PhysRevB.102.144306} {\bibfield  {journal} {\bibinfo
  {journal} {Phys. Rev. B}\ }\textbf {\bibinfo {volume} {102}},\ \bibinfo
  {pages} {144306} (\bibinfo {year} {2020})}\BibitemShut {NoStop}%
\bibitem [{\citenamefont {Jafari}\ and\ \citenamefont
  {Akbari}(2021)}]{jafari2021floquet}%
  \BibitemOpen
  \bibfield  {author} {\bibinfo {author} {\bibfnamefont {R.}~\bibnamefont
  {Jafari}}\ and\ \bibinfo {author} {\bibfnamefont {A.}~\bibnamefont
  {Akbari}},\ }\href {\doibase 10.1103/PhysRevA.103.012204} {\bibfield
  {journal} {\bibinfo  {journal} {Phys. Rev. A}\ }\textbf {\bibinfo {volume}
  {103}},\ \bibinfo {pages} {012204} (\bibinfo {year} {2021})}\BibitemShut
  {NoStop}%
\bibitem [{\citenamefont {Jafari}\ \emph {et~al.}(2022)\citenamefont {Jafari},
  \citenamefont {Akbari}, \citenamefont {Mishra},\ and\ \citenamefont
  {Johannesson}}]{jafari2022floquet}%
  \BibitemOpen
  \bibfield  {author} {\bibinfo {author} {\bibfnamefont {R.}~\bibnamefont
  {Jafari}}, \bibinfo {author} {\bibfnamefont {A.}~\bibnamefont {Akbari}},
  \bibinfo {author} {\bibfnamefont {U.}~\bibnamefont {Mishra}}, \ and\ \bibinfo
  {author} {\bibfnamefont {H.}~\bibnamefont {Johannesson}},\ }\href {\doibase
  10.1103/PhysRevB.105.094311} {\bibfield  {journal} {\bibinfo  {journal}
  {Phys. Rev. B}\ }\textbf {\bibinfo {volume} {105}},\ \bibinfo {pages}
  {094311} (\bibinfo {year} {2022})}\BibitemShut {NoStop}%
\bibitem [{\citenamefont {Asb{\'o}th}(2012)}]{asboth2012symmetries}%
  \BibitemOpen
  \bibfield  {author} {\bibinfo {author} {\bibfnamefont {J.~K.}\ \bibnamefont
  {Asb{\'o}th}},\ }\href {\doibase 10.1103/PhysRevB.86.195414} {\bibfield
  {journal} {\bibinfo  {journal} {Phys. Rev. B}\ }\textbf {\bibinfo {volume}
  {86}},\ \bibinfo {pages} {195414} (\bibinfo {year} {2012})}\BibitemShut
  {NoStop}%
\bibitem [{\citenamefont {Asb{\'o}th}\ and\ \citenamefont
  {Obuse}(2013)}]{asboth2013bulk}%
  \BibitemOpen
  \bibfield  {author} {\bibinfo {author} {\bibfnamefont {J.~K.}\ \bibnamefont
  {Asb{\'o}th}}\ and\ \bibinfo {author} {\bibfnamefont {H.}~\bibnamefont
  {Obuse}},\ }\href {\doibase 10.1103/PhysRevB.88.121406} {\bibfield  {journal}
  {\bibinfo  {journal} {Phys. Rev. B}\ }\textbf {\bibinfo {volume} {88}},\
  \bibinfo {pages} {121406(R)} (\bibinfo {year} {2013})}\BibitemShut {NoStop}%
\bibitem [{\citenamefont {Naji}\ \emph {et~al.}(2022)\citenamefont {Naji},
  \citenamefont {Jafari}, \citenamefont {Jafari},\ and\ \citenamefont
  {Akbari}}]{naji2022dissipative}%
  \BibitemOpen
  \bibfield  {author} {\bibinfo {author} {\bibfnamefont {J.}~\bibnamefont
  {Naji}}, \bibinfo {author} {\bibfnamefont {M.}~\bibnamefont {Jafari}},
  \bibinfo {author} {\bibfnamefont {R.}~\bibnamefont {Jafari}}, \ and\ \bibinfo
  {author} {\bibfnamefont {A.}~\bibnamefont {Akbari}},\ }\href {\doibase
  10.1103/PhysRevA.105.022220} {\bibfield  {journal} {\bibinfo  {journal}
  {Phys. Rev. A}\ }\textbf {\bibinfo {volume} {105}},\ \bibinfo {pages}
  {022220} (\bibinfo {year} {2022})}\BibitemShut {NoStop}%
\bibitem [{\citenamefont {Dunlap}\ and\ \citenamefont
  {Kenkre}(1986)}]{dunlap1986dynamic}%
  \BibitemOpen
  \bibfield  {author} {\bibinfo {author} {\bibfnamefont {D.}~\bibnamefont
  {Dunlap}}\ and\ \bibinfo {author} {\bibfnamefont {V.}~\bibnamefont
  {Kenkre}},\ }\href {\doibase 10.1103/PhysRevB.34.3625} {\bibfield  {journal}
  {\bibinfo  {journal} {Phys. Rev. B}\ }\textbf {\bibinfo {volume} {34}},\
  \bibinfo {pages} {3625} (\bibinfo {year} {1986})}\BibitemShut {NoStop}%
\bibitem [{\citenamefont {Das}(2010)}]{das2010exotic}%
  \BibitemOpen
  \bibfield  {author} {\bibinfo {author} {\bibfnamefont {A.}~\bibnamefont
  {Das}},\ }\href {\doibase 10.1103/PhysRevB.82.172402} {\bibfield  {journal}
  {\bibinfo  {journal} {Phys. Rev. B}\ }\textbf {\bibinfo {volume} {82}},\
  \bibinfo {pages} {172402} (\bibinfo {year} {2010})}\BibitemShut {NoStop}%
\bibitem [{\citenamefont {Bhattacharyya}\ \emph {et~al.}(2012)\citenamefont
  {Bhattacharyya}, \citenamefont {Das},\ and\ \citenamefont
  {Dasgupta}}]{bhattacharyya2012transverse}%
  \BibitemOpen
  \bibfield  {author} {\bibinfo {author} {\bibfnamefont {S.}~\bibnamefont
  {Bhattacharyya}}, \bibinfo {author} {\bibfnamefont {A.}~\bibnamefont {Das}},
  \ and\ \bibinfo {author} {\bibfnamefont {S.}~\bibnamefont {Dasgupta}},\
  }\href {\doibase 10.1103/PhysRevB.86.054410} {\bibfield  {journal} {\bibinfo
  {journal} {Phys. Rev. B}\ }\textbf {\bibinfo {volume} {86}},\ \bibinfo
  {pages} {054410} (\bibinfo {year} {2012})}\BibitemShut {NoStop}%
\bibitem [{\citenamefont {Nag}\ \emph {et~al.}(2015)\citenamefont {Nag},
  \citenamefont {Sen},\ and\ \citenamefont {Dutta}}]{nag2015maximum}%
  \BibitemOpen
  \bibfield  {author} {\bibinfo {author} {\bibfnamefont {T.}~\bibnamefont
  {Nag}}, \bibinfo {author} {\bibfnamefont {D.}~\bibnamefont {Sen}}, \ and\
  \bibinfo {author} {\bibfnamefont {A.}~\bibnamefont {Dutta}},\ }\href
  {\doibase 10.1103/PhysRevA.91.063607} {\bibfield  {journal} {\bibinfo
  {journal} {Phys. Rev. A}\ }\textbf {\bibinfo {volume} {91}},\ \bibinfo
  {pages} {063607} (\bibinfo {year} {2015})}\BibitemShut {NoStop}%
\bibitem [{\citenamefont {Mondal}\ \emph {et~al.}(2013)\citenamefont {Mondal},
  \citenamefont {Pekker},\ and\ \citenamefont {Sengupta}}]{mondal2013dynamics}%
  \BibitemOpen
  \bibfield  {author} {\bibinfo {author} {\bibfnamefont {S.}~\bibnamefont
  {Mondal}}, \bibinfo {author} {\bibfnamefont {D.}~\bibnamefont {Pekker}}, \
  and\ \bibinfo {author} {\bibfnamefont {K.}~\bibnamefont {Sengupta}},\ }\href
  {\doibase 10.1209/0295-5075/100/60007} {\bibfield  {journal} {\bibinfo
  {journal} {EPL}\ }\textbf {\bibinfo {volume} {100}},\ \bibinfo {pages}
  {60007} (\bibinfo {year} {2013})}\BibitemShut {NoStop}%
\bibitem [{\citenamefont {Divakaran}\ and\ \citenamefont
  {Sengupta}(2014)}]{divakaran2014dynamic}%
  \BibitemOpen
  \bibfield  {author} {\bibinfo {author} {\bibfnamefont {U.}~\bibnamefont
  {Divakaran}}\ and\ \bibinfo {author} {\bibfnamefont {K.}~\bibnamefont
  {Sengupta}},\ }\href {\doibase 10.1103/PhysRevB.90.184303} {\bibfield
  {journal} {\bibinfo  {journal} {Phys. Rev. B}\ }\textbf {\bibinfo {volume}
  {90}},\ \bibinfo {pages} {184303} (\bibinfo {year} {2014})}\BibitemShut
  {NoStop}%
\bibitem [{\citenamefont {Agarwala}\ \emph {et~al.}(2016)\citenamefont
  {Agarwala}, \citenamefont {Bhattacharya}, \citenamefont {Dutta},\ and\
  \citenamefont {Sen}}]{agarwala2016effects}%
  \BibitemOpen
  \bibfield  {author} {\bibinfo {author} {\bibfnamefont {A.}~\bibnamefont
  {Agarwala}}, \bibinfo {author} {\bibfnamefont {U.}~\bibnamefont
  {Bhattacharya}}, \bibinfo {author} {\bibfnamefont {A.}~\bibnamefont {Dutta}},
  \ and\ \bibinfo {author} {\bibfnamefont {D.}~\bibnamefont {Sen}},\ }\href
  {\doibase 10.1103/PhysRevB.93.174301} {\bibfield  {journal} {\bibinfo
  {journal} {Phys. Rev. B}\ }\textbf {\bibinfo {volume} {93}},\ \bibinfo
  {pages} {174301} (\bibinfo {year} {2016})}\BibitemShut {NoStop}%
\bibitem [{\citenamefont {Agarwala}\ and\ \citenamefont
  {Sen}(2017)}]{agarwala2017effects}%
  \BibitemOpen
  \bibfield  {author} {\bibinfo {author} {\bibfnamefont {A.}~\bibnamefont
  {Agarwala}}\ and\ \bibinfo {author} {\bibfnamefont {D.}~\bibnamefont {Sen}},\
  }\href {\doibase 10.1103/PhysRevB.95.014305} {\bibfield  {journal} {\bibinfo
  {journal} {Phys. Rev. B}\ }\textbf {\bibinfo {volume} {95}},\ \bibinfo
  {pages} {014305} (\bibinfo {year} {2017})}\BibitemShut {NoStop}%
\bibitem [{\citenamefont {Iubini}\ \emph {et~al.}(2019)\citenamefont {Iubini},
  \citenamefont {Chirondojan}, \citenamefont {Oppo}, \citenamefont {Politi},\
  and\ \citenamefont {Politi}}]{iubini2019dynamical}%
  \BibitemOpen
  \bibfield  {author} {\bibinfo {author} {\bibfnamefont {S.}~\bibnamefont
  {Iubini}}, \bibinfo {author} {\bibfnamefont {L.}~\bibnamefont {Chirondojan}},
  \bibinfo {author} {\bibfnamefont {G.-L.}\ \bibnamefont {Oppo}}, \bibinfo
  {author} {\bibfnamefont {A.}~\bibnamefont {Politi}}, \ and\ \bibinfo {author}
  {\bibfnamefont {P.}~\bibnamefont {Politi}},\ }\href {\doibase
  10.1103/PhysRevLett.122.084102} {\bibfield  {journal} {\bibinfo  {journal}
  {Phys. Rev. Lett.}\ }\textbf {\bibinfo {volume} {122}},\ \bibinfo {pages}
  {084102} (\bibinfo {year} {2019})}\BibitemShut {NoStop}%
\bibitem [{\citenamefont {Mukherjee}\ \emph {et~al.}(2020)\citenamefont
  {Mukherjee}, \citenamefont {Nandy}, \citenamefont {Sen}, \citenamefont
  {Sen},\ and\ \citenamefont {Sengupta}}]{mukherjee2020collapse}%
  \BibitemOpen
  \bibfield  {author} {\bibinfo {author} {\bibfnamefont {B.}~\bibnamefont
  {Mukherjee}}, \bibinfo {author} {\bibfnamefont {S.}~\bibnamefont {Nandy}},
  \bibinfo {author} {\bibfnamefont {A.}~\bibnamefont {Sen}}, \bibinfo {author}
  {\bibfnamefont {D.}~\bibnamefont {Sen}}, \ and\ \bibinfo {author}
  {\bibfnamefont {K.}~\bibnamefont {Sengupta}},\ }\href {\doibase
  10.1103/PhysRevB.101.245107} {\bibfield  {journal} {\bibinfo  {journal}
  {Phys. Rev. B}\ }\textbf {\bibinfo {volume} {101}},\ \bibinfo {pages}
  {245107} (\bibinfo {year} {2020})}\BibitemShut {NoStop}%
\bibitem [{\citenamefont {Haldar}\ \emph {et~al.}(2021)\citenamefont {Haldar},
  \citenamefont {Sen}, \citenamefont {Moessner},\ and\ \citenamefont
  {Das}}]{haldar2021dynamical}%
  \BibitemOpen
  \bibfield  {author} {\bibinfo {author} {\bibfnamefont {A.}~\bibnamefont
  {Haldar}}, \bibinfo {author} {\bibfnamefont {D.}~\bibnamefont {Sen}},
  \bibinfo {author} {\bibfnamefont {R.}~\bibnamefont {Moessner}}, \ and\
  \bibinfo {author} {\bibfnamefont {A.}~\bibnamefont {Das}},\ }\href {\doibase
  10.1103/PhysRevX.11.021008} {\bibfield  {journal} {\bibinfo  {journal} {Phys.
  Rev. X}\ }\textbf {\bibinfo {volume} {11}},\ \bibinfo {pages} {021008}
  (\bibinfo {year} {2021})}\BibitemShut {NoStop}%
\bibitem [{\citenamefont {Rodriguez-Vega}\ and\ \citenamefont
  {Seradjeh}(2018)}]{rodriguez2018universal}%
  \BibitemOpen
  \bibfield  {author} {\bibinfo {author} {\bibfnamefont {M.}~\bibnamefont
  {Rodriguez-Vega}}\ and\ \bibinfo {author} {\bibfnamefont {B.}~\bibnamefont
  {Seradjeh}},\ }\href {\doibase 10.1103/PhysRevLett.121.036402} {\bibfield
  {journal} {\bibinfo  {journal} {Phys. Rev. Lett.}\ }\textbf {\bibinfo
  {volume} {121}},\ \bibinfo {pages} {036402} (\bibinfo {year}
  {2018})}\BibitemShut {NoStop}%
\bibitem [{\citenamefont {Sreejith}\ \emph {et~al.}(2016)\citenamefont
  {Sreejith}, \citenamefont {Lazarides},\ and\ \citenamefont
  {Moessner}}]{sreejith2016parafermion}%
  \BibitemOpen
  \bibfield  {author} {\bibinfo {author} {\bibfnamefont {G.}~\bibnamefont
  {Sreejith}}, \bibinfo {author} {\bibfnamefont {A.}~\bibnamefont {Lazarides}},
  \ and\ \bibinfo {author} {\bibfnamefont {R.}~\bibnamefont {Moessner}},\
  }\href {\doibase 10.1103/PhysRevB.94.045127} {\bibfield  {journal} {\bibinfo
  {journal} {Phys. Rev. B}\ }\textbf {\bibinfo {volume} {94}},\ \bibinfo
  {pages} {045127} (\bibinfo {year} {2016})}\BibitemShut {NoStop}%
\bibitem [{\citenamefont {Kitaev}(2001)}]{kitaev2001}%
  \BibitemOpen
  \bibfield  {author} {\bibinfo {author} {\bibfnamefont {A.~Y.}\ \bibnamefont
  {Kitaev}},\ }\href {\doibase 10.1070/1063-7869/44/10S/S29} {\bibfield
  {journal} {\bibinfo  {journal} {Physics-Uspekhi}\ }\textbf {\bibinfo {volume}
  {44}},\ \bibinfo {pages} {131} (\bibinfo {year} {2001})}\BibitemShut
  {NoStop}%
\bibitem [{\citenamefont {DeGottardi}\ \emph {et~al.}(2013)\citenamefont
  {DeGottardi}, \citenamefont {Thakurathi}, \citenamefont {Vishveshwara},\ and\
  \citenamefont {Sen}}]{thakurathi2013majorana}%
  \BibitemOpen
  \bibfield  {author} {\bibinfo {author} {\bibfnamefont {W.}~\bibnamefont
  {DeGottardi}}, \bibinfo {author} {\bibfnamefont {M.}~\bibnamefont
  {Thakurathi}}, \bibinfo {author} {\bibfnamefont {S.}~\bibnamefont
  {Vishveshwara}}, \ and\ \bibinfo {author} {\bibfnamefont {D.}~\bibnamefont
  {Sen}},\ }\href {\doibase 10.1103/PhysRevB.88.165111} {\bibfield  {journal}
  {\bibinfo  {journal} {Phys. Rev. B}\ }\textbf {\bibinfo {volume} {88}},\
  \bibinfo {pages} {165111} (\bibinfo {year} {2013})}\BibitemShut {NoStop}%
\bibitem [{\citenamefont {Lieb}\ and\ \citenamefont
  {Robinson}(1972)}]{lieb1972finite}%
  \BibitemOpen
  \bibfield  {author} {\bibinfo {author} {\bibfnamefont {E.~H.}\ \bibnamefont
  {Lieb}}\ and\ \bibinfo {author} {\bibfnamefont {D.~W.}\ \bibnamefont
  {Robinson}},\ }\href {\doibase 10.1007/BF01645779} {\bibfield  {journal}
  {\bibinfo  {journal} {Comm. Math. Phys.}\ }\textbf {\bibinfo {volume} {28}},\
  \bibinfo {pages} {251} (\bibinfo {year} {1972})}\BibitemShut {NoStop}%
\bibitem [{\citenamefont {Bukov}\ \emph {et~al.}(2015)\citenamefont {Bukov},
  \citenamefont {D'Alessio},\ and\ \citenamefont {Polkovnikov}}]{bukov2015}%
  \BibitemOpen
  \bibfield  {author} {\bibinfo {author} {\bibfnamefont {M.}~\bibnamefont
  {Bukov}}, \bibinfo {author} {\bibfnamefont {L.}~\bibnamefont {D'Alessio}}, \
  and\ \bibinfo {author} {\bibfnamefont {A.}~\bibnamefont {Polkovnikov}},\
  }\href {\doibase 10.1080/00018732.2015.1055918} {\bibfield  {journal}
  {\bibinfo  {journal} {Adv. Phys.}\ }\textbf {\bibinfo {volume} {64}},\
  \bibinfo {pages} {139} (\bibinfo {year} {2015})}\BibitemShut {NoStop}%
\bibitem [{\citenamefont {Mikami}\ \emph {et~al.}(2016)\citenamefont {Mikami},
  \citenamefont {Kitamura}, \citenamefont {Yasuda}, \citenamefont {Tsuji},
  \citenamefont {Oka},\ and\ \citenamefont {Aoki}}]{mikami2016}%
  \BibitemOpen
  \bibfield  {author} {\bibinfo {author} {\bibfnamefont {T.}~\bibnamefont
  {Mikami}}, \bibinfo {author} {\bibfnamefont {S.}~\bibnamefont {Kitamura}},
  \bibinfo {author} {\bibfnamefont {K.}~\bibnamefont {Yasuda}}, \bibinfo
  {author} {\bibfnamefont {N.}~\bibnamefont {Tsuji}}, \bibinfo {author}
  {\bibfnamefont {T.}~\bibnamefont {Oka}}, \ and\ \bibinfo {author}
  {\bibfnamefont {H.}~\bibnamefont {Aoki}},\ }\href {\doibase
  10.1103/PhysRevB.93.144307} {\bibfield  {journal} {\bibinfo  {journal} {Phys.
  Rev. B}\ }\textbf {\bibinfo {volume} {93}},\ \bibinfo {pages} {144307}
  (\bibinfo {year} {2016})}\BibitemShut {NoStop}%
\bibitem [{\citenamefont {Sen}\ \emph {et~al.}(2021)\citenamefont {Sen},
  \citenamefont {Sen},\ and\ \citenamefont {Sengupta}}]{sen2021}%
  \BibitemOpen
  \bibfield  {author} {\bibinfo {author} {\bibfnamefont {A.}~\bibnamefont
  {Sen}}, \bibinfo {author} {\bibfnamefont {D.}~\bibnamefont {Sen}}, \ and\
  \bibinfo {author} {\bibfnamefont {K.}~\bibnamefont {Sengupta}},\ }\href
  {\doibase 10.1088/1361-648X/ac1b61} {\bibfield  {journal} {\bibinfo
  {journal} {J. Phys. Condens. Matter}\ }\textbf {\bibinfo {volume} {33}},\
  \bibinfo {pages} {443003} (\bibinfo {year} {2021})}\BibitemShut {NoStop}%
\bibitem [{\citenamefont {Fendley}(2012)}]{fendley2012}%
  \BibitemOpen
  \bibfield  {author} {\bibinfo {author} {\bibfnamefont {P.}~\bibnamefont
  {Fendley}},\ }\href {\doibase 10.1088/1742-5468/2012/11/P11020} {\bibfield
  {journal} {\bibinfo  {journal} {J. Stat. Mech.}\ }\textbf {\bibinfo {volume}
  {1211}},\ \bibinfo {pages} {P11020} (\bibinfo {year} {2012})}\BibitemShut
  {NoStop}%
\bibitem [{\citenamefont {Fendley}(2016)}]{fendley2015}%
  \BibitemOpen
  \bibfield  {author} {\bibinfo {author} {\bibfnamefont {P.}~\bibnamefont
  {Fendley}},\ }\href {\doibase 10.1088/1751-8113/49/30/30LT01} {\bibfield
  {journal} {\bibinfo  {journal} {J. Phys. A}\ }\textbf {\bibinfo {volume}
  {49}},\ \bibinfo {pages} {30LT01} (\bibinfo {year} {2016})}\BibitemShut
  {NoStop}%
\bibitem [{\citenamefont {Kemp}\ \emph {et~al.}(2017)\citenamefont {Kemp},
  \citenamefont {Yao}, \citenamefont {Laumann},\ and\ \citenamefont
  {Fendley}}]{kemp2017long}%
  \BibitemOpen
  \bibfield  {author} {\bibinfo {author} {\bibfnamefont {J.}~\bibnamefont
  {Kemp}}, \bibinfo {author} {\bibfnamefont {N.~Y.}\ \bibnamefont {Yao}},
  \bibinfo {author} {\bibfnamefont {C.~R.}\ \bibnamefont {Laumann}}, \ and\
  \bibinfo {author} {\bibfnamefont {P.}~\bibnamefont {Fendley}},\ }\href
  {\doibase 10.1088/1742-5468/aa73f0} {\bibfield  {journal} {\bibinfo
  {journal} {J. Stat. Mech.}\ }\textbf {\bibinfo {volume} {2017}},\ \bibinfo
  {pages} {063105} (\bibinfo {year} {2017})}\BibitemShut {NoStop}%
\bibitem [{\citenamefont {Yates}\ \emph {et~al.}(2020)\citenamefont {Yates},
  \citenamefont {Abanov},\ and\ \citenamefont {Mitra}}]{yates2020}%
  \BibitemOpen
  \bibfield  {author} {\bibinfo {author} {\bibfnamefont {D.~J.}\ \bibnamefont
  {Yates}}, \bibinfo {author} {\bibfnamefont {A.~G.}\ \bibnamefont {Abanov}}, \
  and\ \bibinfo {author} {\bibfnamefont {A.}~\bibnamefont {Mitra}},\ }\href
  {\doibase 10.1103/PhysRevLett.124.206803} {\bibfield  {journal} {\bibinfo
  {journal} {Phys. Rev. Lett.}\ }\textbf {\bibinfo {volume} {124}},\ \bibinfo
  {pages} {206803} (\bibinfo {year} {2020})}\BibitemShut {NoStop}%
\end{thebibliography}%

\section*{Appendix A: Numerical calculation of $\mathbb{A}^{-1}$ in case of degeneracies} 
\label{App:A}

It was mentioned in Sec.~\ref{sec2b} that $\mathbb{A}$ is made out of the eigenvectors of the Floquet eigenvectors $\frac{1}{\sqrt 2} \mathbf{v}_{j}^{\pm}$ along its rows. 
Specifically, we have $\mathbb{A}_{2j,m} = \frac{1}{\sqrt{2}} (\mathbf{v}^{+}_{j})_m$ and $\mathbb{A}_{2j+1,m} = \frac{1}{\sqrt{2}} (\mathbf{v}^{-}_{j})_m $, where $j = 
0,1, \ldots N-1$ and $m = 0,1, \ldots 2N-1$. Further, since the Floquet operator $U$ is entirely real in the Majorana language, the eigenvectors $\mathbf{v}^{+}_{j}$ and $\mathbf{v}^{-}_{j}$ corresponding to the eigenvalues $e^{-i\theta_{j}}$ and $e^{i\theta_{j}}$ 
respectively can be taken to be complex conjugate pairs. In addition, if all the eigenvectors are orthonormalized, the calculation for $\mathbb{A}^{-1}$ becomes 
simple. In that case we have $(\mathbb{A}^{-1})_{l,2j} = \sqrt{2} (\mathbf{v}^{-}_{j})_l $, and $(\mathbb{A}^{-1})_{l,2j+1} = \sqrt{2} (\mathbf{v}^{+}_{j})_l $, where $j = 0,1, \ldots N-1$ and $l = 0,1, \ldots 2N-1$. All 
the analytical calculations shown in the paper assume that these properties of the eigenvectors are satisfied. However, these do not necessarily hold in the case of a degeneracy, which happens if a zero quasienergy 
topological state is present. The numerical diagonalization procedure then produces a pair of eigenvectors which are in general neither complex 
conjugates nor orthonormal to each other. We describe below a method to obtain two orthonormalized vectors which are complex conjugates of each other starting from any 
two vectors having degenerate eigenvalues.

Suppose that $\mathbf{v}_1$ and $\mathbf{v}_2$ are two arbitrary degenerate eigenvectors. We then define
\begin{align}
\mathbf{u}_1 &= \frac{\mathbf{v}_1 + \mathbf{v}^{*}_1}{\langle \mathbf{v}_1 + \mathbf{v}^{*}_1 \mid \mathbf{v}_1 + \mathbf{v}^{*}_1 \rangle}, \non \\
\mathbf{u}_2 &= \frac{\mathbf{v}_2 + \mathbf{v}^{*}_2}{\langle \mathbf{v}_2 + \mathbf{v}^{*}_2 \mid \mathbf{v}_2 + \mathbf{v}^{*}_2 \rangle}, 
\end{align}
By construction, $\mathbf{u}_1$ and $\mathbf{u}_2$ are two real and 
normalized vectors. We now perform the Gram-Schmidt orthogonalization procedure using these two vectors to obtain a pair of orthonormal real vectors $\mathbf{w}_1$ and $\mathbf{w}_2$.
\begin{align}
\mathbf{w}_1 &= \mathbf{u}_1, \non \\
\mathbf{w}^{'}_2 &= \mathbf{u}_1 - \frac{\langle \mathbf{w}_1 \mid \mathbf{u}_2 \rangle}{\langle \mathbf{w}_1 \mid \mathbf{w}_1 \rangle} \mathbf{w}_1, \non \\
\mathbf{w}_2 &= \frac{\mathbf{w}^{'}_2}{\langle \mathbf{w}^{'}_2 \mid 
\mathbf{w}^{'}_2 \rangle}.
\end{align}
The final complex-conjugate pair of orthonormal vectors is then defined as
\begin{align}
\mathbf{\tilde{v}}_1 &= \frac{\mathbf{w}_1 + i \mathbf{w}_2}{\langle \mathbf{w}_1 + i \mathbf{w}_2 \mid \mathbf{w}_1 + i \mathbf{w}_2 \rangle}, \non \\
\mathbf{\tilde{v}}_2 &= \frac{\mathbf{w}_1 - i \mathbf{w}_2}{\langle \mathbf{w}_1 - i \mathbf{w}_2 \mid \mathbf{w}_1 - i \mathbf{w}_2 \rangle}.
\end{align}
We use $\mathbf{\tilde{v}}_1$ and $\mathbf{\tilde{v}}_2$ instead of $\mathbf{v}_1$ 
and $\mathbf{v}_2$ in our calculations whenever there is a degeneracy due to 
a zero quasienergy topological edge state.
\vspace*{.8cm}

\section*{Appendix B: Infinite-temperature two-point correlators for Majorana operators}
\label{App:B}

Here we will fill in the steps of the calculation from Eq.~\eqref{am} to Eq.~\eqref{2point}. For convenience, let us rewrite Eq.~\eqref{am} as
\begin{equation}
a_{m}(pT) ~=~ \sum_{j=0}^{N-1} ~( (\mathbb{A}^{-1})_{m,2j} ~e^{-ip \theta_j} \alpha_{j} + (\mathbb{A}^{-1})_{m,2j+1} ~e^{ip \theta_j}\alpha^{\dagger}_{j}).
\end{equation}
Substituting this in the two-point correlator $\braket{a_m(pT) a_n(qT)}$, we find that the calculation reduces to finding the expectation values of $\braket{\alpha_j 
\alpha^{\dagger}_{k}}$ and $\braket{\alpha^{\dagger}_{j} \alpha_{k}}$ in the 
infinite-temperature ensemble of many-body states, where $\alpha_{j}$ and 
$\alpha^{\dagger}_{k}$ are the particle creation and annihilation operators in the basis of single-particle states. We note that the expectation values 
$\braket{\alpha^{\dagger}_{j} \alpha^{\dagger}_{k}}$ and $\braket{\alpha_{j} \alpha_{k}}$ vanish. In terms of many-body eigenstates $\ket{\psi_{p}}$, we have
\begin{align}
\braket{\alpha_j \alpha^{\dagger}_{k}} ~=~ \frac{1}{2^{N}} ~\sum_{p =1}^{2^{N}} ~\langle \psi_{p} \mid \alpha_j \alpha^{\dagger}_{k} \mid \psi_{p} \rangle.
\end{align}
For this expression to be non-zero, we must first have $k = j$. Next, there should be some many-body Fock states, where the $j$-th single-particle state is empty. However, for this case, the rest of the $N-1$ single-particle states can be either 
filled or empty. Therefore there are $2^{N-1}$ many-body states which will contribute equally to the expectation value of $\alpha_j \alpha^{\dagger}_{j}$. So we can write 
\begin{align}
\braket{\alpha_j \alpha^{\dagger}_{k}} ~=~ \frac{1}{2^{N}} \sum_{p =1}^{2^{N}} \langle \psi_{p} \mid \alpha_j \alpha^{\dagger}_{k} \mid \psi_{p} \rangle ~=~ \frac{2^{N-1}}{2^{N}} \delta_{jk} ~=~ \frac{\delta_{jk}}{2}. \label{corr1}
\end{align}
Using a very similar argument we can show that
\begin{equation} \braket{\alpha^{\dagger}_{j} \alpha_{k}} ~=~ \frac{\delta_{jk}}{2}. \label{corr2} \end{equation}
Substituting Eqs.~\eqref{corr1} and \eqref{corr2} into the two-point correlator $\braket{a_m(pT) a_n(qT)}$, we obtain the 
expression for the Majorana two-point correlator given in Eq.~\eqref{2point}.

\end{document}